\documentclass[thmsa,a4paper,notitlepage,12pt]{article}
\usepackage{amssymb}

\usepackage{sw20lart}

\input{tcilatex}
\begin{document}

\author{J. Lacki\thanks{%
Histoire et Philosophie des Sciences, Universit\'{e} de Gen\`{e}ve.}, H.
Ruegg\thanks{%
D\'{e}partement de Physique Th\'{e}orique, Universit\'{e} de Gen\`{e}ve.}
and V. L. Telegdi\thanks{%
Caltech, Pasadena and CERN, Geneva.}}
\title{The Road to Stueckelberg's Covariant Perturbation Theory as Illustrated by
Successive Treatments of Compton Scattering}
\date{}
\maketitle

\begin{abstract}
We review the history of the road to a manifestly covariant perturbative
calculus within quantum electrodynamics from the early semi-classical
results of the mid-twenties to the complete formalism of Stueckelberg in
1934. We chose as our case study the calculation of the cross-section of the
Compton effect. We analyse Stueckelberg's paper extensively. This is our
first contribution to a study of his fundamental contributions to the
theoretical physics of twentieth century.
\end{abstract}

\tableofcontents

\section{Introduction}

One often considers the birth of quantum electrodynamics (QED) as one of the
most exciting and challenging chapters of the history of modern physics.
Though not a fundamentally new theory, like general relativity or quantum
mechanics, QED, as an effort to obtain a consistent relativistic treatment
of the interaction of matter with radiation, gave birth to most fertile
concepts of modern microphysics such as quantum fields, gauge symmetries and
renormalization techniques\footnote{%
For a philosophical analysis of the emergence of these concepts, see Cao
1997.}. These developments have already been covered in many excellent
studies and publications\footnote{%
For instance the penetrating accounts by A. Pais, \textit{Inward Bound. Of
Matter and Forces in the Physical Universe} (Pais 1986) and S. S. Schweber's 
\textit{Q.E.D. and the Men who made it, (} Schweber 1994)\textit{, }both
invaluable sources for contemporary history of particle physics.}. There is
however another aspect of the history of QED, equally symptomatic of a
definite tendency that culminated with the widely used Feynman diagrams. We
deal with a theory defined, maybe for the first time in such explicit terms,
by its associated perturbation expansion. Indeed, most of the tools and
techniques used were introduced in an attempt to give theoretical meaning
and practical usefulness to the successive terms of perturbative series.

Although classical electrodynamics had existed and had been successfully
applied for more than half a century, its closed-form fundamental laws
(Maxwell equations) were not adequate to yield a fully successful quantum
version by a straightforward process of quantization. Many difficulties had
to be overcome. On the one hand, there was the need to accommodate the
continuous field equations to the fundamentally discontinuous quantum
processes of energy and momentum exchange between photons and matter
particles. On the other hand, one faced the difficulties, already present in
the classical theory, of formulating a consistent relativistic theory of
several interacting particles.

The first step on the road to a full-fledged quantum theory was the
quantization of the radiation field itself. The 1927 papers of Dirac opened
the way. He showed how to quantize the radiation field as an assembly of
non-interacting bosons and proposed a perturbation scheme enabling one to
deal with its time-dependent coupling to matter. His method of quantization
made possible the first quantum treatment of emission-absorption processes
and provided a decisive step towards a genuine quantum theory of
photon-electron interaction. Until the advent of the modern S-matrix
approach within quantum field theory\footnote{%
For an analysis of the S-matrix approach see Cushing 1990.} at the end of
the forties, the spirit of this method was very influential\footnote{%
The other important source of inspiration was the general theory of
quantized fields by Heisenberg and Pauli 1929 and 1930; Fermi 1932
simplified the quantization of the electromagnetical field.}, supplemented
by Dirac's equally famous relativistic electron equation, as the main
theoretical tool of investigation in the study of matter-radiation
interaction.

Following Dirac's pioneering theory, one could distinguish several lines of
research. The main one dealt with a consistent formulation of a relativistic
theory of particles interacting with and through the radiation field. Before
the advent of quantum mechanics, the problem was tackled in a
''correspondence principle'' way as illustrated for instance by the
achievement of Kramers and Heisenberg (1925)\footnote{%
For an account of the old quantum theory, see e.g. Hund 1967; Jammer 1966,
and Mehra and Rechenberg 1982-87, vol.1.}. The quantum evolution equations
of matrix and wave mechanics were the foundations of semi-classical
treatments, like those of Dirac (1926b and 1927a), of Gordon (1926) (for
spin 0 particles), and of Klein and Nishina (1929) (spin 1/2 electrons). The
fundamental insights of Dirac (Dirac 1932) enabled further progress,
resulting in the interaction picture (Dirac, Fock, Podolsky 1932, see
however already Dirac 1927b) and the multiple-time formalism. Of course, the
problem couldn't find a complete solution until it was realized that both
radiation and matter had to be expressed as quantized fields. The
point-particle singularities were then ''dissolved'' into field excitations
and the problem of coupling consistently matter to radiation transformed
into that of field-field coupling. This modern standpoint was initiated by
the work of Pauli and Heisenberg (Heisenberg and Pauli 1929 and 1930) and
preceded by Jordan's insights (Jordan 1927a,b). Along this fundamental
thread, one had to provide a systematic way of generating and organizing the
successive terms of the perturbative series without loosing the manifest
relativistic invariance and other symmetries (gauge invariance)
characteristic of the full equations. We shall focus our attention on these
issues, leaving aside the further problem of the infinities arising in the
higher-order processes, such as the self-energy or vacuum polarization,
which constitute another important chapter of the history of QED\footnote{%
For historical accounts of the problem of infinities in the thirties, see e.
g. Pais 1986, chapter 16, Schweber1994, chapter 1,2; Brown 1993. For
personal recollections of actors see for example Wentzel 1960 and Weisskopf
1983.}.

In what follows we shall attempt to sketch an outline of the various stages
in the attainment of a manifestly covariant perturbation calculus of QED
processes. As an illustration for our study we shall choose the case of
Compton and related effects to lowest order\footnote{%
For a complete historical study of the Compton effect, see Stuewer 1975. The
importance of these processes was pointed out by Brown (1993), who after
discussing divergence difficulties makes the following comment: ''What
really established QED as a usable theory of processes at high energy (as
well at the lower energies available in the laboratory) was the calculation
of the so-called ''shower processes'': bremsstrahlung (i.e. continuous x-ray
production), pair production by a high-energy photon in the Coulomb field of
a nucleus, and electron pair annihilation [and of course Compton
scattering].'', p. 9. A historical account of the related experiments is
given in Cassidy 1981.}. Here, by Compton effect, we mean mainly the
scattering of a photon on a free electron; indeed in the twenties, the
Compton effect covered also the case of the scattering of high-energy
photons on bound electrons, where the latter could be considered as free. We
shall discuss the early results of Dirac, Gordon, and Klein and Nishina,
then shall consider the generalized Klein-Nishina type formulas of Waller
(1930) and Tamm (1930), and finally shall end up with the results of
Stueckelberg (1934). The contribution of Stueckelberg shall be examined in
closer detail, as we believe it constitutes the first complete and easily
generalizable instance of a manifest relativistically invariant perturbative
calculus. It is of interest in many respects, both technical and
epistemological. The ''modernity''\ of this contribution is striking and
characteristic of Stueckelberg's original turn of mind. From this point of
view, the present study can be considered as a first stage in an attempt at
drawing the attention of historians to Stueckelberg's achievements during
those years.

Our choice of the landmarks on the road towards Stueckelberg's 1934 paper is
accordingly motivated by the need to illustrate both the key challenges as
well as the main theoretical inputs which motivated Stueckelberg's approach.
Thus, we start with a brief reminder of the ''infamous'' Bohr, Kramers and
Slater proposal (Bohr, Kramers, Slater 1924), which can be considered as the
final expression of the fundamental impotence of the ''old'' quantum physics
as regards the emerging problem of matter-radiation interaction followed by
a brief discussion of the Heisenberg-Kramers dispersion formulas. We next
present two early Dirac papers on the Compton effect first treated with the
matrix formalism and then with wave mechanics, a progression characteristic
of the whole of quantum studies of the time (Dirac 1926b, 1927a). Gordon's
1926 contribution is crucial in understanding the Klein-Nishina 1929 paper,
both papers sharing the same strategy and scope. Dirac's 1927 field
quantization and time-dependent perturbation theory (Dirac 1926c, 1927b) are
the essential techniques used subsequently by Waller and Tamm.
Stueckelberg's 1934 paper took also advantage from Dirac's reflection (Dirac
1932) on the backreaction problem, which lead among other thing to the
interaction picture of the later joint work of Fock, Dirac and Podolsky. We
end up the paper with a discussion of the little interest met by
Stueckelberg's work at his time.

Throughout the paper we have tried to adopt a unified (modern) notation with
some exceptions when there was a danger of betraying the spirit of the
original paper. We report in the footnotes the most important changes with
respect to the original versions, and discuss our conventions in the
Appendix .

\section{Preliminaries: the end of the old quantum theory, the BKS crisis
and the Heisenberg-Kramers dispersion formula}

\subsection{Introduction}

By the mid-twenties, the old quantum theory of the founding fathers, Planck,
Einstein, Bohr and Sommerfeld was spectacularly extended by powerful new
techniques which provided a formal framework for many of the ''quantum
rules''\ used, with much brio, in the blossoming field of atomic physics.
Quantum mechanics, first in its matrix form, shortly followed by the wave
version, was born. It offered the promise of a unified treatment of many
problems and a way of solving long-standing difficulties. Indeed, in spite
of undisputed successes of the old theory, persistent difficulties were
taking increasing importance. Thus, after Bohr's successful theory of the
atom, there remained the problem of explaining the dynamics of quantum
transitions between atomic levels and more generally that of the interaction
of radiation with matter. The classical continuous wave picture of
radiation, after being challenged by Einstein in 1905 with his light-quantum
interpretation of the photo-electric effect , was again questioned in 1922
this time by Compton who studied X-ray scattering. It was known since the
work of D. C. H. Florance in 1910 and J. A. Gray in 1913 that the secondary
radiation produced by irradiating a metal with gamma rays is ''softer''
(less penetrating) than the primary radiation. After his experiments in
1921, Compton, who realized that there was a shift of frequency, thought
that a wave-model interpretation could still hold (Compton 1921). Facing the
evidence of his subsequent experiments, he progressively changed his mind
and proposed an explanation (Compton 1923) in terms of collision processes
between individual electrons and light-quanta\footnote{%
Debye reached the same conclusions at approximately the same time see Debye
1923.}. Cloud-chamber experiments by C. T. R. Wilson and W. Bothe (Wilson
1923, Bothe 1923) showed tracks of the recoil electrons which were
consistent with the light-quantum hypothesis\footnote{%
For a review of the theory and experiments, see Compton and Allison (1935).}%
. Still, the latter was far from being accepted and several first-rank
physicists kept trying to accommodate a continuous wave type explanation to
the discrete reality, using more or less modified forms of classical
electrodynamics.

One of the fiercest opponents to the light-quantum nature of radiation was
Niels Bohr. In an attempt to push the old wave scheme to its limits, Bohr,
Kramers and Slater (BKS) published a paper (Bohr Kramers and Slater 1924)%
\footnote{%
See Van der Waerden 1968 for a study and an english translation of the
paper. The reader will also find in Max Dresden's (1987) biography of
Kramers an insightful account of this and other episodes of the old quantum
theory.} where they followed up on an earlier idea of Slater (Slater 1922).
The latter suggested to view the atom in a stationary state as an assembly
of ''virtual'' harmonic oscillators with eigenfrequencies determined by the
transitions between stationary states. In genuine ''theoretical despair'',
Bohr, Kramers and Slater went as far as to question the validity of the
energy conservation principle at the level of individual atomic processes,
and replaced it by a milder version, holding only at the statistical level,
a move apparently due to Bohr and Kramers and originally rejected by Slater.
Part of Bohr's motivation was then to save the wave model of radiation,
which he believed essential for the explanation of the Compton effect, since
the latter implied an increase of the wave-length, the measurement of which
presupposed, in his opinion, the wave picture\footnote{%
See Murdoch 1987.}. Secondary radiation thus resulted in his view from a
continuous process of emission of coherent waves from each illuminated atom.

The empirical refutation of the BKS proposal came quickly. Indeed, in the
BKS theory, the electron recoil (which paradoxically still made sense in
this approach) would occur with finite probability in \textit{any}
direction, whereas according to the light-quantum hypothesis, it would 
\textit{strictly} depend on that of the incident radiation. The validity of
the conservation principles in the individual collisions was established
beyond doubt by Bothe and Geiger (1925) on one hand, and by Compton and
Simon (1925) on the other.

Be it as it may, Compton scattering vindicated the power of conservation
principles and joined the ranks of other well-established processes
involving light quanta and electrons, such as Bremsstrahlung, or the
photo-electric effect, which were all in 1924 still lacking a fully
satisfactory quantum treatment because of the absence of a genuine quantum
starting point.

\subsection{The Heisenberg-Kramers dispersion relations and the first
attempts}

In the quest for a quantitative approach, an important step was however
taken by Kramers and Heisenberg [Kramers and Heisenberg 1924] who
established a dispersion formula describing the features of the radiation
reemitted by an atom after illumination by primary radiation\footnote{%
See again Van der Waerden 1968.}. The authors were able to achieve a
quantitative result but their method excluded the possibility to be applied
to more energetic regimes, those characteristic for instance of light
scattering off unbound electrons. Beyond its intrinsic interest, the
Heisenberg-Kramers dispersion formula became \textit{\ un passage oblig\'{e}}
in every subsequent study, a standard to be reattained and a testing ground
for any new technique, only to be replaced in this role by the Klein-Nishina
formula at the end of the twenties. The approach of Kramers and Heisenberg
was based purely on the correspondence principle and constitutes a brilliant
example of the surprisingly powerful method the latter constituted in such
expert hands. Together with similar approaches by Smekal 1923, Breit 1926,
Wentzel 1927, and others, it can be considered as the outmost excursion of
the old theory into a territory which could be fully explored only with
radically new tools which were soon to come.

The Kramers-Heisenberg paper starts with an analysis of the then current
standpoint on the problem of dispersion of radiation by atoms. After
recalling the usual treatment of the coherent secondary radiation,
understood (classically) as the result of the harmonic motion of the
electrons excited by the external radiation, the authors mention the
difficulty of analyzing this effect within the quantum theoretical picture
of electronic transitions between stationary states:

\begin{quotation}
We are thus faced with the problem of describing the scattering and
dispersion effects of the atom in terms of the quantum theoretical picture
of atomic structure. According to this picture, the appearance of a spectral
line is not linked with the presence of elastically oscillating electrons,
but with transitions from one stationary state to another. Bohr's
correspondence principle however gives us a significant hint on how to
describe the reaction of the atom to the radiation field, using classical
concepts. In a recently published paper by Bohr, Kramers and Slater, the
authors sketched in rough outline how such a description can be accomplished
in a relatively simple manner. What is above all typical for this theory, is
the assumption that the reaction of the atom to the radiation field should
primarily be understood as a reaction of an atom existing in a given
stationary state; it is also assumed that transitions between two stationary
states are of very short duration, and that the detailed nature of these
transitions will not play any role in the description of the optical
phenomena. \footnote{%
Kramers-Heisenberg (1925), p. 682. We rely here and later on the English
translation in Van der Waerden (1968), pp. 223-252.}
\end{quotation}

Folowing this path, it is assumed that, under the influence of an external
radiation, given by an oscillating electric field with amplitude $\mathbf{E}
_{0}$and frequency $\nu $%
\begin{equation}
\mathbf{E}(t)=\func{Re}\left\{ \mathbf{E}_{0}e^{i2\pi \nu t}\right\} ,
\label{kh1}
\end{equation}
as long as the atom remains in a stationary state, say, of energy $E_{k}$ it
reemits spherical waves, corresponding to an oscillating dipole moment 
\begin{equation}
\mathbf{D}(t)=\func{Re}\left\{ \mathbf{d}e^{i2\pi \nu t}\right\} ,
\label{kh2}
\end{equation}
where the vector $\mathbf{d}$ is proportional to $\mathbf{E}_{0}\mathbf{.}$
However, if one takes further into account the (spontaneous) transitions to
states of lower energy $E_{l}$, the atom radiates, in addition to the
coherent emission (\ref{kh2}), spherical waves with frequencies given by
Bohr's conditions $\nu _{q}=(E_{k}-E_{l})/h.$ Interferences between the
coherent and the spontaneous radiation should then lead to a modification of
the secondary radiation. The form of these modifications as affecting the $%
\nu -$ dependence had been investigated by Kramers (1924) relying on
classical dispersion theory and the correspondence principle. There,
spontaneous emission was analyzed in terms of classical oscillators with
eigen-frequencies $\nu _{q}$ and the actual (quantum) behaviour obtained
taking the limit of high quantum numbers. The paper of Kramers and
Heisenberg further extended the analysis of Kramers:

\begin{quotation}
It is the purpose of this paper to show how the correspondence idea, when
pursued more closely, leads to the surprising result that the assumption (%
\ref{kh2}) for the reaction of an atom to incident radiation is too
restricted, and will in general have to be extended by adding a series of
terms as follows: 
\begin{equation}
\mathbf{D}(t)=\func{Re}\left\{ \mathbf{d}e^{i2\pi \nu t}+\sum_{k}\mathbf{d}
_{k}e^{i2\pi (\nu +\nu _{k})t}+\sum_{l}\mathbf{d}_{l}e^{i2\pi (\nu -\nu
_{l})t}\right\}  \label{kh8}
\end{equation}

where $h\nu _{k}$ or $h\nu _{l}$ denote the energy difference between two
stationary states of the atom, one of which is always identical with the
momentary state of the atom; the vectors $\mathbf{d}_{k}$ and $\mathbf{d}%
_{l} $ again depend on $E$ (in the form of a linear vector function) and on $%
\nu $ . Expressed in words, the result can be stated as follows\textit{:
Under the influence of irradiation with monochromatic light, an atom not
only emits coherent spherical waves of the same frequency as that of the
incident light: it also emits systems of incoherent spherical waves, whose
frequencies can be represented as combinations of the incident frequency
with other frequencies that correspond to possible transitions to other
stationary states}. Such additional systems of spherical waves must
evidently occur in the form of scattered light; but they cannot make any
contribution to the dispersion and absorption of the incident light.%
\footnote{%
See van der Waerden translation (1968), p. 227. The Kramers-Heisenberg
result, valid in the long-wave approximation, was generalized to low energy
x-rays and bound electrons by Wentzel 1927, using non-relativistic quantum
mechanics. See also Waller 1929.}
\end{quotation}

The analysis of Kramers and Heisenberg does not rely on the concept of light
quanta, which as we have mentionned, was still considered by some as
unwarranted. However, their result is related to some earlier work by Smekal
(1923) who explicitely used the corpuscular hypothesis. Smekal's paper is
remarkable, not only because of his use of the corpuscular hypothesis, but
also because it conjectured a new effect, incoherent scattering of light by
atoms\footnote{%
One would nowadays speak of inelastic scattering.}. This is what Kramers and
Heisenberg tell us above. This effect, nowadays called Raman scattering, was
confirmed experimentally in 1928\footnote{%
See Raman (1928), Raman and Krishnan (1928), and also Landsberg and
Mandelstam 1928.}.

A more detailed discussion of the Heisenberg-Kramers would be here out of
place; we wanted merely to illustrate the spirit of an approach using the
correspondence principle, the latter, when coupled to the general theory of
multiply periodic systems (not discussed here\footnote{%
For an excellent survey, one can consider the book of Friedrich Hund, 
\textit{Geschichte der Quantentheorie}, Hoschultaschenb\"{u}cher,
Bibliographisches Institut, Mannheim 1967. See also Dresden (1987).}), being
then the sole systematic theoretical tool in the obtention of quantum
results before the breakthrough of quantum mechanics.

\section{Semi-classical attempts to Compton scattering}

\subsection{Introduction}

In the first attempts at studying matter-radiation interaction, as in the
problem of illuminating an atom with radiation, the standpoint was to
consider, the re-emitted radiation as a result of some internal process
involving the electrons within a well-defined regime, described as a
classical multiply-periodic system to which the quantization rules could
then further be applied with the guidance of the correspondence principle.
The situation changed drastically with the advent of quantum mechanics and
the availability of genuine quantum evolution equations. Then, the
interaction of electrons with radiation could be, at least in principle,
handled directly. In the first stage, where the radiation field wasn't
quantized yet, one had to use semi-classical methods deriving the features
of the secondary radiation from the classical Maxwell formulas supplemented
by the quantum expressions for charge and current densities.

In what follows, we shall review several contributions which we chose to
group together because they all share a peculiar interplay of classical and
quantum techniques. None of the authors (including Klein and Nishina) were
working consistently in a purely quantum mechanical setting, until the
contributions of Waller and especially that of Tamm in 1930, which were
already part of the next stage, starting with the field quantization
breakthrough of Dirac in 1927. This interplay illustrates how young the
quantum formalism was in those years and how dominant the influence of the
correspondence principle still was.

\subsection{Dirac: first uses of the new quantum mechanics}

We start our story with a publication of Dirac (Dirac 1926b, received 29
April 1926), who applied Heisenberg's matrix formalism to provide \textit{a
first treatment of the Compton scattering} within the setting of the newly
born quantum mechanics\footnote{%
It is interesting to mention also a paper by Schr\"{o}dinger (1927) where he
obtains the conservation law of the Compton scattering using the analogy
between radiation ''scattering'' off the square of the wave function, and
radiation scattered off a pressure wave, as solved by Brillouin.}. In this
paper, as in the preceding ones (Dirac 1925 and 1926a), considerable effort
is devoted to the development of a universal q-calculus, wherein Born,
Heisenberg and Jordan's original formalism is systematized and generalized
(Born, Heisenberg and Jordan 1926). His treatment tries moreover to build
relativity into the matrix formalism, a rather cumbersome task because of
the necessarily privileged role of time in the canonical formalism. The key
idea is to use in those terms of the matrix elements which are related to an
oscillation of angular frequency $\omega =2\pi \nu $, instead of the non
covariant expressions $\exp (i\omega t)$, the relativistic expressions $\exp
(i\omega t^{\prime }),$ where 
\begin{equation}
t^{\prime }=t-\frac{(l_{1}x_{1}+l_{2}x_{2}+l_{3}x_{3})}{c}  \label{dirtime}
\end{equation}
Here, $(l_{1},l_{2},l_{3})$ is the unit vector in the direction of
observation and $x_{i}\quad i=1,2,3$ the coordinates of the electron.
Clearly, $(l_{1},l_{2},l_{3})$ will eventually correspond to the direction
of the scattered (secondary) photon. The 1926 paper achieves this program
using only the tools of canonical commutation relations of operators%
\footnote{%
The paper was apparently not easy to understand, to witness Klein (1968) p.
79: ''One day Ehrenfest brought us [Uhlenbeck and Goudsmit] a new paper by
Dirac containing a quantum dynamical theory of the Compton effect which we
tried to read, but without success''. The same Ehrenfest wrote to Dirac (in
Kragh 1990, p. 53): ''Einstein is currently in Leiden (until Oct. 9th$).$ In
the few days we have left, he, Uhlenbeck and I are struggling together for
hours at a time studying your work, for Einstein is eager to understand it.
But we are hitting at a few difficulties, which - because the presentation
is so short - we seem absolutely unable to overcome.''}. Dirac will publish
a second paper (1927a, received 22 Nov. 1926) where he reaches the same goal
with the help of wave mechanics. The physical treatment of both papers
shares the same overall strategy. However, the use of wave mechanics makes
the calculations easier because, according to Dirac:

\begin{quotation}
This [matrix mechanics] method is rather artificial; particularly so since
on the quantum theory there is no unique way of writing the Hamiltonian
equation [...] (owing to the ambiguity in the order of the factors in $H$),
and it has to be proved that all reasonable ways of doing so lead to the
same results. A more natural and more easily understood method of obtaining
the matrices is provided by Schr\"{o}dinger's wave mechanics\footnote{%
Dirac (1927a), p. 500.}.
\end{quotation}

Indeed, instead of using quantum equations for canonical operators
compatible with the variable $t^{\prime }$, one can in the wave formalism
simply work at the level of the solutions $\psi $ of the wave equation.
Assuming known a set of such solutions $\psi _{\alpha }$parametrized by some
quantum numbers (collectively denoted $\alpha )$ the matrix elements of any
dynamical variable $C$ are obtained through the expansion 
\begin{equation}
C\psi _{\alpha }=\sum_{\alpha ^{\prime }}C_{\alpha \alpha ^{\prime }}\psi
_{\alpha ^{\prime }}  \label{direxp}
\end{equation}
where the matrix elements $C_{\alpha \alpha ^{\prime }}$ can then be
expressed as functions of $t^{\prime }.$ The off-diagonal matrix elements
are further interpreted as related to the transition from state $\alpha $ to
state $\alpha ^{\prime },$ as initiated in Heisenberg (1925), Born,
Heisenberg and Jordan (1926), and in Dirac's own papers, as for instance
Dirac (1926a). This is precisely the road that Dirac will follow in his
second paper as we discuss it now\footnote{%
The reader will find a general discussion of Dirac's first paper in Mehra
and Rechenberg 1982-87, vol. 4, p. 196-213.}.

One considers, in the presence of an electromagnetic potential , the
relativistic equation for a charged particle 
\begin{equation}
(p^{\nu }-\frac{e}{c}A^{\nu })(p_{\nu }-\frac{e}{c}A_{\nu })=-m^{2}c^{2},
\label{dircommat_1}
\end{equation}
where the four vector $A=(\mathbf{A},A_{0}),\mathbf{A}$ is the vector and $%
A_{0}$ the scalar potential, $e$ the electron charge, $c$ the speed of light
(the summation convention over $\nu =0,1,2,3$ is applied throughout). This
yields, after the (Schr\"{o}dinger) substitutions 
\begin{equation}
p_{i}\rightarrow -i\hbar \frac{\partial }{\partial x_{i}}\equiv \hat{p}%
_{i};cp_{0}=E\rightarrow i\hbar \frac{\partial }{\partial t}\equiv \hat{E}
\label{schršdsub}
\end{equation}
the wave equation 
\begin{equation}
\left\{ m^{2}c^{2}-\frac{\hat{E}^{2}}{c^{2}}+\hat{p}_{1}^{2}+(\hat{p}%
_{2}+a^{\prime }\cos k_{1}(ct-x_{1}))^{2}+\hat{p}_{3}^{2}\right\} \psi =0
\label{waveq}
\end{equation}
where Dirac specializes to the case $A_{1}=A_{3}=A_{0}=0,$and $A_{2}=a\cos
k_{1}(ct-x_{1})$ with $k_{1}=\omega /c$, the only non zero component of the
wave vector $\mathbf{k}=(\omega /c,0,0)$ giving the direction of propagation
of the electromagnetic wave\footnote{%
Dirac's notation for $\omega /c$ is $\nu $ and his $h$ is actually $\hbar .$}%
; moreover, $a^{\prime }=ae/c$. The solutions to (\ref{waveq}) will be
obtained under the the assumption that the amplitude $a$ is small so that
higher order terms in $a$ can be neglected. Dirac first performs the
(canonical) transformations (in order not to overload the notations, we drop
here and in the sequel the carot distinguishing the operators $\hat{p}$ from
the c-numbers $p)\footnote{%
In Dirac's paper the new variables are primed.}$%
\begin{eqnarray}
\tilde{x}_{1}=ct-x_{1}; &&p_{1}=-\tilde{p}_{1}+l_{1}\frac{\tilde{E}}{c} 
\nonumber \\
\tilde{x}_{2}=x_{2}; &&p_{2}=\tilde{p}_{2}+l_{2}\frac{\tilde{E}}{c}
\label{cantransf} \\
\tilde{x}_{3}=x_{3}; &&p_{3}=\tilde{p}_{3}+l_{3}\frac{\tilde{E}}{c} 
\nonumber \\
\tilde{t}=t-(l_{1}x+l_{2}x_{2}+l_{3}x_{3})/c; &&E=\tilde{E}-c\tilde{p}_{1} 
\nonumber
\end{eqnarray}
These transformations enable to separate the variables because the wave
equation does not involve now explicitly $\tilde{x}_{2},\tilde{x}_{3}$ and $%
t^{\prime }$: 
\begin{equation}
\begin{array}{c}
(m^{2}c^{2}+2\frac{\tilde{E}}{c}\left\{ (1-l_{1})\tilde{p}_{1}+l_{2}\tilde{p}%
_{2}+l_{3}\tilde{p}_{3}+l_{2}a^{\prime }\cos k_{1}\tilde{x}_{1}\right\} + \\ 
\tilde{p}_{2}^{2}+\tilde{p}_{3}^{2}+2a^{\prime }\tilde{p}_{2}\cos k_{1}%
\tilde{x}_{1})\psi =0.
\end{array}
\label{waveqsub}
\end{equation}
Consequently one can set the general solution to be\footnote{%
Dirac writes the eigenvalues $\mathtt{\tilde{p}}_{2},\mathtt{\tilde{p}}_{3},%
\mathtt{\tilde{E}}$ of the operators $\tilde{p}_{2},\tilde{p}_{3}$ and $%
\tilde{E}$ as $\alpha _{2},\alpha _{3}$ and $\alpha _{4}.$} 
\begin{equation}
\psi _{(\mathtt{\tilde{p}}_{2},\mathtt{\tilde{p}}_{3},\mathtt{\tilde{E}}%
)}=e^{i\mathtt{\tilde{p}}_{2}\tilde{x}_{2}/\hbar }e^{i\mathtt{\tilde{p}}_{3}%
\tilde{x}_{3}/\hbar }e^{-i\mathtt{\tilde{E}}\tilde{t}/\hbar }\chi (\tilde{x}%
_{1}).  \label{separ}
\end{equation}
here$,\mathtt{\tilde{p}}_{2},\mathtt{\tilde{p}}_{3},\mathtt{\tilde{E}}$ are
the eigenvalues of the operators $\tilde{p}_{2},\tilde{p}_{3}$ and $\tilde{E}%
.$ The substitution of this ansatz into (\ref{waveqsub}) enables to
determine the function $\chi (\tilde{x}_{1})$: 
\begin{eqnarray}
\chi (\tilde{x}_{1}) &=&\exp \frac{-i}{(1-l_{1})\hbar }\left\{ (l_{2}\mathtt{%
\tilde{p}}_{2}+l_{3}\mathtt{\tilde{p}}_{3}+\frac{cb}{2\mathtt{\tilde{E}}})%
\tilde{x}_{1}+\frac{a^{\prime }}{k_{1}}(l_{2}+\frac{c\mathtt{\tilde{p}}_{2}}{%
\mathtt{\tilde{E}}})\sin k_{1}\tilde{x}_{1}\right\}  \label{final} \\
b &=&m^{2}c^{2}+\mathtt{\tilde{p}}_{2}^{2}+\mathtt{\tilde{p}}_{3}^{2} 
\nonumber
\end{eqnarray}
We have now a set of solutions $\psi _{(\mathtt{\tilde{p}}_{2},\mathtt{%
\tilde{p}}_{3},\mathtt{\tilde{E}})}$ each parametrized by the quantum
numbers $\mathtt{\tilde{p}}_{2},\mathtt{\tilde{p}}_{3},\mathtt{\tilde{E}}.$
corresponding to the motion of the electron in the external radiation. To
study the features of the secondary radiation, one has now to determine the
relevant observables. Since classically the secondary radiation is related
to the oscillation of the electron, the latter are chosen to be the
amplitudes of vibration along (a basis) of axes chosen perpendicular to the
direction of observation $(l_{1},l_{2},l_{3})$. Dirac sets\footnote{%
Multiplied by the charge they represent the components of the polarisation
of the system, a terminology used by Dirac.}: 
\begin{eqnarray*}
X &=&l_{3}x_{1}-\frac{l_{2}l_{3}}{1-l_{1}}x_{2}+(\frac{l_{2}^{2}}{1-l_{1}}%
-l_{1})x_{3} \\
Y &=&l_{2}x_{1}+(\frac{l_{3}^{2}}{1-l_{1}}-l_{1})x_{2}-\frac{l_{2}l_{3}}{%
1-l_{1}}x_{3}
\end{eqnarray*}
One can express further $X$ and $Y$ (as acting on the initial state) with
the help of the constants of motion which enables to make explicit their $%
t^{\prime }$ dependence: 
\begin{eqnarray}
(1-l_{1})X &=&c_{3}+\frac{2\mathtt{\tilde{p}}_{3}\mathtt{\tilde{E}}}{b}%
c_{4}+l_{3}c\tilde{t}+\frac{2(1-l_{1})\mathtt{\tilde{p}}_{3}\mathtt{\tilde{E}%
}}{b}\tilde{t}-\frac{2ca^{\prime }\mathtt{\tilde{p}}_{2}\mathtt{\tilde{p}}%
_{3}}{k_{1}b\mathtt{\tilde{E}}}\sin k_{1}\tilde{x}_{1}.  \label{polmat} \\
(1-l_{1})Y &=&c_{2}+\frac{2\mathtt{\tilde{p}}_{2}\mathtt{\tilde{E}}}{b}%
c_{4}+l_{2}c\tilde{t}+\frac{2(1-l_{1})\mathtt{\tilde{p}}_{2}\mathtt{\tilde{E}%
}}{b}\tilde{t}+(\frac{ca^{\prime }}{k_{1}\mathtt{\tilde{E}}}-\frac{%
2ca^{\prime }\mathtt{\tilde{p}}_{2}^{2}}{k_{1}b\mathtt{\tilde{E}}})\sin k_{1}%
\tilde{x}_{1}  \nonumber
\end{eqnarray}
The knowledge of the constants of motion $c_{2},c_{3},c_{4}$ above is not
material here so that we refer reader to Dirac's paper, p. 503 for their
exact expression.

We know that only the oscillating terms contribute to the emitted radiation.
Inspection of (\ref{polmat}) shows that we have to concentrate on $\sin k_{1}%
\tilde{x}_{1}.$ According to the scheme explained before, an off-diagonal
matrix element of this term will contain the information pertaining to the
secondary radiation corresponding to the transition from initial state ($%
\mathtt{\tilde{p}}_{2},\mathtt{\tilde{p}}_{3},\mathtt{\tilde{E})}$ to a
final one ($\mathtt{\tilde{p}}_{2}^{\prime },\mathtt{\tilde{p}}_{3}^{\prime
},\mathtt{\tilde{E}}^{\prime }\mathtt{)}.$ To obtain the general matrix
element, one acts on the (initial) state $\psi _{(\mathtt{\tilde{p}}_{2},%
\mathtt{\tilde{p}}_{3},\mathtt{\tilde{E})}}$with the operator $\exp ik_{1}%
\tilde{x}_{1}.$and observes that (compare with (\ref{separ})) 
\begin{eqnarray*}
&&\exp ik_{1}\tilde{x}_{1}\cdot \psi _{(\mathtt{\tilde{p}}_{2},\mathtt{%
\tilde{p}}_{3},\mathtt{\tilde{E})}} \\
&=&\exp \frac{i}{\hbar }\left\{ (\tilde{x}_{2}\mathtt{\tilde{p}}_{2}+\tilde{x%
}_{3}\mathtt{\tilde{p}}_{3}-\mathtt{\tilde{E}}\tilde{t})-(\frac{l_{2}\mathtt{%
\tilde{p}}_{2}+l_{3}\mathtt{\tilde{p}}_{3}+\frac{cb}{2\mathtt{\tilde{E}}}}{
(1-l_{1})}+\hbar k_{1})\tilde{x}_{1}\right\} \\
&\equiv &\exp (-i\omega ^{\prime }\tilde{t})\exp \frac{i}{\hbar }\left\{ (%
\tilde{x}_{2}\mathtt{\tilde{p}}_{2}+\tilde{x}_{3}\mathtt{\tilde{p}}_{3}-%
\mathtt{\tilde{E}}^{\prime }\tilde{t})-\frac{(l_{2}\mathtt{\tilde{p}}
_{2}+l_{3}\mathtt{\tilde{p}}_{3}+\frac{cb}{2\mathtt{\tilde{E}}^{\prime }})%
\tilde{x}_{1}}{(1-l_{1})}\right\}
\end{eqnarray*}
where we have set $\omega ^{\prime }=2\pi \nu ^{\prime }.$ In this
calculation, one has again neglected higher terms in $a^{\prime }$. We have
then the fundamental result 
\[
\exp (ik_{1}\tilde{x}_{1})\cdot \psi _{(\mathtt{\tilde{p}}_{2},\mathtt{%
\tilde{p}}_{3},\mathtt{\tilde{E})}}=\exp (-i\omega ^{\prime }\tilde{t})\psi
_{(\mathtt{\tilde{p}}_{2}^{\prime },\mathtt{\tilde{p}}_{3}^{\prime },\mathtt{%
\tilde{E}}^{\prime }\mathtt{)}} 
\]
with the only non-vanishing matrix element for $(\mathtt{\tilde{p}}_{2},%
\mathtt{\tilde{p}}_{3},\mathtt{\tilde{E})\rightarrow }(\mathtt{\tilde{p}}%
_{2}^{\prime },\mathtt{\tilde{p}}_{3}^{\prime },\mathtt{\tilde{E}}^{\prime }%
\mathtt{)}$ being $\exp (-i\omega ^{\prime }\tilde{t}).$ Notice that this
matrix element is taken between electron states which correspond \textit{both%
} to solutions of involving the \textit{initial} radiation field. Here, 
\begin{eqnarray}
\mathtt{\tilde{p}}_{2}^{\prime } &=&\mathtt{\tilde{p}}_{2}  \label{matel} \\
\mathtt{\tilde{p}}_{3}^{\prime } &=&\mathtt{\tilde{p}}_{3}  \nonumber \\
\mathtt{\tilde{E}}^{\prime }+\hbar \omega ^{\prime } &=&\mathtt{\tilde{E}} 
\nonumber \\
\frac{cb}{2\mathtt{\tilde{E}}^{\prime }} &=&\frac{cb}{2\mathtt{\tilde{E}}}
+(1-l_{1})\frac{\hbar \omega }{c}  \nonumber
\end{eqnarray}
Concerning $\mathtt{\tilde{p}}_{1},$ it can be obtained, relying on the
analogy with the free case, as the coefficient of $\tilde{x}_{1}$ in the eq.
(\ref{final}) at the price of neglecting $a^{\prime }\sin k_{1}\tilde{x}
_{1}. $ This yields 
\[
\mathtt{\tilde{p}}_{1}=-\frac{(l_{2}\mathtt{\tilde{p}}_{2}+l_{3}\mathtt{%
\tilde{p}}_{3}+\frac{cb}{2\mathtt{\tilde{E}}})}{(1-l_{1})} 
\]
To justify this step Dirac offers the following classical ''justification''.
The neglecting of $a^{\prime }$ enables to give meaning to the variation of $%
p_{1}$ because, \textit{in order to give a meaning to the recoil momentum of
the electron we must neglect the oscillations of the electron due to the
incident radiation.}\footnote{%
Dirac (1927a), p. 506.}\textit{. }Using \ref{matel}, one gets now 
\[
\mathtt{\tilde{p}}_{1}^{\prime }=\mathtt{\tilde{p}}_{1}-\frac{\hbar \omega }{%
c} 
\]
The equations above give the relations between the quantum numbers of
initial and final states. These equations correspond to energy-momentum
conservation; in the original variables, recalling eq.(\ref{cantransf}) and (%
\ref{separ}), one finds 
\begin{eqnarray}
\mathtt{p}_{1}^{\prime }+l_{1}\frac{\hbar \omega ^{\prime }}{c} &=&\mathtt{p}%
_{1}+\frac{\hbar \omega }{c}  \label{matelold} \\
\mathtt{p}_{2}^{\prime }+l_{2}\frac{\hbar \omega ^{\prime }}{c} &=&\mathtt{p}%
_{2}  \nonumber \\
\mathtt{p}_{3}^{\prime }+l_{3}\frac{\hbar \omega ^{\prime }}{c} &=&\mathtt{p}%
_{3}  \nonumber \\
\mathtt{E}^{\prime }+\hbar \omega ^{\prime } &=&\mathtt{E}+\hbar \omega 
\nonumber
\end{eqnarray}

In the rest system, (\ref{matelold}) yield the Compton relation 
\begin{equation}
\frac{\omega }{\omega ^{\prime }}=1+\frac{\hbar \omega }{mc^{2}}(1-\cos
\theta ).  \label{comptr2}
\end{equation}
which is equivalent in terms of the wave-length $\lambda $ to 
\begin{equation}
\lambda ^{\prime }=\lambda +\frac{h}{mc}(1-\cos \theta )  \label{comptr1}
\end{equation}
Here, $\theta $ is the scattering angle.

The features of the emitted radiation are then obtained in the following
way. Recall that $X$ represents a component of the polarization and its
matrix elements have been determined (see (\ref{polmat}) and (\ref{matel})).
Relying on Heisenberg's prescriptions, and his own results\footnote{%
See Dirac (1926). Dirac discusses there the quantization of
multiply-periodic quantum systems through the use of uniformizing variables $%
J$ and $\omega $. For such systems, the observables $C$have the general
expansion 
\[
\sum_{\alpha }C_{\alpha }(J_{1},J_{2},...)e^{i(\alpha \omega )}=\sum_{\alpha
}e^{i(\alpha \omega )}C_{\alpha }^{\prime }(J_{1},J_{2},...) 
\]
where the sum is over integers. The difference of order between the
coefficients $C_{\alpha }$, resp. $C_{\alpha }^{\prime }$ and the
exponentials above can be related to the two possible transitions between a
given pair of stationary states$.$ In the case of polarization, the product
of the coefficients $C$ gives the intensity. See Dirac's paper for more
details.}, Dirac identifies the product $X_{\mathtt{pp}^{\prime }}X_{\mathtt{%
p}^{\prime }\mathtt{p}}$ (here we denote collectively the quantum numbers by 
\texttt{p)} as a quarter of the square of the amplitude of vibration in the $%
X$-direction. The square of the total amplitude \TEXTsymbol{\vert}$C|^{2}$is
accordingly the sum of that for $X$ and that for the (perpendicular) $Y$. It
can be related to the intensity (energy flux) at distance $r$ using the
classical formula 
\begin{equation}
I=\frac{e^{2}\omega ^{4}}{8\pi c^{3}r^{2}}|C|^{2}  \label{clasinten}
\end{equation}
This formula is the classical dipole approximation for the radiation emitted
by an oscillating density of charge and current. Since for $h=0$ one has in (%
\ref{matel}) $\mathtt{\tilde{E}}^{\prime }=\mathtt{\tilde{E}}$, the quantum
expression for the intensity turns out to be equal to $(\mathtt{\tilde{E}}/%
\mathtt{\tilde{E}}^{\prime }$) times its classical value (see Dirac's first
paper p. 405-406 and the second p. 506 for the detailed expressions). In
particular, for an electron initially at rest, Dirac obtains for the
intensity of the secondary radiation at distance $r$ from the emitting
electron the expression 
\begin{equation}
I=I_{0}\frac{e^{4}}{m^{2}c^{4}r^{2}}(\frac{\omega ^{\prime }}{\omega }%
)^{3}\sin ^{2}\phi  \label{compt1}
\end{equation}
where $I_{0}=a^{2}\omega ^{2}/8\pi c$ is the intensity of the primary
radiation$,$ and $\phi $ is the the angle between the direction of
observation and the initial polarization. In terms of the (differential)
scattering cross section\footnote{%
The differential scattering cross section $\frac{d\sigma }{d\Omega }$ is
nowadays defined as the ratio of the number of scattered particles into the
unit solid angle over the number of incident particles. This is related to
the energy radiated per unit time, unit solid angle and unit incident
intensity, because the energy is the number of light quanta multiplied by $%
\hbar \omega .$ To make contact with the preceding equation for the
intensity one has to multiply it by $r^{2}$, divide by the intensity of the
incident (plane) wave, and multiply by $\frac{\omega ^{\prime }}{\omega }$},
Dirac's result is then: 
\[
\frac{\omega ^{\prime }}{\omega }\frac{d\sigma }{d\Omega }=\frac{e^{4}}{%
m^{2}c^{4}}(\frac{\omega ^{\prime }}{\omega })^{3}\sin ^{2}\phi 
\]

As noted by Dirac, the result (\ref{compt1}) is just $(\frac{\omega ^{\prime
}}{\omega })^{3}$ times its value according to the classical theory. For
unpolarized incident radiation, one has to take the average of $\sin
^{2}\phi $ over all directions of polarization$,$ which produces a factor of 
$(1+\cos ^{2}\theta )/2,$ i.e.

\begin{equation}
\frac{d\sigma }{d\Omega }=\frac{e^{4}}{m^{2}c^{4}}(\frac{\omega ^{\prime }}{%
\omega })^{2}(1+\cos ^{2}\theta )/2
\end{equation}
or, using (\ref{comptr2}) 
\begin{equation}
\frac{d\sigma }{d\Omega }(\theta )=\frac{e^{4}}{2m^{2}c^{4}}\frac{1+\cos
^{2}\theta }{(1+\alpha (1-\cos \theta ))^{2}}  \label{compt1prim}
\end{equation}
where $\alpha =\hbar \omega /mc^{2}$. The results (\ref{compt1}) and (\ref
{compt1prim}) were already obtained in Dirac 1926b. After comparing this
result with the available experimental data of Compton (Compton 1923), Dirac
observes:

\begin{quotation}
[...] the experimental values are all less than the values given by the
present theory, in roughly the same ratio (75 per cent.), which shows that
the theory gives the correct law of variation of intensity with angle, and
suggests that in absolute magnitude Compton's values are 25 per cent. too
small\footnote{%
Dirac 1926b, p. 423.}.
\end{quotation}

The source of this discrepancy will be clarified by Klein and Nishina 1929.

\subsection{Gordon: quantized current of a scalar electron as source of the
retarded classical potential}

At the end of the second paper of Dirac, a footnote added in proof mentions
a paper by Walter Gordon dealing with the same subject. Indeed, before
Dirac's second paper appeared, Gordon had treated the same problem, albeit
in a more detailed way (Gordon 1926, received 29 Sept. 1926)\footnote{%
It appears that Klein and Uhlenbeck were also working on the wave approach
to Compton scattering but didn't reach a quantitative result before Gordon.
(Klein 1968, p. 79). This seems as well to have been the case with Pauli,
see Rechenberg and Mehra 1982-87, vol. 5, p. 830.}. As Dirac previously, and
following Schr\"{o}dinger's prescription, he derived the relativistic
quantum wave equation for a spinless charged particle coupled to the
classical electromagnetic potential\footnote{%
Gordon writes $\Phi $ instead of $A$} $A_{\mu }$: 
\begin{equation}
\left[ \left( -i\hbar \partial _{\mu }-\frac{e}{c}A_{\mu }\right) \left(
-i\hbar \partial ^{\mu }-\frac{e}{c}A^{\mu }\right) +m^{2}c^{2}\right] \psi
=0  \label{klein-gordon}
\end{equation}
This invariant wave equation is named today after him and Oscar Klein, who
had obtained the same expression in a publication that appeared a few months
earlier (Klein 1926)\footnote{%
One should at this point also mention the results of Schr\"{o}dinger himself
(Schr\"{o}dinger 1926d) and Fock (1926). Further references can be found in
Schweber (1961), p. 54. and Kragh 1984. The naming of the scalar
relativistic equation after Klein and Gordon appears then quite arbitrary.}.
An important feature of Gordon's paper is his understanding of gauge
transformations at the wave function level. He properly realized that to the
transformation $A_{\alpha }\rightarrow A_{\alpha }+\partial _{\alpha }f$
there corresponds the multiplicative phase factor $\psi \rightarrow
e^{\left( \frac{i}{\hbar }\frac{e}{c}f\right) }\psi $, where $f$ is an
arbitrary function\footnote{%
The same had already been realized by Fock (1926).}.

In his derivation of the secondary radiation, the strategy followed by
Gordon sets the standard for subsequent works. It is, together with the
previous approach by Dirac, a testimony to the power of the quantum
formalism which takes care of the situation way beyond what might have been
expected at first sight. The electron is considered as perturbed from its
free evolution by the primary radiation. Assuming its perturbed motion as
known, the features of the secondary radiation are then obtained from the
classical radiation formula with retardation 
\begin{equation}
A_{\mu }^{\prime }\left( \mathbf{y}\right) =\frac{1}{c}\dint \frac{s_{\mu }(%
\mathbf{x},t-\frac{|\mathbf{x}-\mathbf{y}|}{c})}{|\mathbf{x}-\mathbf{y}|}d%
\mathbf{x},\quad d\mathbf{x\equiv }dx_{1}dx_{2}dx_{3}  \label{austrahl}
\end{equation}
where the densities of charge and current of the perturbed electron enter
through the quantum expressions 
\begin{equation}
s_{\alpha }=\frac{1}{i}\left( \psi ^{*}\partial _{\alpha }\psi -\psi
\partial _{\alpha }\psi ^{*}-\frac{2i}{\hbar }\frac{e}{c}A_{\alpha }\psi
\psi ^{*}\right)  \label{kgcurrent}
\end{equation}
These densities are obtained from the wave equation in a way similar to that
used by Schr\"{o}dinger in his pioneering work on the non-relativistic case.
Actually, there is an implicit factor of $e$ in front of this expression
which is ''hidden'' in the normalization of $\psi $ as will be explained
shortly, see eqs. (\ref{packet}) and (\ref{cnorm}).

Notice how the strategy followed by Gordon differs from that of Dirac
examined earlier. Gordon uses the classical radiation formula with
retardation into which he inserts the quantum expression for the current. It
is through the latter that the quantum features will enter the stage, namely
the proper correlation between initial and final states, and the resulting
secondary radiation, as will be seen shortly. Dirac, on the other hand,
inserted by hand the covariant phase $(ct-\mathbf{kx)}$ by a change of
variables, and used a final dipole approximation formula, where he injected
the squared quantum amplitude for the polarization. In his case, the
information about the final state and the resulting secondary radiation is
obtained from the action of the polarization observable upon the initial
state. In view of the fact that Gordon obtained the same final result as
Dirac, the two approaches prove equivalent albeit their heuristics is
different.

Gordon's ''two-step''\ scheme of computation (first solve the
radiation-perturbed electron evolution, then compute the resulting classical
radiation and interpret the latter as Compton radiation) leans upon an
approximation not easily expressible within the framework of more modern
(fully quantized) approaches (the same holds also for Dirac's approach
discussed earlier). Nevertheless, its success is remarkable, especially so
in the Klein-Nishina computation to be discussed shortly, where, to the
order considered, a correct formula is obtained in spite of using a
classical electromagnetic field. The reasons of this ``two-step'' technique
are clear. Until the advent of Dirac's field quantization (1927), there was
no way to treat the interaction of the radiation field with matter quantum
mechanically. The quantum ingredient could only be introduced using the
quantum evolution equations for matter subjected to (classical) radiation
and radiating back classically through the associated charge and densities
(quantum) expressions (this is dubbed by the various authors as a \textit{\
korrespondenzm\"{a}ssig }procedure\footnote{%
see for instance Klein (1926), p. 416.}). The power of the quantum formalism
is manifest in the fact that it is effectively sufficient to use a quantum
evolution equation for the electron (and keeping radiation classical) to
achieve a result where the process is handled in terms of an effective
matrix element between initial and final quantum states, thus introducing an
element of discontinuity in a otherwise apparently continuous process. In
this picture, the \textit{discrete} shifts in the frequency (and wave
vector) of the radiation can be understood as induced by the quantum
evolution of the electron wave function itself. Notice that these features
are not explicitly present in the original formulation of the problem.
There, a purely classical electromagnetic radiation affects the motion of
the electron which radiates according to the classical laws, all this along
a purely continuous course of events: by the very nature of the approach, no
such notions as initial or final states can find their way naturally in this
context.

To see how the discontinuity related to a transition from initial to final
states emerges out of this seemingly continuous context, let us define with
Gordon the radiation field as $A_{\mu }=a_{\mu }\cos \varphi $ where%
\footnote{%
Instead of $k,$ Gordon uses the notation $l.$} $\varphi =\left( \mathbf{kx}%
-\omega t\right) \equiv k\cdot x$ with $x_{0}=ct$ and $k_{0}=\omega /c$ The
description of the perturbed electron motion is obtained by Gordon with the
additional approximation that the terms $A^{2}$ in (\ref{klein-gordon}) can
be neglected, in the same way as in Dirac. He thus ends up solving the
truncated equation 
\begin{equation}
\partial \cdot \partial \psi -\frac{2i}{\hbar }\left( \frac{e}{c}a\cdot
\partial \psi \right) \cos \varphi -\frac{m^{2}c^{2}}{\hbar ^{2}}\psi =0
\label{klein-gordon-trunc}
\end{equation}
A few remarks are in order here. The neglect of the $A^{2}$ terms may appear
surprising as far as as the electrodynamics of scalar particles is
considered. Indeed, one can show (using for example the modern propagator
approach; see also section 7.2) that there is a natural gauge choice where
the total contribution to the scattering amplitude in the laboratory frame
(initial electron at rest) comes precisely from this particular term%
\footnote{%
See e. g. Bjorken and Drell 1964, pp. 194-195. This is further explained in
section 6.2. of the present paper.}. However, In his paper, Gordon does not
commit himself explicitely to any particular gauge or frame. To understand
the situation, one has to remember again the peculiar nature of the
semi-classical approach used here. In this picture, the current eq. (\ref
{kgcurrent}) contains already the coupling to the primary photon\footnote{%
We thank prof. Roberto Casalbuoni for a valuable discussion concerning this
point.}, as evidenced by the term proportionnal to $e,$ namely 
\begin{equation}
\frac{2i}{\hbar }\frac{e}{c}A_{\alpha }\psi \psi ^{*}.  \label{kgcoupling}
\end{equation}
In the course of the calculation, the ''coupling'' to the secondary photon
(to use modern expressions) will be obtained indirectly and at this point it
is best to keep track of the powers of the coupling constant (proportional
to $e)$ as we shall see shortly.

Gordon can solve (\ref{klein-gordon-trunc}) choosing the ansatz 
\begin{equation}
\psi =\exp (\frac{i}{\hbar }W),  \label{ansatz}
\end{equation}
with the modified phase\footnote{%
This is the solution of the truncated equation (\ref{klein-gordon-trunc});
the solution of the full equation would involve terms quadratic in the
potential.} 
\begin{equation}
W=p\cdot x+\frac{e}{c}\frac{p\cdot a}{p\cdot k}\sin \varphi  \label{ansatz1}
\end{equation}
provided than the momenta $p_{\mu }$ obey the mass-shell equation: 
\begin{equation}
p\cdot p+m^{2}c^{2}=\mathbf{p}^{2}-\left( \frac{E}{c}\right)
^{2}+m^{2}c^{2}=0.  \label{masshell}
\end{equation}
Gordon considers then an arbitrary superposition of solutions 
\begin{equation}
\psi =\int z(p)C(p)e^{(\frac{2\pi i}{h}W)}d\mathbf{p,}  \label{packet}
\end{equation}
where $d\mathbf{p=}dp_{1}dp_{2}dp_{3}.$ It contains two factors: $z(p),$
that Gordon identifies as a weighting of the individual solutions (\ref
{ansatz}), and $C(p)$ that he considers a normalization. The reason for this
splitting will appear shortly. Gordon builds then the bilinear expressions
entering the current four-vector (\ref{kgcurrent}) 
\begin{eqnarray}
\psi \psi ^{*} &=&\int \int e^{\frac{i}{\hbar }(W(p)-W(p^{\prime
}))}z(p)z(p^{\prime })C(p)C(p^{\prime })d\mathbf{p}d\mathbf{p}^{\prime }
\label{kgcurrent_pieces} \\
\psi ^{*}\partial _{\mu }\psi &=&\frac{i}{\hbar }\int \int (p_{\mu }+\frac{e%
}{c}\frac{p\cdot a}{p\cdot k}k_{\mu }\cos \varphi )e^{\frac{i}{\hbar }%
(W(p)-W(p^{\prime }))}z(p)z(p^{\prime })C(p)C(p^{\prime })d\mathbf{p}d%
\mathbf{p}^{\prime }  \nonumber \\
&&\mathrm{and\quad c.c.}  \nonumber
\end{eqnarray}
As can be seen, these expressions are of the form 
\[
O=\sum_{ij}O_{ij} 
\]
where the labels $i$ and $j$ run over the set of the solutions of (\ref
{klein-gordon-trunc}). Hence, considering superpositions (\ref{packet}), one
can obtain the current (\ref{kgcurrent}) as a sum of ''transition currents''
related to a \textit{quantum transition from eigenstate }$i$\textit{\ to
eigenstate }$j.$ This is just how the initial and final states of the
electron will enter the stage.

The value of normalization factor $C(p)$ is obtained by Gordon using a
comparison with the classical case and the arbitrary (but then familiar
requirement) that 
\[
\int z^{2}(p)d\mathbf{p}=1, 
\]
hence the splitting\footnote{%
Actually, neither Gordon nor Dirac knew to normalize the initial and final
states in the continuum. Therefore both had to use the classical limit in
order to get the normalization in an indirect way.}. The value of $C(p)$
turns then out to be 
\begin{equation}
C^{2}(p)=\frac{ec^{2}}{4\pi h^{2}E}  \label{cnorm}
\end{equation}

where $E$ is given by (\ref{masshell}). Notice here the presence of another $%
e$ factor in the numerator. This is the one alluded to in the comment after
eq. (\ref{kgcurrent}). Combined with the one already present in the term eq.
(\ref{kgcoupling}), it makes for the correct $e^{2}$ dependence of the
result. This justifies \textit{a posteriori} the neglect of the quadratic
terms not only in eq. (\ref{klein-gordon-trunc}), but also in the next steps.

Indeed, Gordon is now in position to obtain the features of the secondary
radiation using the preceding expressions to build up the current in (\ref
{austrahl}): He first expands the exponentials in eq. (\ref{kgcurrent_pieces}
) to first order in $a$, then, assuming that the charge distribution is of
negligable size with respect to the distance $|\mathbf{y}|\equiv r$ of the
observation point $\mathbf{y}$\textbf{\ }, approximates $|\mathbf{x}-\mathbf{%
y}|=r-\mathbf{\hat{k}}^{\prime }\mathbf{\cdot x,}$where $\mathbf{\hat{k}}%
^{\prime }$ is the unit vector\footnote{%
Instead of $\mathbf{\hat{k}}$ Gordon uses $\mathbf{\xi .}$} in the direction 
$\mathbf{y}.$ He arrives thus at the dipole expression

\begin{equation}
A_{\mu }^{\prime }=\frac{ec}{2h^{3}r}\func{Re}\dint \frac{Z(P)Z^{\prime
}(P^{\prime })}{\sqrt{E\Delta E^{\prime }\Delta ^{\prime }}}T_{\mu }e^{\frac{%
i}{\hbar }(\mathbf{P}-\mathbf{P}^{\prime })\mathbf{x}-i\omega ^{\prime }(t-%
\frac{r}{c})}d\mathbf{P}d\mathbf{P}^{\prime }d\mathbf{x}  \label{ausstrahl_1}
\end{equation}
with 
\begin{eqnarray}
T_{\mu } &=&\frac{e}{c}\left[ \frac{1}{\hbar }(\frac{p\cdot a}{p\cdot k}-%
\frac{p^{\prime }\cdot a}{p^{\prime }\cdot k})(p_{\mu }+p_{\mu }^{\prime })+(%
\frac{p\cdot a}{p\cdot k}+\frac{p^{\prime }\cdot a}{p^{\prime }\cdot k}%
)k_{\mu }-2a_{\mu }\right]  \label{ausstrahldef} \\
\omega ^{\prime } &=&\frac{E-E^{\prime }+\hbar \omega }{\hbar }  \nonumber
\end{eqnarray}
The integral above is written in terms of new integration variables 
\begin{eqnarray*}
\mathbf{P} &=&\mathbf{p}+\hbar \mathbf{k}-\frac{E+\hbar \omega }{c}\mathbf{%
\hat{k}}^{\prime } \\
\mathbf{P}^{\prime } &=&\mathbf{p}^{\prime }-\frac{E^{\prime }}{c}\mathbf{\ 
\hat{k}}^{\prime }
\end{eqnarray*}
which explains the jacobians\footnote{%
One recognizes $\Delta $ as the Doppler factor corresponding to the initial
velocity $v$ of the electron.} 
\[
\Delta =1-\frac{c}{E}\mathbf{p}\cdot \mathbf{\hat{k}}^{\prime }\mathbf{=}1-%
\frac{v}{c}\mathbf{\cos }\varphi 
\]
as well as 
\[
Z^{2}(P)\Delta (p)=z^{2}(p), 
\]
and similarly for the corresponding primed quantities. We recognize here $%
\Delta $ as the Doppler factor depending on the angle between the velocity $%
v $ of the electron and the direction of observation.

The features interpretable in terms of a transition from an initial to a
final state are thus ensured by the bilinearity of the above expression.
Now, the final state values for the frequency and momenta are further
obtained thanks to the appearance in (\ref{ausstrahl_1}) of the Dirac $%
\delta -$ function hidden in its Fourier form 
\[
\int e^{\frac{i}{\hbar }\sum (P_{k}-P_{k}^{\prime })x_{k}}d\mathbf{x}
=(h)^{3}\prod\limits_{k}\delta (P_{k}-P_{k}^{\prime }) 
\]
which ensure conservation of energy-momentum yielding the equality 
\begin{eqnarray*}
p+\hbar k &=&p^{\prime }+\hbar k^{\prime },\quad \mathrm{with} \\
\mathbf{k}^{\prime } &=&\frac{\omega ^{\prime }}{c}\mathbf{\hat{k}}^{\prime
};\quad k_{0}^{\prime }=\frac{\hbar \omega ^{\prime }}{c}
\end{eqnarray*}
so that $\hbar k$ and $\hbar k^{\prime }$ are the respective 4-momenta of
the initial and final photon. After obtaining his final expression for the
radiation 
\begin{equation}
A_{\mu }^{\prime }=\frac{ec}{2r}\dint \frac{Z(P)Z^{\prime }(P)}{\sqrt{
E\Delta E^{\prime }\Delta ^{\prime }}}T_{\mu }\cos \omega ^{\prime }(t-\frac{
r}{c})dP  \label{kgausstrahl}
\end{equation}
Gordon makes the significative comment:

\begin{quotation}
Since $Z^{2}(P)dP$ and $Z^{\prime 2}(P)dP$ are the weights of the two state
spaces and are combined together, the radiation potential corresponding to
the single transition is given by\footnote{%
Da dann $Z^{2}(P)dP$ und $Z^{\prime 2}(P)dP$ die Gewichte der beiden
Zustandsgebiete sind, die mitenander kombinieren, so geh\"{o}rt zum
einzelnen ''\"{U}bergang`` das Austrahlungspotential 
\[
\frac{ec}{2r}\frac{1}{\sqrt{E\Delta E^{\prime }\Delta ^{\prime }}}T_{\alpha
}\cos \omega ^{\prime }(t-\frac{r}{c}).
\]
Gordon 1926, p. 130.} 
\[
\frac{ec}{2r}\frac{1}{\sqrt{E\Delta E^{\prime }\Delta ^{\prime }}}T_{\alpha
}\cos \omega ^{\prime }(t-\frac{r}{c})
\]
\end{quotation}

This is how in this correspondence principle way the elements of the now
so-familiar picture of electron-photon process emerge albeit in a
semi-classical context yet not involving Dirac's field quantization nor
perturbation theory. To make contact with modern treatments, the reader can
contract the gauge invariant expression (\ref{ausstrahldef})with the
polarization vector of the secondary photon, which in a sense provides the
second external photon line supplementing the one already present in the
current (\ref{kgcurrent}), (to be compared with for instance eq. (9.30) of
Bjorken and Drell\footnote{%
Bjorken and Drell 1964, p. 194.} and section 6.2 of the present paper).

For the final formula giving the intensity of the secondary radiation Gordon
obtains ($\frac{\omega ^{\prime }}{\omega }$ is the ratio of the final and
the initial frequencies) 
\[
I=\left( \frac{\omega ^{\prime }}{\omega }\frac{\Delta }{(1-\frac{v}{c}%
\mathbf{\ \cos }\delta )}\right) ^{3}I_{\text{kl}} 
\]
which is the Dirac's result (\ref{compt1}) once one substitues for $I_{\text{
kl}}$, the Thompson formula and assumes the electron at rest, $v=0$, so that
the ratio of the Doppler factors $\Delta /(1-\frac{v}{c}\mathbf{\ \cos }
\delta )=1$ ($\delta $ is the angle between the velocity and the wave vector
of the inital radiation).

Before turning to the Klein-Nishina paper (and later to Waller and Tamm), we
want to mention a contribution of Klein 1927 submitted at the very end of
1926 and entitled \textit{Elektrodynamik und Wellenmechanik vom Standpunkt
des Korrespondenzprinzips}. In this paper, Klein also uses the relativistic
scalar equation and studies its various applications to the dispersion of
light by atoms and free electrons. As indicated in the title, Klein
discusses quite extensively the spirit of the correspondence principle
approach. Anyone interested in the heuristics of the latter will find in
Klein's contribution a valuable discussion of this method. Another feature
of Klein's paper is an interesting perturbative scheme which the latter
proposes to deal with time-dependent problems. It fits somehow between the
original proposal of Schr\"{o}dinger and announces that of Dirac\footnote{%
See the next section.}. Klein's scheme is however practical only in the case
of equations linear (first order) in time derivative, so that, albeit in
possession of the scalar relativistic equation, Klein considers only its
non-relativistic limit (Schr\"{o}dinger equation)when discussing the
dispersion of light scattered off atoms, or uses instead an ansatz for the
case of Compton scattering.

\subsection{Klein and Nishina: spin 1/2 electrons}

The Klein-Nishina paper (Klein and Nishina 1929, received 30 Oct. 1928) is
the first to apply the freshly discovered spinor equation of Dirac (Dirac
1928, received 2 Jan. 1928) to Compton scattering. It reads, using the
original set of $\sigma _{i}$ and $\rho _{i},$ $i=1,..,3$ introduced by Dirac%
\footnote{%
Here, the 4x4 $\sigma $ matrices are obtained putting the 2x2 Pauli matrices
on the diagonal.}, 
\begin{equation}
\left\{ \frac{E+eV}{c}+\rho _{1}\left( \mathbf{\sigma },\mathbf{p}+\frac{e}{c%
}\mathbf{A}\right) +\rho _{3}mc\right\} \psi =0  \label{dequation}
\end{equation}
Alternatively, one can use the nowadays more familiar $\gamma $ matrices
setting 
\begin{eqnarray*}
\rho _{1}\sigma _{i} &=&\gamma _{0}\gamma _{i}\quad i=1,2,3 \\
\rho _{2}\sigma _{i} &=&\gamma _{i},\quad i=1,2,3 \\
\rho _{3} &=&-i\gamma _{0}
\end{eqnarray*}
in which case one can rewrite (\ref{dequation}) as\footnote{%
See the original paper of Dirac. The conventions used here are a departure
from those of Dirac (and consequently from those of Klein-Nishina and
Waller, who follow Dirac) in that he uses the euclidean metric in which $%
\left\{ \gamma _{\mu },\gamma _{\nu }\right\} =2\delta _{\mu \nu },$ with $%
\mu ,\nu =1,...,4,$ whereas we chose to work with $\gamma _{0}=i\gamma _{4}$
which amounts to use the metric $g_{\mu \nu }=(-1,1,1,1).$} 
\begin{equation}
\left\{ i\left( -\gamma _{0}\frac{E+eV}{c}+(\mathbf{\gamma },\mathbf{p}+%
\frac{e}{c}\mathbf{A})\right) +mc\right\} \psi =0  \label{direq}
\end{equation}
As already mentioned, the use of this relativistic equation which takes into
account the electron spin constitutes the main novelty of the paper. For the
rest, it closely follows the correspondence principle strategy of Gordon.
Instead of the expressions for the current densities (\ref{kgcurrent})
pertaining to the scalar (Klein-Gordon) equation, Klein and Nishina use
instead the Dirac densities 
\begin{eqnarray}
\rho &=&e\bar{\psi}\gamma _{0}\psi ^{\prime }  \label{dcurrent} \\
\mathbf{J} &=&ec\bar{\psi}\mathbf{\gamma }\psi ^{\prime }  \nonumber
\end{eqnarray}
Here $\bar{\psi}$ stands for the (one-row) spinor obtained from $\psi $ via
hermitian conjugation and multiplication by the matrix $\gamma _{0}$: $\bar{%
\psi}=(\psi ^{*})^{t}\gamma _{0}$. $\psi $ is taken as solution of the
second-order equation (the scalar potential $V$ is set to zero)

\begin{equation}
\left\{ \frac{\hbar ^{2}}{c^{2}}\partial _{t}^{2}+\left( -i\hbar \mathbf{\
\nabla }+\frac{e}{c}\mathbf{A}\right) ^{2}+m^{2}c^{2}\right\} \psi +\frac{
e\hbar }{c}\left( \mathbf{\sigma B}\right) \psi +\frac{ie\hbar }{c}\rho
_{1}\left( \mathbf{\sigma E}\right) \psi =0  \nonumber
\end{equation}
obtained from (\ref{dequation}) taking the square \footnote{%
This use of the second order equations may appear puzzling. It is possible
that Klein and Nishina used the latter instead of (\ref{dequation}) to keep
close to the original discussion of Gordon.}. Klein and Nishina assume a
vector potential of the form\footnote{%
Klein and Nishina's notation for the potential is $U.$} 
\[
\mathbf{A=\mathbf{a}}e^{-i(\mathbf{kx-}ct\mathbf{)}}+\mathbf{a}^{*}e^{i(%
\mathbf{kx-}ct\mathbf{)}}=2\func{Re}(\mathbf{a}e^{-ik\cdot x});\quad \mathbf{%
k}=\frac{\omega }{c}\mathbf{\hat{k},}k_{0}=\frac{\omega }{c}, 
\]
and make the following ansatz for the solution:

\begin{equation}
\psi =\left\{ 1+g(p)e^{-ik\cdot x}+\tilde{g}(p)e^{ik\cdot x}\right\} \psi
_{0}  \label{knansatz1}
\end{equation}

where $\psi _{0}$ is the free solution for $\mathbf{A}=0\footnote{%
Klein and Nishina write $v(p)$ instead of $u(p)$, the latter denoting in
their paper the adjoint spinor $v(p)^{\dagger }.$ In order not to make the
discussion unnecessarily cluttered with notations, we do not discuss here
the conventions on the basis of spinor solutions to the free Dirac equation.}
$%
\begin{equation}
\psi _{0}=u(p)e^{-\frac{i}{\hbar }(p\cdot x\mathbf{)}}.
\end{equation}
The ansatz (\ref{knansatz1}) is the spinor version of the ansatz of Gordon (%
\ref{ansatz}) and (\ref{ansatz1}). Indeed, the functions $g$ and $\tilde{g}$
are given by 
\begin{eqnarray}
g(p) &=&\frac{e}{\hbar c}\frac{1}{pk}\left\{ \mathbf{pa}+\frac{1}{2}\hbar (%
\mathbf{\sigma \eta )+}\frac{i}{2}\mathbf{\hbar }\rho _{1}(\mathbf{\sigma
\epsilon )}\right\}  \label{knansatz2} \\
\tilde{g}(p) &=&-\frac{e}{\hbar c}\frac{1}{pk}\left\{ \mathbf{pa}^{*}+\frac{1%
}{2}\hbar (\mathbf{\sigma \eta }^{*}\mathbf{)+}\frac{i}{2}\mathbf{\hbar }%
\rho _{1}(\mathbf{\sigma \epsilon }^{*}\mathbf{)}\right\}  \nonumber
\end{eqnarray}
where $\mathbf{E}=\mathbf{\epsilon }e^{-ik\cdot x}+\mathbf{\epsilon }%
^{*}e^{ik\cdot x}$, and $\mathbf{B}=\mathbf{\eta }e^{-ik\cdot x}+\mathbf{%
\eta }^{*}e^{ik\cdot x}$. In the case of linear polarization (the case
studied by Gordon), one has $\mathbf{\epsilon }^{*}\mathbf{=\epsilon ,}$
which implies $\mathbf{a}^{*}\mathbf{=-a.}$ Then, from (\ref{knansatz1}),
and (\ref{knansatz2}), one sees that 
\[
\psi =\left\{ 1+\frac{e}{\hbar c}\frac{\mathbf{pa}}{pk}2\cos kr-\frac{e}{%
\hbar c}\frac{i}{pk}\left[ \hbar (\mathbf{\sigma \eta )+}i\mathbf{\hbar }%
\rho _{1}(\mathbf{\sigma \epsilon )}\right] \sin kr\right\} \psi _{0} 
\]

where we recognize the linear part of the solution (\ref{ansatz}) and (\ref
{ansatz1}) of Gordon supplemented by spin terms. As in the latter's case,
quadratic terms in the potential are thus further neglected. The
''initial-final'' picture emerges also in a way analogous to that of Gordon,
Klein and Nishina following essentially the same steps;\ they obtain for
instance as the analog of (\ref{kgausstrahl}) the result:

\[
A(p,p^{\prime })=\frac{(h)^{3}}{r}\frac{1}{\sqrt{\Delta \Delta ^{\prime }}}%
\{e^{i\omega ^{\prime }(t-\frac{r}{c})}\bar{u}(p)[\mathbf{\gamma }
g(p^{\prime })+\tilde{g}(p)^{\dagger }\mathbf{\gamma }]u^{\prime }(p^{\prime
})+c.c 
\]

In order to obtain explicit formulas for the intensity of the scattered
radiation, Klein and Nishina still have to go through the tedious steps of
the evaluation of products consisting typically of initial and final spinors 
$\bar{u}(p)$ and $u^{\prime }(p^{\prime })$ sandwiching various combinations
of spin matrices. The technicalities involved in these calculations don't
bring any further physical insight and we shall not comment them here%
\footnote{%
Let's remark however that the technical tricks making the reduction of the
spinor algebra much easier are due to Casimir 1933 and Wannier 1935.}. Klein
and Nishina finally obtain the following celebrated expression for the
scattering cross section for unpolarized incident light, and \textit{in the
rest frame of the initial electron} 
\begin{equation}
\frac{d\sigma }{d\Omega }(\theta )=\frac{e^{4}}{2m^{2}c^{4}}(\frac{1+\cos
^{2}\theta }{(1+\alpha (1-\cos \theta ))^{2}})\left[ 1+\alpha ^{2}\frac{%
(1-\cos \theta )^{2}}{(1+\cos ^{2}\theta )(1+\alpha (1-\cos \theta ))}\right]
\label{kn_4}
\end{equation}
where $\theta $ is the scattering angle, and $\alpha =\hbar \omega /mc^{2}$.
It differs from the result obtained earlier by Dirac (and also Gordon) by
terms of second order in $\alpha $ and higher (compare with (\ref{compt1prim}%
)). We know nowadays that this expression constitutes the correct result to
the lowest order. It is indeed remarkable that Klein and Nishina were able
to obtain it within such a semi-classical context. The reason of this
success is due to the quantum mechanism, already mentioned above, which
automatically takes care of shifting the state of the photon ensuring
conservation of energy-momentum, and to the fact that at the tree level,
where multiple effects of emission/absorption are neglected, it is
sufficient to consider the radiation field as classical. The following
appreciation may serve as a closing word\footnote{%
Ya. S. \bigskip Smorodinskii 1987, p. 823.}

\begin{quote}
.[In calculating] the scattering of an electromagnetic wave by a charged
particle, Maxwell's classical equations were used, but the transition
current was substituted for the classical current on their right-hand side.
The justification of this operation was that if in the ordinary formula for
the current the wave function and its conjugate were replaced by their
expansions in Fourier integrals the resulting double integral could be
interpreted as the sum (integral) of all possible contributions (with
different frequencies) to the total radiation. This semiclassical method in
no way took into account the quantum nature of the electromagnetic field and
by its own nature did not involve calculations with intermediate states, the
Compton effect being treated in a certain sense as a first-order effect.
This fact was fortunate for the authors, since they did not encounter a
paradox that surprised Tamm and Waller.\footnote{%
The ''paradox'' in question is the necessity to sum over negative energy
intermediate states of the electron as will be seen in the next section.}
\end{quote}

\section{Dirac's perturbation theory, field-quantization and intermediate
states}

In this chapter, we shall examine the key role of the perturbation theory
and field quantization introduced by Dirac (and used subsequently him by
Waller and Tamm). The two developments are related since Dirac quantizes the
radiation field within the context of his perturbation theory for Bose
particles. The choice of the perturbation technique (as distinct from the
choice of the approximation, or better, physical treatment of the process)
is an important theme surfacing again and again in our story. The
interpretation following a given perturbative scheme has a major role in the
way the whole process is pictured and understood.We first start with a brief
recall of the situation prior Dirac's theory.

One of the first formulations of the time-dependent perturbation theory
within quantum mechanics goes logically back to the ''Drei-M\"{a}nner-Arbeit
'' of Born, Heisenberg and Jordan (1926). It is formulated in terms of the
matrix formalism and relies on the analogy with the classical canonical
formalism. It enables the authors to derive as an application the
Heisenberg-Kramers dispersion relations.

With the advent of wave mechanics, the community took possession of a new,
mathematically more familiar tool. The perturbation theory in wave mechanics
superseded then the matrix version. Schr\"{o}dinger developed the
perturbation theory for time independent Hamiltonians in his third paper
(Schr\"{o}dinger 1926c). It is essentially an application of Sturm theory.
The time-dependent perturbation theory appears in the last paper of
Schr\"{o}dinger's tetralogy (Schr\"{o}dinger 1926d), where he introduces his
celebrated time-dependent equation: 
\begin{equation}
\frac{i}{\hbar }\partial _{t}\psi =\left( \triangle -\frac{1}{\hbar ^{2}}V(%
\mathbf{x},t)\right) \psi ,  \label{schrpt_1}
\end{equation}
where $\triangle $ is the Laplacian. As an example, Schr\"{o}dinger set out
to solve the problem of the evolution in a time-dependent potential of the
form $V=V_{0}\left( \mathbf{x}\right) +v\left( \mathbf{x},t\right) $. For
the important case of perturbation by external radiation, i.e. $v\left( 
\mathbf{x},t\right) =A(\mathbf{x})\cos (\omega t),$ he obtained at
first-order the solution 
\begin{eqnarray}
\psi \left( \mathbf{x},t\right) &=&u_{k}\left( \mathbf{x}\right) \exp \left( 
\frac{iE_{k}}{\hbar }t\right)  \label{schrpt_7} \\
&&+\frac{1}{2}\sum_{n=1}^{\infty }a_{kn}u_{n}(\mathbf{x})\left[ \frac{\exp
\left( \frac{i(E_{k}+\hbar \omega )}{\hbar }t\right) }{\left(
E_{k}-E_{n}+\hbar \omega \right) }+\frac{\exp \left( \frac{i(E_{k}-\hbar
\omega )}{\hbar }t\right) }{\left( E_{k}-E_{n}-\hbar \omega \right) }\right]
,  \nonumber
\end{eqnarray}
where the constant coefficients $a_{kn}=\left[ A\right] _{kn}$ are the
matrix elements of $A$ taken in the basis $u_{n}$ of the solutions of the
free $A=0$ case. His technique consisted in eliminating the time dependence
and use then the time-independent theory of his preceding paper. He
interpreted the expression above in terms of a time-dependent potential
driving the system out of its pure $u_{k}$ -oscillating mode and superposing
on it secondary oscillations of frequencies $\hbar ^{-1}E_{k}\pm \omega .$
The heuristics of such a formula were clearly of particular meaning for the
problem of dispersion (scattering of radiation off an atom). Indeed,
computing the resulting electric moment, Schr\"{o}dinger obtained a result
that he declared ''as formally quite identical with Kramers' formula for
secondary radiation''\footnote{%
Schr\"{o}dinger (1926d), p. 120.}.

\subsection{Dirac's perturbation theory for bosons}

In the paper \textit{On the Theory of Quantum Mechanics }(Dirac 1926c)%
\footnote{%
See the discussion of Bromberg 1977.}\textit{, }where the Einstein $B$
coefficient of stimulated emission and absorption is derived from first
principles, Dirac uses for the first time his perturbation theory for
Hamiltonians with explicit time dependence. It is a significant and decisive
step, its importance going much beyond the purely formal content. Indeed, it
offers the advantage to lend itself to quite general situations but, mainly,
together with Dirac's interpretation, it paves the way to field quantization
(1927b)\footnote{%
See e.g. Jost 1972, Bromberg 1977 and Cao 1997.} and naturally construes
elementary processes as transitions between ''free states''. Dirac's method
was to be adopted in most of the following works and its spirit conserved in
the modern techniques.

Dirac develops a systematic approach to deal with arbitrary (time-dependent)
perturbations in the following way. Assume that we have a (complete) set of
solutions $\varphi _{k}$ of a problem associated with the Hamiltonian $H_{0}$
, typically a non-interacting atom or particle;\ the general solution then
reads: $\varphi =\sum_{i}c_{i}\varphi _{i}$. Then consider a new Hamiltonian 
$H=H_{0}+H_{\text{int}}$. To solve the corresponding problem, Dirac proposes
to expand its solution $\psi $ on the set of the free ones:

\begin{equation}
\psi =\sum_{k}a_{k}(t)\varphi _{k}  \label{dpt_0}
\end{equation}
The a$_{k}$ 's are the expansion coefficients that remain to be determined%
\footnote{%
One should notice at this point that in his study of the relativistic scalar
equation Klein (1927) developed a perturbative scheme that used already the
expansion on free solutions. His technique is however less general and can
be seen as a special case of Dirac's. Klein's approach was derived for the
classical approximation of the scalar relativistic equation, involving only
first order time derivative.}. Dirac remarks further, that instead of
considering the a$_{k}$ 's to give the probabilities of being in the state $%
k $, one can as well think of them here as giving the number of individual
(disturbed) systems occupying the state $k$. Talking of atoms, Dirac says
(p. 674):

\begin{quotation}
We shall consider the general solution of [the undisturbed] equation to
represent an assembly of the undisturbed atoms in which $|c_{n}|^{2}$ is the
number of atoms in the $n$-th state, and shall assume that $\psi $ represent
in the same way an assembly of the disturbed atoms, $|a_{k}|^{2}$ being the
number in the $k$-th state at any time $t$.
\end{quotation}

The exact evolution equations for the $a_{k}$ 's are:

\begin{equation}
i\hbar \frac{d}{dt}a_{k}=\sum_{n}a_{n}\left[ H_{\text{int}}\right]
_{kn}\quad ,  \label{dpt_1}
\end{equation}
where $\left[ H_{\text{int}}\right] _{kn}$ are matrix elements of $H_{\text{
int}}$ (in the $\varphi $ basis). All that remains to be done is to solve
the above equations since the whole time-dependence of the solution is
contained in the $a$ 's. This can, and in most cases must be done,
perturbatively. For instance, assuming that at $t=0$, the system is entirely
in the unperturbed state, say $a_{0}(0)=1,$ so that $a_{k}=0;k\neq 0$, then
one can get, inserting on the right-hand side of (\ref{dpt_1}) the initial
values, at first order 
\begin{equation}
a_{k}=\frac{\left[ H_{\text{int}}\right] _{k0}(e^{i(E_{0}-E_{k})t/\hbar }-1)%
}{(E_{0}-E_{k})}  \label{dpt_2}
\end{equation}
If the matrix elements $\left[ H_{\text{int}}\right] _{k0}$ vanish (so that
there is no direct contribution to $a_{k})$ then the method of approximation
has to be refined\footnote{%
The formula (\ref{dpt_1}) as well as (\ref{dpt_2}) are not written or
discussed in their general form in the 1926c paper. There Dirac considers
specifically the first and second order solutions to the problem of
dispersion of radiation by atoms, where the expansion (\ref{dpt_0}) is for
the wave function of the atom. The case of the intermediate states will be
discussed by Dirac later in his paper 1927c. Our treatment here follows
Heitler 1936, p. 89.}. If, for some indices $n^{\prime }$, $\left[
H_{int}\right] _{kn^{\prime }}\neq 0\neq \left[ H_{int}\right] _{n^{\prime
}0},$ then one can get the following second order contribution passing
through ''intermediate states'' $n^{\prime }:$%
\begin{equation}
a_{k}=\sum_{n^{\prime }}\frac{\left[ H_{int}\right] _{kn^{\prime }}\left[
H_{int}\right] _{n^{\prime }0}}{(E_{0}-E_{n^{\prime }})}\left[ \frac{%
(e^{i(E_{0}-E_{k})t/\hbar }-1)}{(E_{0}-E_{n^{\prime }})}-\frac{%
(e^{i(E_{n^{\prime }}-E_{k})t/\hbar }-1)}{(E_{n^{\prime }}-E_{k})}\right]
\label{dpt_3}
\end{equation}

The formulas above constitute the main results of perturbation theory that
we shall find used by many authors in the next sections.

There is however much more to Dirac's technique. In his celebrated paper 
\textit{The quantum theory of emission and absorption of radiation }(Dirac
1927b, received 2 Febr. 1927) the whole discussion is resumed and the
equations (\ref{dpt_1}) are cast in a Hamiltonian form. Thus, the $a_{k}$ 's
and the complex conjugated $a_{k}^{*}$ 's acquire the status of the
dynamical variables of some new \textit{abstract} problem. The latter is
classical, but Dirac, with his fascinating intuition, asks us to consider
the quantum version of this new problem. He writes (p. 250):

\begin{quotation}
...we can describe the effect of a perturbation on an assembly of
independent systems by means of canonical variables and Hamiltonian
equations of motion. The development of the theory which naturally suggests
itself is to make these canonical variables q-numbers satisfying the usual
quantum conditions instead of c-numbers, so that their Hamiltonian equations
of motion become true quantum equations. The Hamiltonian function will now
provide a Schr\"{o}dinger wave equation, which must be solved and
interpreted in the usual manner. The interpretation will give not merely the
probable number of systems in any state, but the probability of any given
distribution of the systems among the various states, this probability
being, in fact, equal to the square of the modulus of the normalized
solution of the wave equation that satisfies the appropriate initial
conditions. We could of course calculate directly from elementary
considerations the probability of any given distribution when the systems
are independent, as we know the probability of each system being in any
particular state. We shall find that the probability calculated directly in
this way does not agree with that obtained from the wave equation except in
the special case when there is only one system in the assembly. In the
general case it will be shown that the wave equation leads to the correct
value for the probability of any given distribution when the systems obey
Einstein-Bose statistics instead of being independent.
\end{quotation}

So, now, the $a_{k}^{*}$ 's and the $a_{k}$ 's become q-numbers, namely the
now familiar creation and annihilation operators of the corresponding
(undisturbed) boson states\footnote{%
Actually Dirac will quantize the canonical variables 
\begin{eqnarray*}
b_{r} &=&a_{r}e^{-\frac{i}{\hbar }W_{r}t} \\
b_{r}^{*} &=&a_{r}^{*}e^{\frac{i}{\hbar }W_{r}t}
\end{eqnarray*}
where $W_{r}$ is the energy of state $r,$ setting 
\[
\lbrack b_{r},b_{s}^{*}]=\delta _{rs} 
\]
This enables to factor out the trivial time dependence due to free evolution.%
}. A huge step has been taken. Dirac will show next that his
second-quantized procedure (now the name is much more meaningful) properly
takes into account Bose-Einstein statistics. Dirac discusses this point
using instead of the operators $a_{k}$ , $a_{k}^{*}$ the canonical variables
corresponding to the operators $N_{k}=a_{k}^{*}a_{k}$ and their canonical
conjugated $\theta _{k}.$ This makes the wave function of this
second-quantized problem depend on the occupation numbers of the states of
the bosons. Indeed, the wave function is defined on the manifold of the $%
N_{k}$ variables, or more properly, the manifold defined by the spectrum of
the $N_{k}$ operators which are the positive integers\footnote{%
For further details about the relation between a choice of state-space basis
and the variables entering the wave-functions, see Dirac's ''transformation
theory'' as developped in Dirac's ''The Physical Interpretation of Quantum
Dynamics'', \textit{Proc. Roy. Soc. (London) A, }vol. 113 (1927), pp.
621-641, received 2 Dec. 1926.}).

So far, Dirac's analysis concerns an assembly of bosons the perturbation of
which results from a potential expressed solely in terms of boson variables.
There is as yet no ''coupling'' to some other system. In the same paper
(1927b), Dirac will consequently generalize his technique, coupling the
bosons to a given external system, typically an atom. When the bosons are
light quanta, and although the interaction matrix elements are\textit{\ a
priori} not known, Dirac is already able at this point to derive the correct
dependence of the probabilities for stimulated and spontaneous emission
(Einstein's $A$ and $B$ coefficients) upon the intensity of the incident
radiation. This is because the algebra of boson operators contains
automatically the right dependence on the occupation numbers.

With Dirac's perturbation theory and the remarkable use he will make of it
in his quantization of the radiation field, we enter definitely into a new
era. The basic ingredients of the machinery of QED are now set,\ but several
important problems are not yet resolved. One feature that remains to be
included is a proper way to take into account the reaction of the emitted
radiation on the source itself, called \textit{R\"{u}ckwirkung}, which we
translate here as backreaction. This problem is related to the more
fundamental problem of a relativistic and interacting theory of particles
and fields. For now, we want to examine some early applications of Dirac's
perturbation/quantization method which recast Compton scattering in the now
familiar formulation of a transition from initial to final states through
successive absorption/emission (or emission/absorption) of an intermediate
photon, but where the reaction of the radiation on the electron is still not
considered.

\subsection{Dirac's quantization of the radiation field}

A spectacular application of the formalism above is Dirac's quantized theory
of the interaction of matter with radiation . In the 1927b paper this is
achieved in the following way: the radiation field is first quantized as an
assembly of independent bosons, then formally coupled to an external system
according to the scheme of the preceding section. To determine explicitely
this coupling to matter, Dirac considers the ''wave'' point of view, where
this time he expands the vector potential $\mathbf{A}$ into its Fourier
components (which he supposes of large but finite number $\mathbf{A=}\sum_{r}%
\mathbf{A}_{r}$). In this ''wave point of view'', the coupling is
explicitely given by 
\begin{equation}
c^{-1}\sum_{r}\mathbf{\dot{X}A}_{r}  \label{dircoup}
\end{equation}
involving the time-derivative of the total ''polarization'' $\mathbf{X}$ of
the (atom) system in the direction of $\mathbf{\ A}_{r}$, where relativistic
effects have been neglected\footnote{%
The term polarization might appear ambiguous. One may offer the following
justification. In the Hamiltonian 
\begin{equation}
\frac{1}{2m}(\mathbf{p}+\frac{e}{c}\mathbf{A)}^{2}
\end{equation}
\par
Dirac writes down the $\frac{e}{mc}\mathbf{pA}$ term as $\frac{1}{c}\frac{d}{
dt}(e\mathbf{x)A}$ using $\mathbf{p=}\frac{d}{dt}(m\mathbf{x)}$ with $%
\mathbf{x}$ the ''position'' of the charge $e.$ Thus $\mathbf{X=}(e\mathbf{%
x).}$ The $\mathbf{A}^{2}$term is neglected and the term $\frac{1}{2m}%
\mathbf{p}^{2}$included in the Hamiltonian for matter.}. After some
manipulation, Dirac shows the equivalence of this point of view with the
previous ''bosonic'' one. This is satisfying, as the Fourier expansion
amounts to the decomposition of the vector potential $\mathbf{A}$ in a basis
of plane waves, which are the solutions for the free radiation (see eq. \ref
{dpt_0}). The payoff is that the matrix elements of the interaction are now
explicitely determined. Eventually, Dirac ends up with the following
formalism: One first considers the Hamiltonian of matter and free bosons $%
H_{0}=H_{\text{matter}}+\sum_{r}W_{r}N_{r}$ where $H_{\text{matter}}$ is
originally the Hamiltonian describing atomic states, $W_{r}=\hbar \omega
_{r} $ are the energies of the free photon states. $N_{r}$ is the occupation
number operator of the $r$ -th state;\ together with its canonically
conjugated variable $\theta _{r,}$they are related to the algebra of
creation and annihilation operators by. 
\begin{eqnarray}
b_{r} &=&N_{r}^{1/2}e^{-i\theta _{r}/\hbar }  \label{dr_2} \\
b_{r}^{*} &=&N_{r}^{1/2}e^{i\theta _{r}/\hbar }  \nonumber \\
\lbrack b_{r},b_{s}^{*}] &=&\delta _{r,s}  \nonumber
\end{eqnarray}
In the coupling term (\ref{dircoup}), the norm $A_{r}$ of each component $%
\mathbf{A}_{r}$ is expressed now as $A_{r}=a_{r}\cos \theta _{r}/\hbar ,$;
using the relation between the intensity of the wave and the energy of the
corresponding photons one obtains \footnote{%
A plane wave corresponds to an energy flux given by $(c/8\pi
)a_{r}^{2}\omega _{r}^{2}$. If one introduces the density of states $\rho
_{r}$ about state $r$ per unit frequency range and solid angle (and definite
polarisation, not made explicit), the intensity per unit frequency range of
the radiation in the neighbourhood of the component $r$ is accordingly $%
I_{r}=(\pi /2c)a_{r}^{2}\nu _{r}^{2}\rho _{r}.$ On the other hand, in the
''boson'' picture, $I_{r}=N_{r}h\nu _{r}^{3}/c^{2};$ this relations is
obtained studying the density of states about the component $r$
characterized by a wave vector $\mathbf{k}_{r}\ $such that $\left| \mathbf{k}%
_{r}\right| ^{2}=\left( \frac{\omega _{r}}{c}\right) ^{2}.$} 
\[
A_{r}=a_{r}\cos \theta _{r}/\hbar =2\sqrt{\frac{\hbar \nu _{r}}{c\rho _{r}}}%
\sqrt{N_{r}}\cos \theta _{r}/\hbar , 
\]
At this point, following the quantization rules established before, Dirac
writes 
\[
\sqrt{N_{r}}\cos \theta _{r}/\hbar =N_{r}^{1/2}e^{i\theta _{r}/\hbar
}+e^{-i\theta _{r}/\hbar }(N_{r})^{1/2}=N_{r}^{1/2}e^{i\theta _{r}/\hbar
}+(N_{r}+1)^{1/2}e^{-i\theta _{r}/\hbar } 
\]
so that 
\[
A_{r}=2(\hbar \nu _{r}/c\rho _{r})^{1/2}\{N_{r}^{1/2}e^{i\theta _{r}/\hbar
}+(N_{r}+1)^{1/2}e^{-i\theta _{r}/\hbar }\} 
\]
and finally the interaction Hamiltonian becomes 
\begin{equation}
H_{\text{int}}=2(\hbar ^{1/2}c^{-3/2})\sum_{r}\sqrt{\frac{\nu _{r}}{\rho _{r}%
}}\dot{X}_{r}\left\{ N_{r}^{1/2}e^{i\theta _{r}/\hbar
}+(N_{r}+1)^{1/2}e^{-i\theta _{r}/\hbar }\right\} .  \label{dr_1}
\end{equation}
This interaction shall be replaced by a relativistically accurate expression
(see note above) in a subsequent paper by Dirac, entitled \textit{Quantum
Theory of dispersion}, (Dirac 1927c, received 4 April 1927). For the time
being, because the coupling above is linear in the field, Dirac notes (Dirac
1927c, p. 712):

\begin{quotation}
The interaction term [eq. (\ref{dr_1})] of the Hamiltonian function [...]
does not give rise to any direct scattering process, in which a
light-quantum jumps from one state to another of the same frequency but
different direction of motion (i.e. the corresponding matrix element $%
v_{mm^{\prime }}=0$ ). All the same, radiation that has apparently been
scattered can appear by a double process in which a third state, $n$ say,
with different proper energy from $m$ and $m^{\prime }$ , plays a part.
[...] The scattered radiation thus appears as the result of the two
processes [...] one of which must be an absorption and the other an
emission, in neither of which is the total proper energy even approximately
conserved.
\end{quotation}

On the other hand, the full interaction Hamiltonian which does not neglect
the terms quadratic in $A$ does yield contributions to direct scattering
processes.

The 1926c and 1927b,c papers of Dirac set the basic language and concepts
characteristic of the modern conception of the elementary processes as
sequences of absorption/emission of light quanta, passing through a sequence
of intermediate states. Let us quote him again, this time with a lucid
discussion of the apparent violation of conservation laws (Dirac 1927b, p.
265):

\begin{quotation}
Also the theory enables one to understand how it comes about that there is
no violation of the law of the conservation of energy when, say, a
photo-electron is emitted from an atom under the action of extremely weak
incident radiation. The energy of interaction of the atom and the radiation
is a q-number that does not commute with the first integrals of the motion
of the atom alone or with the intensity of the radiation Thus one cannot
specify this energy by a c-number at the same time that one specifies the
stationary state of the atom and the intensity of the radiation by
c-numbers. In particular, one cannot say that the interaction energy tends
to zero as the intensity of the incident radiation tends to zero. There is
thus always an unspecifiable amount of interaction energy which can supply
the energy for the photo-electron.
\end{quotation}

\subsection{Waller: Compton scattering with quantized radiation}

The success of the first applications of Dirac's technique (computation of
Einstein's $A$ and $B$ coefficients in 1927b, rederivation of the Kramers'
and Heisenberg's formulas and study of the case of resonance in 1927c) will
motivate others to apply it to further problems. However almost two years
pass before the first to react, Ivar Waller (Waller 1929, received 21 July
1929 and Waller 1930, received the 12 Febr. 1930), at that time at the
University of Upsala, and Igor Tamm (Tamm 1930, received the 7 Apr. 1930) in
Moscow, publish their results\footnote{%
One should also note here a paper by Wentzel (1929) where the latter uses as
well a quantized radiation field but in a non relativistic approximation.}.

The first contribution of Waller, \textit{Die Streuung kurzwelliger
Strahlung durch Atome nach der Diracschen Strahlungstheorie}, dates from
1929 (Waller 1929). There, Waller considers again the problem of radiation
scattering off an atom, neglecting relativistic effects at the level of the
electron bound in the Coulomb potential. He uses on the other hand the full
coupling $\frac{1}{2m}(p+$ $\frac{e}{c}A)^{2}$ without neglecting the $A^{2}$
terms. After studying transitions at the level of the discrete spectrum, he
considers also its continuous part (ionization) and ends up with a short
account of the relativistic case, where he quotes some preliminary results
obtained using Dirac's equation for the electron. He explicitly mentions the
need of taking into account the ''negative energy'' intermediate states, but
further analysis is missing. It will appear in full extent in Waller's next
paper, \textit{Die Streuung von Strahlung durch gebundene und freie
Elektronen nach der Diracschen relativistischen Mechanik }(Waller 1930,
received 12 Feb. 1930). Technically, Waller's 1930's paper is a
straightforward application of the techniques put forward by Dirac. As
previously Klein and Nishina, Waller considers the interacting equation 
\begin{equation}
\left\{ i\left( -\gamma _{0}(E+eV)+\left( \mathbf{\gamma },\mathbf{p}+e%
\mathbf{A}\right) \right) +mc^{2}\right\} \psi =0  \label{w_0}
\end{equation}
After introducing the quantized potential $\mathbf{A}$ and following the
steps of Dirac, Waller obtains an equation for the wave-function $\Phi
(J;...,N_{r})$ depending collectively on atom (electron) variables $J$ , and
light-quanta occupation numbers $N_{r}$ of states $r$) 
\begin{eqnarray}
&&\left[ i\hbar \frac{d}{dt}-E(J^{\prime })-\sum_{r}N_{r}\hbar \omega
_{r}\right] \Phi (J^{\prime };...,N_{r};...)  \label{w_1} \\
&=&\sum_{J}\sum_{r}\sqrt{N_{r}}M^{r}(J^{\prime }J)\Phi (J;...,N_{r}-1;...) 
\nonumber \\
+ &&\sqrt{(N_{r}+1)}\tilde{M}^{r}(J^{\prime }J)\Phi (J;...,N_{r}+1;...) 
\nonumber
\end{eqnarray}
Here, we have used the following notation for the matrix elements of the
interaction corresponding to the $r$-th component: 
\begin{eqnarray}
M^{r}(J^{\prime }J) &\equiv &\mathbf{\epsilon }_{r}\int \bar{\psi}%
_{J^{\prime }}(\mathbf{x})\mathbf{\gamma }\psi _{J}(\mathbf{x})e^{-\frac{i}{%
\hbar }\mathbf{k}_{r}\mathbf{x}}d\mathbf{x}  \label{w_2} \\
\tilde{M}^{r}(J^{\prime }J) &\equiv &\mathbf{\epsilon }_{r}\int \bar{\psi}%
_{J^{\prime }}(\mathbf{x})\mathbf{\gamma }\psi _{J}(\mathbf{x})e^{+\frac{i}{
\hbar }\mathbf{k}_{r}\mathbf{x}}d\mathbf{x},  \nonumber
\end{eqnarray}
$\psi _{J}(\mathbf{x})$ are eigensolutions for the motion of the electron in
an external (atomic) potential $V$ and $E(J)$ the corresponding energies.
Depending on the value of $V$, one can deal with the scattering off a bound
electron or off a free one ($V=0)$. One has further $\mathbf{\epsilon }_{r}=%
\sqrt{(\frac{e^{2}\hbar \nu _{r}}{c\rho _{r}})}\mathbf{e}_{r},$ where $%
\mathbf{e}_{r},$is the polarization of the $r$-th (Fourier) component of the
radiation, and $\mathbf{k}_{r}=\hbar \frac{\omega _{r}}{c}\mathbf{\hat{k}}%
_{r},$the momentum of the $r$-th state associated light quantum\footnote{%
For the numerical factor in front of $\mathbf{e}_{r}$, see the discussion of
Dirac's 1927b in the previous section. Our notations differ from Waller's,
namely one has $M\mathbf{\rightarrow A,}$ $\tilde{M}\mathbf{\rightarrow B,}$ 
$\mathbf{\epsilon \rightarrow \mu .}$and moreover we have included the
scalar product of the current $\bar{\psi}_{J^{\prime }}(\mathbf{x})\mathbf{%
\gamma }\psi _{J}(\mathbf{x})$ with the polarization $\mathbf{e}_{r}$%
directly in the definition of the matrix elements $M$ and $\tilde{M}$. Other
changes involve Waller's use of primed quantities as denoting the
eigenvalues of the corresponding operators, a notation introduced by Dirac
(see for instance The Physical Interpretation of the Quantum Dynamics, 
\textit{Proc. Roy. Soc. (London) A}, vol. 113 (1927), pp. 621-641. For the
sake of making the formulas more transparent, we have dropped this usage.}.

(\ref{w_1}) is of the form 
\[
\left[ i\hbar \partial _{t}-W(m))\right] \Phi (m)=\sum_{m^{\prime
}}(m|M|m^{\prime })\Phi (m^{\prime }) 
\]
where $(m|M|m^{\prime })$ is the $m$ ,$m^{\prime }$ matrix element of $M$,
and this time the variables $m$ label the states of the whole system. In
particular, the emission and absorption processes correspond to the elements 
\begin{eqnarray}
(J^{\prime };...,N_{r}+1,...|M|J;...,N_{r},...) &=&\sqrt{(N_{r}+1)}%
M^{r}(J^{\prime }J)  \label{w_3} \\
(J^{\prime };...,N_{r}-1,...|M|J;...,N_{r},...) &=&\sqrt{N_{r}}\tilde{M}%
^{r}(J^{\prime }J)  \nonumber
\end{eqnarray}
Assuming that at time $t=0$ the system is in the state labelled by $%
(J;...0,0,..,N_{s},...0)$ and after applying Dirac's second-order
perturbation theory formula (see \ref{dpt_3}), Waller obtains the following
expression for the transition amplitude $a$ to the state $(J^{\prime
};...0,1_{r},..,N_{s}-1,...0):$%
\begin{eqnarray}
&&a(J^{\prime };...0,1_{r},..,N_{s}-1,...0)=  \label{w_4} \\
- &&(e^{i\hbar \beta t}-1)\beta ^{-1}\sqrt{N_{s}}\sum_{J^{\prime \prime }}%
\frac{\left[ M^{r}(J^{\prime }J^{\prime \prime })\right] \left[ \tilde{M}
^{s}(J^{\prime \prime }J)\right] }{E(J)-E(J^{\prime \prime })+\hbar \omega
_{s}}  \nonumber \\
- &&(e^{i\hbar \beta t}-1)\beta ^{-1}\sqrt{N_{s}}\sum_{J^{\prime \prime }}%
\frac{\left[ \tilde{M}^{s}(J^{\prime }J^{\prime \prime })\right] \left[
M^{r}(J^{\prime \prime }J)\right] }{E(J)-E(J^{\prime \prime })-\hbar \omega
_{r}}  \nonumber
\end{eqnarray}
Here $\beta =E(J^{\prime })+\hbar \omega _{r}-E(J)-\hbar \omega _{s}.$ The
formula clearly exhibits the two contributions corresponding respectively to
intermediate states $(J^{^{\prime \prime }};...0,..,N_{s}-1,...0)$ and $%
(J^{^{\prime \prime }};...0,1_{r},..,N_{s},...0).$ The probability of
scattering a photon of polarization $\mathbf{e}_{r}$ into the solid angle $%
d\Omega _{r}$ with the simultaneous transition of the electron to state $%
J^{\prime }$ is then given by 
\begin{equation}
\rho _{r}d\Omega _{r}\int \left| a(J^{\prime
};...0,1_{r},..,N_{s}-1,...0)\right| ^{2}d\nu _{r}=\hbar
^{-2}t|\sum_{J^{\prime \prime }}|^{2}\rho _{r}N_{s}d\Omega _{r}  \label{w_5}
\end{equation}
where the sum in the absolute value bars is the same as the one from above,
and the factor $\rho _{r}$ is the density of states about state $r$. For an
incident intensity $I_{0}$ the power $P_{r}$ radiated into the solid angle
in the direction $\mathbf{\hat{k}}_{r},$ and with polarisation $\mathbf{\
\epsilon }_{r}$ is then given by

\begin{eqnarray*}
P_{r} &=&\frac{I_{0}e^{4}\omega _{r}^{2}}{\omega _{s}^{2}}|\sum_{J^{\prime
\prime }}\frac{\left[ M^{r}(J^{\prime }J^{\prime \prime })\right] \left[ 
\tilde{M}^{s}(J^{\prime \prime }J)\right] }{E(J)-E(J^{\prime \prime })+\hbar
\omega _{s}} \\
&&+\sum_{J^{\prime \prime }}\frac{\left[ \tilde{M}^{s}(J^{\prime }J^{\prime
\prime })\right] \left[ M^{r}(J^{\prime \prime }J)\right] }{E(J)-E(J^{\prime
\prime })-\hbar \omega _{r}}|^{2}
\end{eqnarray*}
Waller remarks that in the derivation of the formula above the back-reaction
of the radiation on the scattering system has not been taken into account,
therefore one can obtain the same result without quantizing the field.

One can now specialize to the free case (in the bound case, Waller neglects
the relativistic and spin effects). Here, Waller takes for the $\psi _{J}(x)$
the free plane-wave spinor solutions 
\begin{equation}
\psi ^{\sigma }(\mathbf{p})=u^{\sigma }(\mathbf{p})e^{-\frac{i}{\hbar }%
(E^{\sigma }t-\mathbf{px})}  \label{w_6}
\end{equation}
with the conventions that for $\sigma =1,2$ one deals with positive energy
solutions $E=c\sqrt{m^{2}c^{2}+\mathbf{p}^{2}}$ and $\sigma =3,4$ with
negative ones: the normalizations of the single-row spinors are 
\begin{eqnarray}
\sum_{l=1}^{4}u_{l}^{*\sigma }(\mathbf{p})u_{l}^{\sigma ^{\prime }}(\mathbf{p%
}) &=&h^{-3}\delta (\sigma -\sigma ^{\prime });  \label{w_7} \\
\sum_{\sigma =1}^{4}u_{l}^{*\sigma }(\mathbf{p})u_{l^{\prime }}^{\sigma }(%
\mathbf{p}) &=&h^{-3}\delta (l-l^{\prime })  \nonumber
\end{eqnarray}
further 
\[
\int \psi ^{\sigma }(\mathbf{p})\psi ^{\sigma ^{\prime }}(\mathbf{p}^{\prime
})d\mathbf{x}=\delta (\sigma -\sigma ^{\prime })\delta (\mathbf{p}-\mathbf{p}%
^{\prime }) 
\]
Expanding the wave-function for the electron and radiation using a
3-dimensional Fourier transform, 
\begin{equation}
\Psi (x,N)=\int \sum_{\sigma }\Phi ^{\sigma }(\mathbf{p,}N)u^{\sigma }(%
\mathbf{p})e^{-\frac{i}{\hbar }(\mathbf{px})}d\mathbf{p}  \label{walfour}
\end{equation}
and inserting into (\ref{w_0}) with $V=0$, one recovers (\ref{w_1}), where $%
\mathbf{p}$, together with the index $\sigma ,$ play now the role of the
variable $J.$ Going through the steps already discussed above, assuming that
the electron initial state is a (positive energy) superposition with a
momentum distribution $\alpha ^{\sigma }(\mathbf{p})$, and that there are $%
N_{s}$ initial photons, the amplitude formula (\ref{w_4}) reads, resuming
our usual conventions for the labelling of\textit{\ \ in} and \textit{out}
states 
\[
a^{\sigma ^{\prime }}(\mathbf{p}^{\prime };1_{r},..,N_{s}-1)\equiv a^{\sigma
^{\prime }}(\mathbf{p}^{\prime },\mathbf{k}^{\prime })=-\sqrt{N}\sum_{\sigma
=1}^{2}B^{\sigma ^{\prime }\sigma }\alpha ^{\sigma }(\mathbf{p}^{\prime }+%
\mathbf{k}^{\prime }-\mathbf{k})(e^{\frac{i}{\hbar }\beta t}-1)/\beta 
\]

with $\beta =E^{\sigma ^{\prime }}(\mathbf{p}^{\prime })+\hbar \omega
^{\prime }-E^{\sigma }(\mathbf{p})-\hbar \omega $ and where 
\begin{eqnarray}
B^{\sigma ^{\prime }\sigma } &=&h^{6}\sum_{\rho }\frac{\left[ \mathbf{\
\epsilon }^{\prime }\mathbf{\Pi }_{\rho ;\mathbf{p}+\mathbf{k}}^{\sigma
^{\prime };\mathbf{p}^{\prime }}\right] \left[ \mathbf{\epsilon \Pi }
_{\sigma ;\mathbf{p}}^{\rho ;\mathbf{p}+\mathbf{k}}\right] }{E^{\sigma }(%
\mathbf{p})-E^{\rho }(\mathbf{p}+\mathbf{k})+\hbar \omega }  \label{w_24} \\
+ &&h^{6}\sum_{\rho }\frac{\left[ \mathbf{\epsilon \Pi }_{\rho ;\mathbf{p}-%
\mathbf{k}^{\prime }}^{\sigma ^{\prime };\mathbf{p}^{\prime }}\right] \left[ 
\mathbf{\epsilon }^{\prime }\mathbf{\Pi }_{\sigma ;\mathbf{p}}^{\rho ;%
\mathbf{p}-\mathbf{k}^{\prime }}\right] }{E^{\sigma }(\mathbf{p})-E^{\rho }(%
\mathbf{p}-\mathbf{k}^{\prime })-\hbar \omega ^{\prime }}  \nonumber
\end{eqnarray}
with the notation for the current 
\begin{equation}
\mathbf{\Pi }_{\sigma \mathbf{p}}^{\rho \mathbf{p}^{\prime }}=\bar{u}^{\rho
}(\mathbf{p}^{\prime })\mathbf{\gamma }u^{\sigma }(\mathbf{p}).  \label{w_25}
\end{equation}
Again, the structure of the intermediate states is explicit;\ notice in
particular that the index $\rho $ runs over all four values, which takes
into account negative energy states as well. The probability for a
transition to a final state with the electron momentum $\mathbf{p}^{\prime }$
and spin $\sigma ^{\prime },$ and the photon $\mathbf{k}^{\prime }$ and $%
\mathbf{e}^{\prime }$ is found linear in time 
\begin{equation}
\rho ^{\prime }d\mathbf{p}^{\prime }d\Omega ^{\prime }\int \left| a^{\sigma
^{\prime }}(\mathbf{p}^{\prime };\mathbf{k}^{\prime })\right| ^{2}d\nu
^{\prime }=t\frac{\rho ^{\prime }N}{\hbar ^{2}\Delta ^{\sigma }(\mathbf{p,%
\hat{k}}^{\prime })}\left| \sum_{\sigma }B^{\sigma ^{\prime }\sigma }\alpha
^{\sigma }(\mathbf{p})\right| ^{2}d\mathbf{p}^{\prime }d\Omega ^{\prime }
\label{w_23}
\end{equation}

To obtain the right hand side above, one trades the integration over $\nu
^{\prime }$ for the one over $\beta ,$ with the Jacobian. 
\begin{equation}
\frac{d\beta }{d\nu ^{\prime }}=h\Delta ^{\sigma }(\mathbf{p,\hat{k}}
^{\prime })\equiv h\left[ 1-\frac{c}{E^{\sigma }(\mathbf{p})}\mathbf{p\hat{k}%
}^{\prime }\right] ;\quad \mathbf{p=p}^{\prime }+\mathbf{k}^{\prime }-%
\mathbf{k}
\end{equation}
and notices that the resulting integral has a sharp maximum around $\beta
=0, $ which ensures energy conservation. To get an intensity formula, one
has further to integrate \ref{w_23} over the final electron momentum, and
sum over (final) spin states. Instead of integrating over $\mathbf{p}
^{\prime }\mathbf{\ }$one can as well use $\mathbf{p,}$ with the Jacobian 
\begin{equation}
\frac{d\mathbf{p}^{\prime }}{d\mathbf{p}}=\frac{\Delta ^{\sigma }(\mathbf{p,%
\hat{k}}^{\prime })}{\Delta ^{\sigma ^{\prime }}(\mathbf{p}^{\prime },%
\mathbf{\hat{k}}^{\prime })}=\frac{\nu ^{\prime }E^{\sigma ^{\prime }}(%
\mathbf{p}^{\prime })}{\nu E^{\sigma }(\mathbf{p})}\frac{\Delta ^{\sigma }(%
\mathbf{p,\hat{k}}^{\prime })}{\Delta ^{\sigma ^{\prime }}(\mathbf{p,\hat{k}}%
)}.
\end{equation}
The cross section formula for a positive energy final electron state is then 
\[
\frac{d\sigma }{d\Omega }=\frac{e^{4}}{\nu ^{2}}\frac{E(\mathbf{p}^{\prime })%
}{mc^{2}}\sum_{\sigma ^{\prime }=1}^{2}\int |\sum_{\sigma =1}^{2}B^{\sigma
^{\prime }\sigma }\alpha ^{\sigma }(\mathbf{p})|^{2}d\mathbf{p} 
\]
In the sequel of his calculation, Waller assumes the case of an electron
initially at rest ($\alpha ^{\sigma }(\mathbf{p})$ is vanishing except in
the vicinity of $\mathbf{p}=0\mathbf{).}$ To obtain the total intensity (the
K-N\ formula), it remains to work out the products of the currents (\ref
{w_25}).and add the intensities for a choice of two orthogonal polarizations 
$\mathbf{e}^{\prime }$.

Waller devotes special care to the discussion of the role of the negative
energy states in the summations over intermediate states in (\ref{w_24}), $%
\rho =3$ and $4$. Thus, for the free electron case, he shows that it is
crucial to take them into account in order to obtain the Klein-Nishina
formula. This fact is rendered especially dramatic in the classical limit ( h%
$\nu /mc^{2}\ll 1$) where only negative energy intermediate states
contribute to the classical scattering formula. One has to remember that
Dirac only in the same year proposed his negative-energy sea hypothesis
(Dirac 1930), and that the paradoxes connected with the negative energy
solutions (Klein 1929) were a major source of worry. The paradoxical need of
the negative states as intermediate states of the scattering process must
have certainly added to the confusion.

\subsection{Tamm: matter and radiation quantized}

Even if the semi-classical methods of Gordon and Klein-Nishina seem fully
sufficient to derive the correct expressions for the Compton scattering at
the order considered, Tamm's 1930's contribution goes even further than the
latter in that it considers, in addition to a quantized radiation field, a
quantized matter field, following the lines set up by Heisenberg and Pauli
(Heisenberg and Pauli 1929). Tamm rederives thus the Klein-Nishina formula
in a fully consistent quantum way, which he considers to be a proof of the
validity of the semi-classical approach (p. 545):

\begin{quotation}
The scattering of the radiation by free electrons has already been studied
many times. However, one mainly has quantized only the electron motion and,
using the correspondence principle, determined the scattered radiation from
the calculated distribution of the electron current. Not only is this
procedure logically unsatisfactory, but it is ambiguous\footnote{%
Compare for instance the treatment of Smekal's ''combination scattering'' of
radiation off atoms, on one hand as in E. Schr\"{o}dinger (Ann. d. Phys.,
vol. 81 (1926), p. 109), and as in O. Klein (Zs. f. Phys., vol. 41 (1927),
p. 407) on the other (note in the original text).}. It is therefore of
interest to start the problem anew applying systematically the
quantum-mechanical approach (quantization of the electromagnetic field and
of the $\psi $ wave) within the framework of Dirac's wave equation for the
electron. One rederives in this way the scattering formula of Klein and
Nishina which legitimates the correspondence principle approach used by the
latter.\footnote{%
Die Streuung der Strahlung durch freie Elektronen ist schon vielfach
untersucht worden. Man hat aber dabei meistenteils nur die
Elektronenbewegung gequantelt und die Streustrahlung korrespondenzm\"{a}ssig
aus der berechneten Verteilung des Elektronenstroms bestimmt. Dieses
Verfahren ist aber nicht nur logisch unbefriedigend, sondern auch nicht
eindeutig. Es schien deshalb von Interesse zu sein, das Problem von neuem in
einer konsequenten quantenmechanischen Weise (Quantelung des
elektromagnetischen Feldes und der $\psi $-Wellen) unter Zugrundelegung der
Diracschen Wellengleichung des Elektrons zu behandeln. Man gelangt in dieser
Weise wieder zu der von Klein und Nishina abgeleiteten Streuformel, wodurch
die Folgerichtigkeit der von diesen Forschern durchgef\"{u}hrten
korrespondenzm\"{a}ssigen Behandlung des Problems best\"{a}tigt wird.}
\end{quotation}

In his paper, Tamm recognizes as well the necessity to take into account the
negative energy intermediate states in the derivation of the KN scattering
formula. The last part of his paper is devoted to the computation of the
spontaneous transition of an electron from a positive to a negative energy
state with the resulting emission of two photons. This process, resp. its
inverse, is interpreted by Tamm along the lines of Dirac's hole theory
(Dirac 1930) as pair annihilation and creation respectively; the negative
energy Dirac electron is intepreted further as a proton (p. 547):

\begin{quotation}
This processes can be treated consequently according to Dirac's theory only
when one will succeed in understanding theoretically the interaction of
negative-energy electrons, the vanishing of the field of the electron and
the ''hole'' in the act of annihilation etc. With the present status we
could calculate the spontaneous electron transition from a level of positive
energy to one of negative energy only when totally neglecting these
circumstances. If one interprets the unoccupied negative energy level as a
proton and the transition just mentionned as an annihilation then the result
can be summarized as follows: the annihilation occurs in a collision of the
electron and the proton;\ the effective cross-section relevant to this
process is equal to the classically computed cross-section 
\[
\pi d^{2}=\pi \left( \frac{e^{2}}{mc^{2}}\right) ^{2}
\]
of the elementary charge $e$. Since Dirac's theory is, in the absence of
interactions, symmetrical as concerns electrons and protons, one has to
interpret $m$ in the above expression as specific, but for the time
unspecifiable average of the electron and proton masses\footnote{%
Diese Prozesse k\"{o}nnen erst dann folgerichtig nach der Diracschen Theorie
behandelt werden, wenn es gelingt, die Wechselwirkung der Elektronen
negativer Energie, das Verschwinden des Feldes des Elektrons und des
''Loches'' bei der Zerstrahlung usw. theoretisch zu erfassen. Bei dem
jetzigen Stande der Theorie konnten wir dagegen die Wahrscheinlichkeit des
spontanen Elektronen\"{u}berganges von einem positiven nach einem negativen
Energieniveau nur unter g\"{a}nzlicher Vernachl\"{a}ssigung dieser
Umst\"{a}nde berechnen. Bezeichnet man das unbesetzte negative Energieniveau
als ein Proton und den erwahnten \"{U}bergang als eine Zerstrahlung, so kann
das Ergebnis folgendermassen zasammengefasst werden: die Zerstrahlung findet
bei einem Zusammenstoss des Elektrons und des Protons statt; der f\"{u}r
einen solchen Zusammenstoss massgebende effektive Querschnitt dieser
Teilchen ist gleich dem klassisch berechneten Querschnitt 
\[
\pi d^{2}=\pi \left( \frac{e^{2}}{mc^{2}}\right) ^{2}
\]
der Elementarladung $e$. Da bei der Vernachl\"{a}ssigung der Wechselwirkung
die Diracsche Theorie in bezug auf Elektronen und Protonen symmetrisch ist,
so ist in diesem Ausdruck unter $m$ vermutlich ein bestimmter, vorl\"{a}ufig
nicht n\"{a}her anzugebender Mlittelwert der Masse eines Elektrons und eines
Protons zu verstehen.}.
\end{quotation}

It should be noted that contrary to the Compton effect this annihilation
process does not have a classical counterpart in the context of the
correspondence principle. When commenting his result, Tamm points out that
the lifetime for the hydrogen atom, $10^{-3}$ sec., turns out to be much too
small

Tamm follows the lines set by Dirac, and his obtention of the explicit
scattering formulas is a matter of technical manipulations much of the same
nature as already discussed in the case of Waller. We shall therefore not
dwelve any further on the content of Tamm's paper, also because the issue of
second-quantization of matter waves, its most distinctive feature, is not
relevant to the work of Stueckelberg.

Before closing this section, it is interesting to signal the paper of Seishi
Kikuchi (1931). Using the Heisenberg-Pauli formalism, he shows that \textit{%
at the same time} the photon is scattered, the electron recoils, within
Heisenberg's uncertainty relations. The Heisenberg-Pauli formalism as
applied to matter-radiation interaction has also been studied by Oppenheimer
(1930). Indeed, at the end of his paper, Waller (1930) remarks that
Oppenheimer was able to obtain the equation (\ref{w_1}) from the
Heisenberg-Pauli formalism, at the price of neglecting the self-energy of
the electron.

\section{The emergence of the interaction picture}

As we mentioned in the previous section, the problem of properly taking into
account the backreaction of the emitted radiation on the source is not
solved in the above papers. Here is the problem as analyzed by Dirac (Dirac
1932):

\begin{quotation}
We shall now consider in detail the question of how the information
contained in classical electrodynamics can be taken over into the quantum
theory. We meet at once with the difficulty that the classical theory itself
is not free from ambiguity. To make the discussion precise, let us suppose
we have a single electron interacting with a field of radiation and consider
the radiation resolved into ingoing and outgoing waves. The classical
problem is, given the ingoing radiation and suitable initial conditions for
the electron, determine the motion of the electron and the outgoing
radiation. The classical equations which deal with this problem are of two
kinds, (i) those that determine the field produced by the electron (which
field is just the difference of the ingoing and outgoing fields) in terms of
the variables describing the motion of the electron, and (ii) those that
determine the motion of the electron. Equations (i) are quite definite and
unambiguous, but not so equations (ii). The latter express the acceleration
of the electron in terms of field quantities at the point where the electron
is situated and these field quantities in the complete classical picture are
infinite and undefined. In the usual approximate treatment of the problem
one takes for these field quantities just the contributions of the ingoing
waves. This treatment is necessarily only approximate, since it does not
take into account the reaction on the electron of the waves it emits. We
should expect in an accurate treatment, that the field determining the
acceleration of the electron would be in some way associated with both the
ingoing and outgoing waves. Classical attempts have been made to improve the
theory by assuming a definite structure for the electron and calculating the
effect on one part of it of the field produced by the rest, but such methods
are not permissible in modern physics.\footnote{%
Dirac 1932, p. 457.}
\end{quotation}

In the same paper, which, albeit not well known, constitutes a major example
of his intuition, Dirac conceives of a solution formulated in heuristic
terms. He will give it a formal expression in a subsequent paper with Fock
and Podolsky (Dirac, Fock and Podolsky 1932).

\begin{quotation}
Let us make the assumption \textit{that the passage from the field of
ingoing waves to the field of outgoing waves is just a quantum jump
performed by one field}. This assumption is permissible on account of the
fact, discussed in the preceding section, that all the quantities in
relativistic quantum mechanics are of the nature of probability amplitudes
referring to one ingoing field and one outgoing field, so that we may
associate, say, the right-hand sides of the probability amplitudes with
ingoing fields and the left-hand sides with outgoing fields. In this way we
automatically exclude quantities referring to two ingoing fields or to two
outgoing fields and make a great simplification in the foundations of the
theory.

The significance of the new assumption lies in the fact \textit{that the
classical picture from which we derive our equations of motion must contain
no reference to quantum jumps}. This classical picture must therefore
involve just one field, a field composed of waves passing undisturbed
through the electron and satisfying everywhere Maxwell's equations for empty
space. With this picture the equations of motion for the electron are
perfectly definite and unambiguous. There are no equations of motion for the
field, as the field throughout space-time is pictured as given. Thus the
interaction between electron and field is introduced into the equations in
only one place.

The quantization of the equations of motion derived from this picture may
conveniently be carried out in two stages. Let us first quantize only the
variables describing the electron. We then get just the usual quantum theory
of the motion of an electron in a given classical field, with the difference
that in the present case the field must necessarily be resolvable into plane
waves and must therefore contain nothing of the nature of a Coulomb force.%
\footnote{%
Dirac 1932, p. 458. The italics are his.}
\end{quotation}

So, as we see, the key observation is to recognize that because of the
shortcomings of classical electrodynamics, a genuinely quantum ansatz has to
be taken (the traditional conceptual path through the correspondence
principle is no longer practical), namely that to attribute to the shift
from the ingoing to the outgoing field a quantum jump nature. The effective
classical picture which emerges is that of a free field obeying Maxwell
equations for empty space.

What remains to be done is to find a formal counterpart enabling one to use
free field equations without however neglecting the backreaction. This is
exposed in the paper with Fock and Podolsky and named today the interaction
(or Dirac) picture (Dirac, Fock and Podolsky 1932; to some extent, this was
anticipated in Dirac 1927b). Assume a composite system described by the
Hamiltonian $H_{A}+H_{B}+V$ where the $A$ and $B$ part interact via an
interaction term $V.$ A unitary transformation of the form $O\rightarrow
O^{*}=\exp (i/hH_{B}T)O\exp (-i/hH_{B}T)$ yields a picture where the
transformed dynamical variables of the $B$ system, say $q_{B}^{*}$ , obey
equations of motion for the part $B$ alone, so that the effect of the
interaction term $V$ has been transformed away: 
\[
\partial _{t}q_{B}^{*}=\frac{i}{\hbar }[H_{B},q_{B}^{*}] 
\]
The time variable of the free ( $B$ part only) equations is then to be
considered as a separate time variable for $B.$ If the $B$ part stands for
the radiation Hamiltonian, and part $A$ for the particles, the interaction
picture realizes the required trick of Dirac. The appearance of the
independent time variable $t$ in addition to the collective time $T$ will
directly lead, in a many-body situation, to the multi-time formalism as
exposed later in the same paper. This is because the interaction
transformation yields an individual (Dirac) equation for each separate
particle with its individual time where only the coupling to the radiation
field is present and the other particles absent (the particles interact only
through the radiation field). In the Heisenberg-Pauli theory of quantized
fields (Heisenberg and Pauli 1929), matter is second-quantized, the
particles become an excited collective state of a field, and their
''individual'' character is blurred, so that there is no longer any need to
use multi-particle equations and thus multiple times.

\section{Stueckelberg's 1934 paper\protect\footnote{%
Relativistich invariante St\"{o}rungstheorie des Diracschen Elektrons
(Relativistic Invariant Perturbation Theory of the Dirac Electron), \textit{%
Annalen der Physik}, received 10.Sept.1934.}}

\subsection{Introduction.}

We come now to the final part of our study which concentrates on the 1934
contribution of E. C. G. Stueckelberg: \textit{Relativistic invariant
perturbation theory of the Dirac electron; Part I: radiative scattering and
Bremsstrahlung\footnote{\textit{Part II }actually never appeared.}.}
Stueckelberg was at this time Privatdozent at the University of Zurich with
professor Gregor Wentzel. In the winter of 1934 he was called by the
University of Geneva to substitute for the deceased prof. A. Schidlof. In
1935 he became there associate professor of theoretical physics. His whole
career will be spent at the Universities of Geneva and Lausanne\footnote{%
For a biography see Wenger (1986) and Schweber (1994).}.

In 1934, Stueckelberg, after significative contributions to the quantum
theory of molecular spectra and scattering\footnote{%
Winans and Stueckelberg 1928, Morse and Stueckelberg 1929, and Stueckelberg
1932. In the first paper the authors thank E. U. Condon ''for very helpful
suggestions''.} started working on Q.E.D. At that time, it was obviously a
prominent topic and many among the most renowned physicists were contributing%
\footnote{%
To witness, e.g the publications by Bethe and Fermi 1932, Fermi 1932, Bethe
and Heitler 1934, Heisenberg 1934, Wentzel 1933,1934, Weisskopf 1934.}. In a
letter to the president of the Schulrat of E.T.H. in Z\"{u}rich (8 March
1934), W. Pauli writes:

\begin{quotation}
Dr. St\"{u}ckelberg has stated his desire to get deeper involved with Q.E.D.
and agrees with the nomination of Mr. Weisskopf [as assistant of Pauli, a
position that Stueckelberg himself considered previously].\footnote{%
Enz et al. 1997, p. 57.}
\end{quotation}

This quote seems to indicate that at that time Stueckelberg did not yet feel
sufficiently acquainted with Q.E.D. However, only a couple of months later
he was ready to submit his paper to the \textit{Annalen.}

Stueckelberg felt very indebted to Arnold Sommerfeld whom he approached
twice for extended periods. In a letter of 30 March 1935 he informed the
latter of his nomination in Geneva and continued:

\begin{quotation}
I owe my knowledge in the domain of theoretical physics mainly to the two
years which I could spend in your Institute, once as a student and later as
a National Research Fellow. Furthermore I received my appointment in
Princeton as well as my habilitation in Zurich thanks to your
recommendations. Thus I know that I\ owe it mainly to you, highly esteemed
Herr Geheimrat, to have obtained this position in Geneva.

I regret of course to leave Zurich which like Princeton, was a permanent
source of inspiration to me. I add a reprint of my work on the Dirac
electron [Stueckelberg 1934]. In a note to appear shortly in Helv. Phys.
Acta the Compton scattering of moving electrons is discussed [Stueckelberg
1935a]. These days I\ am busy generalizing the calculation method to the
many body problem [Stueckelberg 1935c].\footnote{%
Stueckelberg went to visit Sommerfeld in Munich the second time (1931-1932)
with an American fellowship. He was on leave from Princeton University where
he held the position of research associate from 1928 to 1932 (See Sopka
1991). Sopka writes further (p.1): ''During the summer of 1928, Stueckelberg
attended the Summer Symposium in Theoretical Physics held annually at the
University of Michigan in Ann Arbor. That year H. A. Kramers was one of the
lecturers. During this summer Stueckelberg was brought up to date on the
relativly new Quantum Mechanics and began his association with Philip M.
Morse.}
\end{quotation}

From 7 to 11 March 1937, A. Sommerfeld was hosted by Stueckelberg in his
house in Geneva following a visit of son Johann Wolfgang in June 1936.%
\footnote{%
We thank Stueckelberg's son, Dr. Georges St\"{u}ckelberg, for this
information.}. Stueckelberg's correspondence with Sommerfeld continued over
the years as is shown by a letter of 31 March 1949 (two years before
Sommerfeld's death), in which Stueckelberg thanks for the reception of
Sommerfelds textbook on Electrodynamics and discusses the substraction
method of Dirac, his work with Rivier and Schwinger's theory.\footnote{%
We thank Dr. K. von Meyenn for communicating us these two letters (dated
1935 and 1949).}

As its title shows, the aim of the 1934 paper is to provide a unified
covariant perturbative treatment of Compton scattering, Bremsstrahlung as
well as annihilation and creation of particle-antiparticle pairs\footnote{%
Stueckelberg used there Dirac's hole theory interpreting negative -energy
states as antiparticles.} (which is covered in the subsequent publication,
Stueckelberg 1935b). In his paper, Stueckelberg thanks G. Wentzel for the
suggestion to work out an invariant perturbation theory using a
four-dimensional Fourier transform\footnote{%
See Pauli's letter to Heisenberg quoted below.}. S. S. Schweber makes the
following comment on the relation between Wentzel and Stueckelberg, and
their emphasis on manifest relativistic covariance:

\begin{quotation}
When Stueckelberg came to the University of Zurich in 1933, Wentzel was
working on eliminating the divergences in relativistic field theories
(Wentzel 1933a,b, 1934a) and he got Stueckelberg interested in quantum field
theory. Stueckelberg began working on the field-theoretical description of
the interaction between particles. It is to be noted that Dirac, Fock, and
Podolsky 's (1932) article on quantum electrodynamics was the point of
departure for both Wentzel's and Stueckelberg's research programs
(Stueckelberg 1934, 1935c 1936). In 1938, again taking the Dirac, Fock
Podolsky paper as his starting point, Stueckelberg gave a formulation of
quantum electrodynamics and of various meson theories in what later became
known as the interaction picture. He stressed its advantages, namely, the
manifest covariance of the ''Schr\"{o}dinger'' equation in that picture, and
the possibility of writing covariant commutation rules for field operators
at different times (Stueckelberg 1938a,b).\footnote{%
Schweber 1994 p. 577-78.}
\end{quotation}

According to his publications, Wentzel was mainly interested in the years
1925-1930 by the problem of light scattering off free and bound electrons,
besides studying atomic spectra with the methods of quantum mechanics,
including the relativistic H-atom\footnote{%
See e.g. Wentzel 1925, 1926, 1927, 1929.}. Wentzel cared for manifest
Lorentz invariance which is suggested by the fact that he used in two
instances the multitime formalism of Dirac, Fock and Podolsky. First, in his
tentative to eliminate the self-energy divergences in classical and quantum
electrodynamics (Wentzel 1933 and 1934a). Second, in his study of the
possible equivalence of a spin 1 photon and of a spin 1/2 particle and its
antiparticle, following an idea of de Broglie (Wentzel 1934b).

As we show in the present paper (sections 6.2 and 6.3) Stueckelberg in 1934
used the interaction picture of Dirac, Fock and Podolsky and again in later
publications (Stueckelberg 1936, 1938). He discussed the multitime formalism
in 1935c and 1938.

The main innovation of the 1934 paper is the introduction of a new
perturbative scheme yielding \textit{manifestly} relativistic expressions
for the matrix elements. This is achieved by performing a four-dimensional
Fourier transformation of the wave-function, thus eliminating space \textit{%
and} time variables\footnote{%
M\o ller 1931 uses also the four-dimensional Fourier-transform but for the
classical retarded potential in a correspondence - theoretic treatment of
electron-electron scattering (see Kragh 1992 and Roqu\'{e} 1992). See also
Bethe-Fermi 1932.}. Compared to Waller 1930, Tamm 1930 or Heitler 1936,
Stueckelberg's procedure is thus the first departure from the ``older
(Dirac) form of the perturbation theory``\footnote{%
See the discussion in Jauch and Rohrlich 1955, p.158.} .The approach
proposed by Stueckelberg is far more powerful, but was not adopted by others
at the time. This lack of interest appears retrospectively as very
unfortunate, as was recognized by V. Weisskopf 1981, resp. 1983, in his
recollections of that period. Weisskopf reminds us of the difficulties
encountered in the higher order corrections to Q.E.D., and remarks:

\begin{quotation}
Already in 1934 [...] it seemed that a systematic theory could be developed
in which these infinities [divergent radiative corrections] are
circumvented. At that time nobody attempted to formulate such a theory
[...].There was one tragic exception [...], and that was Ernst C.G.
Stueckelberg. He wrote several important papers in 1934-38 putting forward a
manifestly invariant formulation of field theory.

This could have been a perfect basis for developing the ideas of
renormalization. Later on, he actually carried out a complete
renormalization procedure in papers with D. Rivier, independently of the
efforts of other authors\footnote{%
Rivier-Stueckelberg 1948, Rivier 1949.}. Unfortunately, his writings and his
talks were rather obscure,and it was very difficult to understand them or to
make use of his methods. He came frequently to Zurich in the years 1934-6,
when I was working with Pauli, but we could not follow his way of
presentation. Had Pauli and I myself been capable of grasping his ideas, we
might well have calculated the Lamb shift and the correction to the magnetic
moment of the electron at the time.\footnote{%
Weisskopf 1981, p. 78, resp. 1983, p. 74.}
\end{quotation}

The virtue of Stueckelberg's approach is to include the findings of his
predecessors, but making manifest the underlying general symmetry features.
These trends characterize Stueckelberg's life-long work and our paper is the
first stage in exploring this broader theme. We shall return to the
assessment of Stueckelberg's contribution and its fate at the end of this
paper. It is time now to get a closer look at his method.

We recall first the successive stages in the calculation of Compton
scattering.

1. Dirac 1926b; Dirac 1927a

The first two papers of Dirac illustrate one of the first uses of quantum
mechanics in the problem of matter-radiation interaction. The dynamical
variables of the charged particle are quantized, and its wave function obeys
the relativistic Schr\"{o}dinger (Klein-Gordon) equation. This equation is
solved with the help of canonical transformations. The electro-magnetic
field is not quantized.

2 Gordon 1926, Klein-Nishina 1929

Their treatment is semi-classical. The electromagnetic field is not
quantized, but the electron satisfies the relativistic Klein-Gordon, resp.
Dirac equation. The electron current, modified by the primary field, is the
source of the final retarded field, calculated using Maxwell's equations.
The treatment is ''continuous'' and no intermediate states appear. There is
however a notion of initial and final states with a discontinous switch from
the former to the latter. The general treatment is in principle
relativistic, manifest in Gordon's case, implicit in the Klein-Nishina
paper, where the main part of the calculations is done in the rest frame of
the electron.

3. Waller 1930 and Tamm 1930.

The radiation field is now quantized (and the electron field as well in
Tamm), and the time-dependent perturbation scheme of Dirac 1927b is applied.
Intermediate states are now necessary because the interaction term linear in
the field prevents a direct transition from the initial to the final state
(see our discussion of Dirac's paper section 4.1 and 4.2). At the
interaction point (vertex), energy is not conserved. Although the electron
obeys the relativistic Dirac equation, the formulas of the perturbation
theory break manifest covariance, and this is reflected by the presence in
the matrix elements of the non-covariant factors $(E-E^{\prime })^{-1}.$
This is because standard perturbation theory expands the full solution in a
basis of unperturbed solutions with time-dependent coefficients, which, in
the case of free motion (plane waves), amounts to perform a Fourier
transform for space variables only. This unsymmetrical treatment of space
and time makes the calculation clumsy. Furthermore, the contributions of
positive and negative electron energies in the intermediate state, necessary
for obtaining the correct classical limit, are considered separately and
reunited only later using spinor identities. The results are obtained in the
rest system of the electron.

\bigskip

In Stueckelberg's 1934 paper, thanks to the use of the four-dimensional
Fourier transform, manifest covariance is kept throughout. The factor $%
(E-E^{\prime })^{-1}$is now replaced by the covariant expression $%
(p^{2}+M^{2})^{-1}$ where $p$ is the four-momentum of the intermediate
state, $M=\frac{mc}{\hbar },$ and $m$ is the electron mass, corresponding to
a virtual particle off mass-shell ($p^{2}+M^{2}\neq 0).$ Energy and
three-momentum are now conserved at the vertices. The contributions of
positive and negative energies (corresponding to virtual electrons and
positrons) are contained in the single propagation function $%
(p^{2}+M^{2})^{-1}$, which corresponds to what later was called Feynman
propagator\footnote{%
In this particular case (Born approximation) $i\epsilon $ is not necessary.}%
, and by Rivier-Stueckelberg 1948, causal function\footnote{%
Feynman 1948 and 1949. Rivier-Stueckelberg 1948 and Rivier 1949.}. All of
Stueckelberg's expressions for matrix elements \textit{are identical to
those obtained nowadays from Feynman diagrams.}

Let us mention two other important features. First, because he doesn't
commit himself to any specific gauge, Stueckelberg's matrix elements are
manifestly gauge invariant, a feature rather unusual for those times (for
instance Fermi in his review 1931, and after him Heitler 1936 worked in the
radiation (or Coulomb) gauge).

Next, in evaluating his expressions, Stueckelberg uses integration over the
complex energy-plane, making explicit use of the singularity structure as a
device for putting external particle states on mass-shell.\footnote{%
Stueckelberg probably learned \textbf{complex integration techniques in
physics} from A.Sommerfeld, whose courses he followed in Munich 1924/1925
and whom he visited, as we saw, in 1930. It is interesting to note that
Wentzel used as well Sommerfeld's technique (Sommerfeld 1916) in his 1926
paper on multiply-periodic systems within the new quantum mechanics (Wentzel
1926). See also Stueckelberg's paper ``Sur l'int\'{e}gration de
l'\'{e}quation $(\sum_{i=1}^{4}$ $\partial _{x_{i}}^{2}-l^{2})Q=-\rho $ en
utilisant la m\'{e}thode de Sommerfeld``, Stueckelberg 1939).}.

\subsection{Compton scattering on scalar particles following Stueckelberg's
method}

As a warm-up exercise, we consider the scattering of photons on charged
scalar particles, using the method developed by Stueckelberg in his 1934
paper on the Compton scattering on spin 1/2 electrons. This is the same
problem which was solved by Dirac 1926b, 1927a, and Gordon 1926 (see section
3). Stueckelberg's method is not only more direct, but also yields a matrix
element for the scattering process which is manifestly Lorentz and gauge
invariant. The starting point is again the relativistic equation for the
scalar wave-function $\phi (x)$ of the ''electron'', interacting with the
electromagnetic field $A_{\mu }$, which is now quantized according to Dirac
1927b. 
\begin{equation}
(-i\partial _{\mu }+eA_{\mu }(x))(-i\partial ^{\mu }+eA^{\mu }(x))\phi (x)+%
\frac{m^{2}c^{2}}{\hbar ^{2}}\phi (x)=0  \label{h1}
\end{equation}

In the sequel, we put $M=\frac{mc}{\hbar }$. Instead of the pseudoeuclidean
metric with $x_{4}=ict$ fashionable in Stueckelberg's time, we use the real
metric $g_{\mu \nu }$ with $g_{ii}=-g_{00}=1$, $i=1,2,3$.

In the interaction picture (see section 6)\footnote{%
See Dirac 1927b and Dirac-Fock-Podolsky 1932.}, $A_{\mu }$ is a free field
with the Fourier expansion\footnote{%
Stueckelberg writes $\Gamma _{k}instead$ of $a_{k}$ and $\sigma ^{k}$for $%
e_{k}$} 
\begin{equation}
A^{\mu }=\sum\limits_{\mathbf{k}}e_{\mathbf{k}}^{\mu }V_{k}\left[
a_{k}e^{i(k\cdot x)}+a_{k}^{\dagger }e^{-i(k\cdot x)}\right]  \label{h2}
\end{equation}

The sum over $\mathbf{k}$ refers to photons in a box $G$ with\footnote{%
Stueckelberg writes $T_{k}$ instead of $eV_{k}.$} 
\begin{equation}
V_{k}^{2}=\frac{2\pi }{Gk_{0}c\hbar }  \label{sbox}
\end{equation}

The 4-momentum $k$ and 4-vector of polarization $e_{\mathbf{k}}$ satisfy the
mass-shell and transversality conditions 
\begin{equation}
k_{\mu }k^{\mu }=k^{2}=0;k_{\mu }e_{\mathbf{k}}^{\mu }=k\cdot
e_{k}=0;e_{k}^{2}=1  \label{h3}
\end{equation}

The annihilation, resp. creation operators $a_{k},a_{k^{\prime }}^{\dagger }$
obey the usual commutation relations 
\begin{equation}
\left[ a_{k},a_{k^{\prime }}^{\dagger }\right] =\delta _{kk^{\prime }}
\label{h4}
\end{equation}

Applied to an eigenstate $|N_{1},...,N_{k},...>$ of the photon number
operator $N$, with eigenvalues $N_{i},$ they yield the well known result%
\footnote{%
Stueckelberg writes $u(N_{1},...,N_{k},...)$ instead of $|$ $%
N_{1},...,N_{k},...>$} 
\begin{eqnarray}
a_{k}|N_{1},...,N_{k},... &\succ &=\sqrt{N_{k}}|N_{1},...,N_{k}-1,...\succ
\label{h5} \\
a_{k}^{\dagger }|N_{1},...,N_{k},... &\succ &=\sqrt{N_{k}+1}%
|N_{1},...,N_{k}+1,...\succ  \nonumber
\end{eqnarray}

Following Stueckelberg the interacting wave-function is expanded on the
photon number eigenstates $|N^{j}\succ $ with coefficients $\varphi (x),$
according to Dirac (1927b) (see section 4.1) 
\begin{equation}
\phi ^{j}(x)=\sum_{j}\varphi ^{j}(x)|N^{j}\succ  \label{h6}
\end{equation}

Here $N^{j}$ denotes all possible photon configurations $%
\{N_{1},N_{2},...,N_{k},...\}$. Now comes the departure of Stueckelberg's
method from those of his predecessors, Waller, Tamm, Heitler, etc. He
introduces the four-dimensional Fourier transformation for the ''electron''
wave function $\varphi (x)\footnote{%
In the following formula, Stueckelberg writes $l\cdot x$ instead of $p\cdot
x.$ We retain the more familiar notation $p,$ but the reader has to remember
that\textit{\ this} $p$ includes the factor $\hbar ^{-1}.$}$%
\begin{equation}
\varphi ^{j}(x)=\int d^{4}pe^{i(p\cdot x)}\chi ^{j}(p)  \label{h7}
\end{equation}

The result will be the elimination in the perturbation expansion of time
together with space\footnote{%
Notice that although the photon states are discretized (i.e. radiation is in
a box), the Fourier expansion above is continuous, the states considered not
being necessarily on mass-shell.}. Introducing (\ref{h2}), (\ref{h6}) and (%
\ref{h7}) into the Klein-Gordon equation (\ref{h1}) yields 
\begin{eqnarray}
&&\sum_{j}\int d^{4}pe^{i(p\cdot x)}(\left[ p^{2}+M^{2}\right] \chi ^{j}(p) 
\nonumber \\
&&+e\sum_{\mathbf{k}}(2p-k)\cdot e_{\mathbf{k}}V_{k}(a_{k}+a_{-k}^{\dagger
})\chi ^{j}(p-k)+  \label{h7bis} \\
&&e^{2}\sum_{\mathbf{k,k}^{\prime }}e_{\mathbf{k}}\cdot e_{\mathbf{k}%
^{\prime }}V_{k}V_{k^{\prime }}(a_{k}+a_{-k}^{\dagger })(a_{k^{\prime
}}+a_{-k^{\prime }}^{\dagger })\chi ^{j}(p-k-k^{\prime }))|N^{j}\succ =0 
\nonumber
\end{eqnarray}

The general approach of Stueckelberg is to use a perturbation expansion in
powers of the charge $e$. The perturbation expansion applies to the \
time-independent functions $\chi ^{j}(p).$ Thus, the zeroth order is given
by the free state equation 
\begin{equation}
\int d^{4}pe^{i(p\cdot x)}\left[ (p^{2}+M^{2})\right] \chi ^{j(0)}(p)=0
\label{h8}
\end{equation}

To get the first order we insert $\chi ^{j(0)}(p-k)$ into the second term of
(\ref{h7bis}) 
\begin{eqnarray}
&&\sum_{j}[\int d^{4}pe^{i(p\cdot x)}(\left[ p^{2}+M^{2}\right] \chi
^{j(1)}(p)+  \label{h8bis} \\
&&e\sum_{\mathbf{k}}(2p-k)\cdot e_{\mathbf{k}}V_{k}(a_{k}+a_{-k}^{\dagger
})\chi ^{j(0)}(p-k))]|N^{j}\succ =0  \nonumber
\end{eqnarray}

For the second approximation,$\chi ^{j(1)}$ is introduced into the second
term and $\chi ^{j(0)}$ into the third one 
\begin{eqnarray}
&&\sum_{j}[\int d^{4}p^{\prime }e^{i(p^{\prime }\cdot x)}\left[ p^{\prime
2}+M^{2}\right] \chi ^{j(2)}(p^{\prime })|N^{j}\succ  \label{h8ter} \\
&=&-\int d^{4}p^{\prime }e^{i(p^{\prime }\cdot x)}\{e\sum_{\mathbf{k}
^{\prime }}(2p^{\prime }-k^{\prime })\cdot e_{\mathbf{k}^{\prime
}}V_{k^{\prime }}(a_{k^{\prime }}+a_{-k^{\prime }}^{\dagger })\chi
^{j(1)}(p^{\prime }-k^{\prime })-  \nonumber \\
&&e^{2}\sum_{\mathbf{k,k}^{\prime }}e_{\mathbf{k}}\cdot e_{\mathbf{k}
^{\prime }}V_{k}V_{k^{\prime }}(a_{k}+a_{-k}^{\dagger })(a_{k^{\prime
}}+a_{-k^{\prime }}^{\dagger })\chi ^{j(0)}(p^{\prime }-k-k^{\prime
})\}]|N^{j}\succ =0  \nonumber
\end{eqnarray}

This expression gives a matrix element symmetric in $k$ and $k^{\prime }$.
For Compton scattering, we take $p,k$ for the 4-momenta of the initial
''electron'' and photon. a$_{-k^{\prime }}^{\dagger }a_{k}$ annihilates the
photon $k$ and creates a photon with momentum $-k^{\prime }.$ The value of
the matrix element of $a_{-k^{\prime }}^{\dagger }a_{k}$ is one. In order to
describe an outgoing photon we change the sign of $k^{\prime }$. With $%
p^{\prime }$ the final ''electron'' momentum the conservation law reads%
\footnote{%
Stueckelberg's notation is here $l^{0}$for $p,$ $l$ for $p^{\prime },$ $-p$
or $m$ for $k^{\prime }$. See the footnote preceding eq. (\ref{h7})} 
\[
p+k=p^{\prime }+k^{\prime } 
\]
so that finally the relation between the initial wave-function $\chi
^{j(0)}(p)$ and the final $\chi ^{j(2)}(p^{\prime })$ is given by 
\begin{equation}
\chi ^{j(2)}(p^{\prime })=\frac{e^{2}}{p^{\prime 2}+M^{2}}V_{k}V_{k^{\prime
}}\Omega \chi ^{j(0)}(p)  \label{h9}
\end{equation}

\begin{eqnarray}
\Omega &=&\frac{(2p^{\prime }+k^{\prime })\cdot e_{\mathbf{k}^{\prime
}}\quad (2p+k)\cdot e_{\mathbf{k}}}{(p+k)^{2}+M^{2}}  \label{h10} \\
+ &&\frac{(2p^{\prime }-k)\cdot e_{\mathbf{k}}\quad (2p-k^{\prime })\cdot e_{%
\mathbf{k}^{\prime }}}{(p-k^{\prime })^{2}+M^{2}}-2e_{\mathbf{k}}\cdot e_{%
\mathbf{k}^{\prime }}  \nonumber
\end{eqnarray}

We see that Stueckelberg's scheme is entirely within momentum space $p.$ The
matrix element $\Omega (p)$ is obviously Lorentz invariant. It is also
invariant under the gauge transformations $e_{\mathbf{k}}\rightarrow e_{%
\mathbf{k}}+\lambda k,e_{\mathbf{k}^{\prime }}\rightarrow e_{\mathbf{k}
^{\prime }}+\lambda ^{\prime }k^{\prime }$ if the external electrons are on
mass-shell. The first two terms with denominators contain the contribution
of intermediate states whereas the last term corresponds to the direct
interaction. It is important to realize that Stueckelberg didn't need to
modify the overall scheme of iteration to achieve this result; such is not
the case with the previous (Dirac theory) where a refinement of the
perturbative method had to be made to obtain a contribution through
intermediate states (see section on Dirac's perturbation theory).\footnote{%
See also Heitler 1936, p. 88.}

Equation (\ref{h10}) is a witness to the modernity of Stueckelberg's
approach, since one had to wait until 1948 to find similar expressions. $%
\Omega $ is actually identical (except for normalizations) to the
corresponding factor given by Bjorken and Drell 1964, eq. (9.30), using
Feynman techniques.

(\ref{h10}) greatly simplifies in the rest system of the initial charged
particle ($\mathbf{p}=0\mathbf{)}$ and in the Coulomb or radiation gauge
where $e_{0}=e_{0}^{\prime }=0$ and hence $p\cdot e_{\mathbf{k}}=p\cdot e_{%
\mathbf{k}^{\prime }}=0.$ In this case $\Omega _{\text{lab.}}=-2e_{\mathbf{k}%
}\cdot e_{\mathbf{k}^{\prime }}.$

The steps above illustrate the iteration method of Stueckelberg.

We now pass to the discussion of Stueckelberg's method for calculating the
cross-section. It starts with the definition of ''on mass-shell
wave-functions''. We first introduce the on mass-shell 4-momentum, denoted
by 
\begin{equation}
\bar{p}=(\mathbf{p},\bar{p}_{0});\bar{p}_{0}^{2}=\mathbf{p}^{2}+M^{2}
\label{ms}
\end{equation}
To solve the free Klein-Gordon equation (\ref{h8}), Stueckelberg makes the
Ansatz 
\begin{equation}
\chi ^{(0)}(p)=\frac{1}{i\pi }\frac{\omega ^{(0)}(p)}{p^{2}+M^{2}}
\label{h11}
\end{equation}
where $\omega ^{(0)}(p)$ is a continuous function which vanishes fast enough
at infinity. Then $\chi ^{(0)}(p)$ satisfies (\ref{h8}) if $\omega ^{(0)}(%
\bar{p})$ obeys 
\begin{equation}
(\bar{p}^{2}+M^{2})\omega ^{(0)}(\bar{p})=0  \label{msp}
\end{equation}
Notice that $p^{2}+M^{2}=\bar{p}_{0}^{2}\mathbf{-}p_{0}^{2}$ which makes
explicit the singularity of (\ref{h11}) in the energy variable $p_{0}.$ The
reason of this special way in which Stueckelberg expresses the solution of (%
\ref{h8}) is the following. It is different from zero only for those values
of $p$ where the $\bar{p}_{0}^{2}=\mathbf{p}^{2}+M^{2}$. Elsewhere, it
vanishes$.$ The form of (\ref{h11}) with its explicit singularity makes it
possible to write both cases in a single \textit{closed form}, which is then
suitable for being reinserted into interation steps. Each order of iteration
will add a new singularity the impact of which being crucial during
contour-integration. The information about the values of $p$ where the
non-trivial solution is valid is then retrieved by complex
contour-integration around the singularity (This feature of Stueckelberg's
method is quite original and contributes to the full automatism of his
perturbative calculus\footnote{%
Nowadays, instead of (\ref{h11}) one writes $\chi ^{(0)}(p)=\delta
(p^{2}+M^{2})\omega ^{(0)}(p)$.}). For instance, in the first order term
there are two poles 
\begin{equation}
\chi ^{j(1)}(p)=\frac{-1}{\left[ p^{2}+M^{2}\right] }\sum_{\mathbf{k}%
}(2p-k)\cdot e_{\mathbf{k}}V_{k}(a_{k}+a_{-k}^{\dagger })\chi
^{j(0)}(p-k))]|N^{j}\succ =0
\end{equation}
. However, to this order, it is impossible to simultaneously satisfy the
mass-shell conditions ($p^{2}+M^{2})=0$, $(p-k)^{2}+M^{2}=0$\ and $\mathbf{k}%
^{2}=0$.

Accordingly, $\chi ^{(0)}(p)$ is integrated over the complex $p_{0}$-plane,
in the positive sense if $p_{0}=\bar{p}_{0},$ negative if $p_{0}=-\bar{p}
_{0}.$ The result is, using Cauchy's formula 
\begin{equation}
\int_{\mathcal{C}}dp_{0}\chi ^{(0)}(p)e^{ip_{0}x_{0}}=\int_{\mathcal{C}%
}dp_{0}e^{ip_{0}x_{0}}\frac{1}{i\pi }\frac{\omega ^{(0)}(p)}{p^{2}+M^{2}}=%
\frac{\omega ^{(0)}(\bar{p})}{\bar{p}_{0}}e^{i\bar{p}_{0}x_{0}}
\label{h12bis}
\end{equation}

This should then be put into the integral 
\begin{equation}
\varphi ^{(0)}(x)=\int d^{4}pe^{i(p\cdot x)}\chi ^{(0)}(p)  \label{h12ter}
\end{equation}

To obtain the final ''electron'' wave-function on mass-shell, we follow
Stueckelberg and define, by analogy with (\ref{h12bis}) 
\begin{equation}
\frac{\omega ^{(2)}(\bar{p}^{\prime },x_{0})}{\bar{p}_{0}^{\prime }}=\int_{%
\mathcal{C}^{\prime }}dp_{0}^{\prime }e^{i(p_{0}^{\prime }-\bar{p}
_{0}^{\prime })x_{0}}\chi ^{(2)}(p^{\prime })  \label{h13}
\end{equation}

As will be seen presently, $\omega ^{(2)}$ is time dependent, while the same
formula for $\omega ^{(0)}$ gives a time-independent expression. The idea is
to define a number $n^{(2)}$ of final ''electrons'' which will turn out to
be linear in $t$. and will allow then an easy computation of the number of
''electrons'' scattered per unit time. This will now be sketched.

Remembering (\ref{h9}) 
\[
\chi ^{(2)}(p^{\prime })=\frac{e^{2}}{p^{\prime 2}+M^{2}}V_{k}V_{k^{\prime
}}\Omega (p)\chi ^{(0)}(p) 
\]
we see that $\chi ^{(2)}$ has now two poles, at $p_{0}^{\prime }=\bar{p}%
_{0}^{\prime }$ and $p_{0}=\bar{p}_{0},$the latter coming from $\chi
^{(0)}(p)$ (there is a third pole in $\Omega (p),$see (\ref{h10}), which
does not contribute for the same reason as above) To integrate over the
complex $p_{0}^{\prime }$-plane one has to substitute $p_{0}=p_{0}^{\prime
}+k_{0}^{\prime }-k_{0}.$ Defining 
\[
s_{0}=\bar{p}_{0}^{\prime }+k_{0}^{\prime }-\bar{p}_{0}-k_{0} 
\]

the integration (\ref{h13}) gives, up to terms of order $s_{0}^{n}(n%
\geqslant 0),$%
\begin{equation}
\frac{\omega ^{(2)}(\bar{p}^{\prime },x_{0})}{\bar{p}_{0}^{\prime }}=-2e^{2}%
\frac{(1-e^{-is_{0}x_{0}})}{s_{_{0}}}V_{k}V_{k^{\prime }}\frac{\Omega (\bar{p%
})\omega ^{(0)}(\bar{p})}{2\bar{p}_{0}^{\prime }2\bar{p}_{0}}  \label{h14}
\end{equation}

Later on, we shall integrate over the pole in $s_{0}$ justifying the neglect
of higher order terms.

In the previous approach (Waller and Tamm) the transition amplitude of a
process under consideration was directly related to the value of the
expansion coefficients (see (\ref{dpt_3}) or more specifically (\ref{w_4})
and (\ref{w_5})) and thus ''ready to use''. Here, to obtain the scattering
cross-sections, one uses the conserved currents of free particles 
\begin{equation}
j_{\mu }(x)=-i(\varphi ^{*}\partial _{\mu }\varphi -\varphi \partial _{\mu
}\varphi ^{*})
\end{equation}

and defines the number of incident charged particles $n^{(0)}$ as 
\begin{equation}
n^{(0)}=\int d^{3}\mathbf{x}j_{0}(x)=-i\int d^{3}\mathbf{x}(\varphi
^{*}\partial _{0}\varphi -\varphi \partial _{0}\varphi ^{*})  \label{h14bis}
\end{equation}

One finds consequently 
\begin{equation}
n^{(0)}=(2\pi )^{3}\int d^{3}\mathbf{p}2\bar{p}_{0}\frac{\omega ^{(0)*}(\bar{
p})}{\bar{p}_{0}}\frac{\omega ^{(0)}(\bar{p})}{\bar{p}_{0}}  \label{h15}
\end{equation}

Similarly, for the final ''electrons'' Stueckelberg uses the same expression
but now with the perturbed solutions 
\begin{equation}
n^{(2)}(x_{_{0}})=(2\pi )^{3}\int d^{3}\mathbf{p}2\bar{p}_{0}\frac{\omega
^{(2)*}(\bar{p})}{\bar{p}_{0}}\frac{\omega ^{(2)}(\bar{p})}{\bar{p}_{0}},
\label{h16}
\end{equation}

With the solution (\ref{h14}) this gives the time dependent expression 
\begin{equation}
n^{(2)}(x_{_{0}})=e^{4}\frac{(1-\cos (s_{0}x_{0}))}{2s_{_{0}}^{2}}
V_{k}^{2}V_{k^{\prime }}^{2}\overline{\Omega ^{*}\Omega }  \label{h17}
\end{equation}

where the expectation value of $\Omega ^{*}\Omega $ is 
\begin{equation}
\overline{\Omega ^{*}\Omega }=\frac{(2\pi )^{3}}{n^{(0)}}\int d^{3}\mathbf{p}
2\bar{p}_{0}\frac{\omega ^{(0)}(p)^{\dagger }}{\bar{p}_{0}}\Omega
^{*}(p)\Omega (p)\frac{\omega ^{(0)}(p)}{\bar{p}_{0}}  \label{h18}
\end{equation}

To get the cross-section one has to integrate over the phase space of final
electrons $d^{3}\mathbf{p}^{\prime }$and of final photons $(2\pi
)^{-3}Gk_{0}^{\prime 2}dk_{0}^{\prime }d\Omega .$ in the solid angle $%
d\Omega .$(not to be confused with the preceeding $\Omega $ ). Furthermore
one assumes that the initial ''electron'' wave packet is sharply peaked at a
initial momentum value.

After changing the variable $k_{0}^{\prime }$ into $s_{0}$ one meets the
integral 
\begin{equation}
\int ds_{0}\left\{ \frac{(1-e^{-is_{0}x_{0}})}{2s_{_{0}}}\right\} ^{2}=\int
ds_{0}\frac{(1-\cos (s_{0}x_{0}))}{2s_{_{0}}}=\frac{2\pi }{4\hbar }x_{0}=%
\frac{\pi }{2\hbar }ct  \label{h19}
\end{equation}

Now, dividing by $t,$\ by the incident photon flux and the density of target
particles, one finds the gauge invariant differential cross-section (all
external particles on mass-shell) 
\begin{equation}
\frac{d\sigma }{d\Omega }=\frac{\alpha ^{2}}{M^{2}}\frac{k_{0}^{\prime 2}}{
k_{0}^{2}}\left\{ \frac{(\bar{p}\cdot e_{k})(\bar{p}^{\prime }\cdot
e_{k^{\prime }})}{\bar{p}\cdot k}-\frac{(\bar{p}^{\prime }\cdot e_{k})(\bar{p%
}\cdot e_{k^{\prime }})}{\bar{p}\cdot k^{\prime }}-e_{k}\cdot e_{k^{\prime
}}\right\} ^{2}.  \label{crossscal}
\end{equation}

Here, $\alpha =e^{2}/\hbar c$ is the fine structure constant, $M^{-1}=\hbar
/mc$ is the Compton wave-length and $\alpha /M=r_{0}$ is the classical
radius of the electron.

\subsection{\protect\bigskip The spinor case}

We are now in a position to review Stueckelberg's 1934 paper for spin 1/2
electrons. The essential complication with respect to the scalar case above
is the necessity to take into account the spin degrees of freedom.
Furthermore, Stueckelberg generalizes slightly the formalism in order to
derive an expression which is then suitable to be applied to Compton
scattering and also to Bremsstrahlung and pair production. The starting
point is now the Dirac equation for the spinor wave-function $\Psi (x)%
\footnote{%
The original notation is $C$ instead of $M$.}$

\begin{equation}
\left[ \frac{1}{i}(\gamma \cdot \partial _{x})+M+eV(x)\right] \Psi (x)=0
\label{eq.1.1}
\end{equation}
where the coupling of the matter current to the electromagnetic field is

\begin{equation}
V(x)=\sum\limits_{\mathbf{k}}V_{\mathbf{k}}(e_{\mathbf{k}}\cdot \gamma
)\left[ a_{\mathbf{k}}e^{i(k\cdot x)}+a_{\mathbf{k}}^{\dagger }e^{-i(k\cdot
x)}\right]  \label{eq.1.4}
\end{equation}
$V_{\mathbf{k}}$ is given by (\ref{sbox}), ($e_{k}\cdot \gamma $) denotes
the scalar product of the photon polarization 4-vector $e_{k}^{\mu },$ and
the Dirac matrices $\gamma _{\mu }$, $\mu =0,1,2,3$ (we now use the
following convention for $\gamma $ matrices: $\gamma _{\mu }\gamma _{\nu
}+\gamma _{\nu }\gamma _{\mu }=-2g_{\mu \nu }$)\footnote{%
Stueckelberg works with the convention 
\[
\gamma _{\mu }\gamma _{\nu }+\gamma _{\nu }\gamma _{\mu }=-2\delta _{\mu \nu
} 
\]
\par
the minus sign of the right hand side being the opposite of that chosen by
Dirac (and all the other authors discussed in our paper).}. The analogue of
the expansion (\ref{h6}) involves now spinor functions $\varphi ^{j}(x):$

\begin{equation}
\Psi (x)=\sum_{j}\varphi ^{j}(x)|N^{j}\succ  \label{eq.1.3}
\end{equation}
and as previously Stueckelberg Fourier expands\footnote{%
See footnote before eq. (\ref{h7}). Compare also this formula to (\ref
{walfour}).}

\begin{equation}
\Psi (x,N)=\sum_{j}\int d^{4}pe^{i(p\cdot x)}u^{j}(p)|N^{j}\succ
\label{eq.2.2}
\end{equation}

with $u^{j}(p)$ the spinor in momentum space\footnote{%
Stueckelberg writes $A^{j}(l)$ instead of $u^{j}(p).$}$.$

The Fourier-transform of the full Dirac equation (\ref{eq.1.1}) can be
written

\begin{equation}
\sum_{j}\int d^{4}pe^{i(p\cdot x)}\left\{ \left[ (\gamma \cdot p)+M\right]
u^{j}(p)+e\sum_{k,i}V_{k}P_{kji}(e_{k}\cdot \gamma )u^{i}(p-k)\right\}
|N^{j}\succ =0  \label{eq.2.4}
\end{equation}

Here $P_{kji}$ is a matrix element of $P_{k}$ the latter standing for one of
the operators $a_{k}$, $a_{-k}^{+}$ , or $1$. This notation allows a unified
treatment of the above mentioned processes\footnote{$P_{k}=1$ will be
relevant for Bremsstrahlung, see below section 6.6}

\begin{equation}
P_{k}|N^{j}\succ =\sum_{i}|N^{i}\succ P_{kij}  \label{eq.2.1}
\end{equation}
and $V_{k}$ is one term in the sum of eq. (\ref{eq.1.4}).

The perturbation expansion is obtained in the same way as in the scalar
case. The first approximation $u^{(1)}$ is found to be (compare with (\ref
{h8bis}))

\begin{equation}
u^{(1)}=-\frac{1}{(\gamma \cdot p)+M}\sum_{k}V_{k}P_{kj0}(e_{k}\cdot \gamma
)u^{(0)}(p-k)  \label{eq.2.11}
\end{equation}
Similarly, for the second order, the result is (compare (\ref{h9}) and (\ref
{h10})):

\begin{equation}
u^{(2)}(p^{\prime })=-\frac{1}{p^{\prime 2}+M^{2}}\sum_{k^{\prime
},k}V_{k^{\prime }}V_{k}(P_{k^{\prime }})_{ji}(P_{k})_{i0}\Omega
(p)u^{(0)}(p)  \label{eq.2.12}
\end{equation}

\begin{equation}
\Omega (p)=\left( (\gamma \cdot p^{\prime })-M\right) \left\{ \frac{%
(e_{k^{\prime }}\cdot \gamma )((\gamma \cdot p^{\prime }-k^{\prime
})-M)(e_{k}\cdot \gamma )}{(p^{\prime }-k^{\prime })^{2}+M^{2}}+\frac{
(e_{k}\cdot \gamma )((\gamma \cdot p^{\prime }-k)-M)(e_{k^{\prime }}\cdot
\gamma )}{(p^{\prime }-k)^{2}+M^{2}}\right\}  \label{eq.2.13}
\end{equation}
with $p=p^{\prime }-k-k^{\prime }$and $(P_{k})_{ij}$ are the matrix elements
of the operators $a_{k}$, $a_{k}^{\dagger }$ and $1$, that is $\sqrt{N_{k}}$
and $\sqrt{N_{k}+1}$for the first two. Here, one gets only the contributions
corresponding to intermediate states as there is no direct transition,
contrary to the scalar case.

The expressions (\ref{eq.2.12}) and (\ref{eq.2.13}) are obviously
Lorentz-invariant and \textit{correspond exactly to the expressions obtained
by ''modern'' Feynman rules}.

We discuss now the mass shell solutions. The spinor analogue of the
zero-order approximation, is given by the same \textit{Ansatz }(see\textit{\ 
}eq. \textit{(}\ref{h11}\textit{))\footnote{$\omega (p)$ is in
Stueckelberg's paper $B(l^{0}).$}}

\begin{equation}
u^{(0)}(p)=\frac{1}{i\pi }\frac{\omega (p)}{p^{2}+M^{2}}  \label{eq.2.7}
\end{equation}
where this time $\omega $ satisfies the free (''mass-shell'') Dirac equation 
\begin{equation}
(\gamma \cdot \bar{p}+M)\omega (\bar{p})=0  \label{eq.2.7bis}
\end{equation}

$\bar{p}$ is again given by (\ref{ms}). The mass-shell condition is now
obtained as the condition of vanishing determinant of the four by four
matrix $(\gamma \cdot p+M)$. Equation (\ref{eq.2.7bis}) admits four linearly
independent solutions. For each sign of the energy there are two spin states 
$\omega ^{r}(\bar{p})$. In eq. (\ref{eq.2.13}), Stueckelberg gets
automatically the projection operator 
\[
\Lambda ^{+}(\bar{p}^{\prime })=\frac{\gamma \cdot \bar{p}^{\prime }-M}{2M}
=\sum\limits_{r=1}^{2}\omega ^{r}(\bar{p}^{\prime })\bar{\omega}^{r}(\bar{p}%
^{\prime }) 
\]

This shows that the summation over final spin states of the electron is
naturally built in. One can keep it like that, or split it into respective
contributions of the two spin states.

The integration over the complex $p_{0}$ plane yields (cf. (\ref{h12bis})
and (\ref{h12ter}))

\begin{eqnarray}
\varphi ^{(0)}(x) &=&\int d^{4}pe^{i(p\cdot x)}u^{(0)}(p)  \label{eqforphi}
\\
&=&\int d^{3}\mathbf{p}e^{i(\mathbf{px})}\frac{\omega (\bar{p})}{\bar{p}_{0}}
e^{-i\bar{p}_{0}x_{0}}  \nonumber
\end{eqnarray}

For the second order case one finds using eq. (\ref{eq.2.12})

\begin{equation}
\frac{\omega ^{(2)}(\bar{p},x_{0})}{\bar{p}_{0}}=\sum_{k^{\prime },k}\frac{
(1-e^{is_{0}x_{0}})}{s_{_{0}}}V_{k}V_{k^{\prime }}(P_{k})_{ji}(P_{k^{\prime
}})_{i0}\frac{\Omega (\bar{p})\omega (\bar{p})}{\bar{p}_{0}^{\prime }\bar{p}
_{0}}  \label{eq.2.15}
\end{equation}
plus higher order terms in $s_{0}=\bar{p}_{0}^{\prime }-\bar{p}
_{0}-k_{0}-k_{0}^{\prime }.$\bigskip

We now specialize to the case of Compton scattering. We choose the initial
photon configuration as $N_{k}=(0,0,...,1_{k},...)$. The problem is then to
find the transition amplitude to the final configuration $N_{k^{\prime
}}=(0,0,...,1_{k^{\prime }},...,0_{k},...)$, i.e. a photon has been absorbed
in state $k$ and another one emitted in state $k^{\prime }.$ For Compton
scattering, $p$,$p^{\prime }$ are then interpreted as the initial, resp.
final electron 4-momenta and $k$, $-k^{\prime },$ the initial, resp. final
photon 4-momenta. This choice of variables makes the symmetry of the result
in $k$ and $k^{\prime }$ explicit.

In the spinor case, Stueckelberg defines the number of electrons in state $%
\varphi (x)$ by (instead of (\ref{h14bis}))\footnote{%
Here $\varphi ^{\dagger }(x)$ is the hermitian conjugate spinor, whereas for
Stueckelberg $\varphi ^{\dagger }(x)$ denotes the adjoint spinor which
nowadays is written $\bar{\varphi}(x)=$ $\varphi ^{\dagger }(x)\gamma _{0}.$}

\begin{equation}
n=\int d^{3}\mathbf{x}\rho =\int d^{3}\mathbf{x}\varphi ^{\dagger
}(x)\varphi (x)  \label{eq.2.10prime}
\end{equation}
In zeroth approximation (no interaction), the number of particles
(electrons) is given by the time independent and Lorentz invariant $n^{(0)}$%
, using Eq. (\ref{eqforphi})

\begin{equation}
n^{(0)}=(2\pi )^{3}\int d^{3}\mathbf{p}\frac{\omega ^{\dagger }(\bar{p})}{%
\bar{p}_{0}}\frac{\omega (\bar{p})}{\bar{p}_{0}}  \label{eq.2.10}
\end{equation}
The expectation value of the operator $\Omega ^{\dagger }\Omega $ of Eq. (%
\ref{eq.2.13}) is (compare with eq. (\ref{h18}))

\begin{equation}
\overline{\Omega ^{\dagger }\Omega }=\frac{(2\pi )^{3}}{n^{(0)}}\int d^{3}%
\mathbf{p}\frac{\omega (\bar{p})^{\dagger }}{\bar{p}_{0}}\Omega ^{\dagger }(%
\bar{p}_{0})\Omega (\bar{p}_{0})\frac{\omega (\bar{p})}{\bar{p}_{0}}
\label{eq.3.5'}
\end{equation}
As before, the number of final electrons is given by the time dependent

\[
n^{(2)}=(2\pi )^{3}\int d^{3}\mathbf{p}\frac{\omega ^{(2)\dagger }(\bar{p}
,x_{0})}{\bar{p}_{0}}\frac{\omega ^{(2)}(\bar{p},x_{0})}{\bar{p}_{0}}, 
\]

By the same mechanism as in the scalar case (see eq. (\ref{h19})), the
integration over $s_{0}$ yields a linear increase of $n^{(2)}$ with time $%
x_{0}.$ The singularity at $s_{0}=0$ expresses the equality of initial and
final energy (the outgoing photon energy being $-k_{0}^{\prime }$ for
Compton scattering; see the comment after eq. (\ref{eq.2.13})).

These are the main ingredients for obtaining the Klein-Nishina formula
(after insertion of the correct expressions for the $V$ 's, and similarly
for the $P_{k}$ 's) see further discussion in sections 6.5 and 6.7.

\subsection{Lorentz and gauge invariant squared matrix element $\overline{
\Omega ^{\dagger }\Omega }$}

A systematic feature of Stueckelberg's 1934 paper is the care he takes to
make the Lorentz and gauge invariance manifest. This is particularly the
case for $\overline{\Omega ^{\dagger }\Omega },$ the square of the
transition matrix element function of the polarizations $e_{k}$ and $%
e_{k^{\prime }}$ of the electromagnetic field. In his calculation,
Stueckelberg considers a final photon which is on mass-shell and transverse,
i.e. $k^{\prime 2}=k^{\prime }\cdot e_{k^{\prime }}=0.$ On the other hand,
he keeps terms proportional to $k^{2}$ and $k\cdot e_{k}$. This allows him
to use his formulas also for Bremsstrahlung (see below). For Compton
scattering he will of course put also $k^{2}=k\cdot e_{k}=0.$

Stueckelberg now calculates $\overline{\Omega ^{\dagger }\Omega }$ with the
help of eq. (\ref{eq.2.13}). We have shown above that the sum over final
spins is taken care of. What remains to be done is to apply the free Dirac
equation for $\omega (\bar{p}),$ the mass shell conditions for the initial
and final electrons and the energy momentum conservation in order to obtain
a simpler expression. In the end he uses the formula

\[
\omega (\bar{p})^{\dagger }\gamma _{0}\gamma _{\mu }\omega (\bar{p})=\frac{%
\bar{p}_{\mu }}{mc}. 
\]

In the case of Compton scattering, Stueckelberg's result was obtained later
by Wannier 1935, taking traces of Dirac matrices$.$

The general formula found by Stueckelberg is the following

\begin{eqnarray}
\frac{p_{0}}{2p_{0}^{\prime }}\overline{\Omega ^{\dagger }\Omega } &=&\frac{
(e_{k^{\prime }}\cdot p)^{2}}{(k^{\prime }\cdot p)^{2}}\left\{ 2(e_{k}\cdot
p^{\prime })^{2}+\frac{1}{2}e_{k}^{2}k^{2}-2(e_{k}\cdot p^{\prime
})(e_{k}\cdot k)\right\}  \label{eq.4.6} \\
&&+\frac{(e_{k^{\prime }}\cdot p^{\prime })^{2}}{(k^{\prime }\cdot p^{\prime
})^{2}}\left\{ 2(e_{k}\cdot p)^{2}+\frac{1}{2}e_{k}^{2}k^{2}+2(e_{k}\cdot
p)(e_{k}\cdot k)\right\}  \nonumber \\
&&-2\frac{(e_{k^{\prime }}\cdot p)(e_{k^{\prime }}\cdot p^{\prime })}{
(k^{\prime }\cdot p)(k^{\prime }\cdot p^{\prime })}\left\{ 2(e_{k}\cdot
p)(e_{k}\cdot p^{\prime })+\frac{1}{2}e_{k}^{2}k^{2}-(e_{k}\cdot k^{\prime
})(e_{k}\cdot k)\right\}  \nonumber \\
&&+2\frac{(e_{k^{\prime }}\cdot p)}{(k^{\prime }\cdot p)}(e_{k^{\prime
}}\cdot e_{k})(e_{k}\cdot 2p^{\prime }-k)  \nonumber \\
&&-2\frac{(e_{k^{\prime }}\cdot p^{\prime })}{(k^{\prime }\cdot p^{\prime })}
(e_{k^{\prime }}\cdot e_{k})(e_{k}\cdot 2p+k)+2(e_{k}\cdot e_{k^{\prime
}})^{2}  \nonumber \\
&&+\frac{1}{2(k^{\prime }\cdot p)(k^{\prime }\cdot p^{\prime })}\left\{
e_{k}^{2}(k^{\prime }\cdot k)^{2}-2(e_{k}\cdot k^{\prime })(e_{k}\cdot
k)(k^{\prime }\cdot k)+(e_{k}\cdot k^{\prime })^{2}k^{2}\right\}  \nonumber
\end{eqnarray}
only Lorentz invariant scalar products appear. In this expression $p^{\prime
}$ and $p$ satisfy the mass-shell conditions $(p^{\prime })^{2}+M^{2}=0$ and 
$(p)^{2}+M^{2}=0$\footnote{%
In order not to overload the notations we skipped in the above formula the
''bars'' on $p$ and $p^{\prime }.$}. Equation (\ref{eq.4.6}) is again
manifestly Lorentz invariant and \textit{coincides} \textit{with the result
which would be obtained with Feynman rules.}

It is also invariant under the two independent gauge transformations%
\footnote{%
The invariance under the first transformation holds even off mass-shell $%
k^{2}\neq 0,$ when the photon is coupled to a conserved current, and the
electrons are on mass.shell (we thank M. Veltman for pointing out this
result to us).}

\begin{eqnarray}
e_{k} &\rightarrow &e_{k}+const.\cdot k \\
e_{k^{\prime }} &\rightarrow &e_{k^{\prime }}+const.\cdot k^{\prime } 
\nonumber
\end{eqnarray}

As Stueckelberg pointedly remarks, checking this invariance prevents
algebraic errors. Indeed, if one specializes (\ref{eq.4.6}) to a transverse
photon $k$, and compares the result with Eq. 11-13 p.231 of Jauch-Rohrlich
1955, one finds an error in the latter\footnote{%
On p. 233 the authors however claim that their equation (11-13) is gauge
invariant. Therefore the missing factor 2 in the last term of the first line
must be a misprint.}. As already mentioned, Stueckelberg's insistence on
gauge invariance was rather unusual at the time of his paper. Where the
gauge is not fixed from the beginning (as Fermi's 1932 or Heitler's 1936
choice of the Coulomb gauge $e_{0}=0$ ), it is merely noticed. Again, the
virtue of Eq. (\ref{eq.4.6}) is its generality. Since the signs of $p_{0}$
and $p_{0}^{\prime }$ are not fixed, the formula can be used, besides for
Compton scattering and Bremsstrahlung, to discuss pair production and
annihilation\footnote{%
Franz 1938 gives the matrix elements for arbitrary polarizations of the
photon \textit{and} the electron, in the electron rest-system. See also
Nishina 1929a,b.}.

\subsection{Pauli 1933 $:$ Klein-Nishina formula for moving electrons}

An obvious application of Stueckelberg's relativistic formalism was to
answer a question raised by Pauli . At that time, discrepancies were found
between the Klein-Nishina formula for Compton scattering and experimental
data for high energy incoming photons (see Schweber 1994 p.82). As we noted
before, all Compton scattering calculations for spin 1/2 electrons done
before Stueckelberg were performed in the rest system of the initial
electron. Pauli asked the question whether the K-N formula was still valid
in the limit where the initial and final light frequencies $\nu $ and $\nu
^{\prime }$ go to infinity, their ratio being kept constant. To this end he
Lorentz-transformed the $K-N$ formula from the rest system of the initial
electron to one with arbitrary velocity $\vec{v}$. Taking then the above
limit, he found that the cross-section depended explicitly on $\vec{v}$. His
very elegant calculation nevertheless took five pages, whereas Stueckelberg
could very easily specialize his formula to this case.

Pauli's formula for Compton scattering of unpolarized photons is (in modern
notation, as in eq. (\ref{crossscal})) 
\begin{equation}
\frac{d\sigma }{d\Omega }=\frac{\alpha ^{2}}{2M^{2}}\left( \frac{mc^{2}}{E}%
\right) ^{2}\left( \frac{\nu ^{\prime }}{\nu }\right) ^{2}\frac{1}{D^{2}}%
\left[ \frac{\nu D}{\nu ^{\prime }D^{\prime }}+\frac{\nu ^{\prime }D^{\prime
}}{\nu D}-\sin ^{2}\theta \right]  \label{p1}
\end{equation}
$\theta $ is the scattering angle, $\nu $ and $\nu ^{\prime }$ the initial
resp. final frequencies of the photon. This formula differs from the
Klein-Nishina one by the necessary changes when going from the initial
electron rest system to an electron moving with velocity $v,$ namely 
\[
mc^{2}\rightarrow E,\quad \frac{\nu ^{\prime }}{\nu }\rightarrow \frac{\nu
^{\prime }}{\nu }\frac{D^{\prime }}{D} 
\]
Here $D$ and $D^{\prime }$ are the Doppler factors $D=1-\frac{v}{c}\cos
\alpha $ and $D^{\prime }=1-\frac{v}{c}\cos \alpha ^{\prime }$, where $%
\alpha $ resp. $\alpha ^{\prime }$ are the angles between the initial
electron and the initial, resp. final photon. This can be written in the
invariant way 
\begin{eqnarray}
p_{0}k_{0}D &=&p\cdot k \\
p_{0}k_{0}^{\prime }D^{\prime } &=&p\cdot k^{\prime }  \nonumber
\end{eqnarray}
Stueckelberg specializes eq. (\ref{eq.4.6}) for $\frac{p_{0}}{2p_{0}^{\prime
}}\overline{\Omega ^{\dagger }\Omega }$ to Compton scattering, i.e. $%
k^{2}=k\cdot e_{k}=0$ and $e_{k}^{2}=1.$ He then chooses the gauge for which 
$p\cdot e_{k}=p\cdot e_{k^{\prime }}=0,$ corresponding to transversality in
the initial electron rest system. Hence he obtains 
\begin{equation}
W\equiv \frac{p_{0}}{2p_{0}^{\prime }}\overline{\Omega ^{\dagger }\Omega }
=2(e_{k}\cdot e_{k^{\prime }})^{2}+\frac{(k\cdot k^{\prime })^{2}}{%
2(k^{\prime }\cdot p)(k^{\prime }\cdot p^{\prime })}  \label{eqforw}
\end{equation}

Using energy-momentum conservation and mass-shell conditions for electrons,
he recognizes that 
\begin{equation}
\frac{(k\cdot k^{\prime })^{2}}{2(k^{\prime }\cdot p)(k^{\prime }\cdot
p^{\prime })}=\frac{1}{2}\left[ \frac{\nu D}{\nu ^{\prime }D^{\prime }}+%
\frac{\nu ^{\prime }D^{\prime }}{\nu D}\right] -1  \label{cinetik}
\end{equation}
Stueckelberg then averages over initial polarizations $e_{k}$ and sums over
final polarizations $e_{k^{\prime }}$. He finds for the first term in eq. (%
\ref{eqforw}) 
\begin{eqnarray}
\frac{1}{2}\sum_{e_{k}}\sum_{e_{k^{\prime }}}2(e_{k}\cdot e_{k^{\prime
}})^{2} &=&1+\cos ^{2}\theta \\
\cos \theta &=&1-\frac{(k\cdot k^{\prime })p^{2}}{(k^{\prime }\cdot
p)(k\cdot p)}
\end{eqnarray}
and the second term is doubled. Therefore 
\begin{equation}
\frac{1}{2}\sum_{e_{k}}\sum_{e_{k^{\prime }}}W=\frac{\nu D}{\nu ^{\prime
}D^{\prime }}+\frac{\nu ^{\prime }D^{\prime }}{\nu D}-\sin ^{2}\theta
\end{equation}
which is the bracket in eq. (\ref{p1}) naturally written in the invariant
form 
\[
\frac{(p\cdot k)}{(p\cdot k^{\prime })}+\frac{(p\cdot k^{\prime })}{(p\cdot
k)}+\left( \frac{(k\cdot k^{\prime })(p\cdot p)}{(k\cdot p)(k^{\prime }\cdot
p)}\right) ^{2}-2\left( \frac{(k\cdot k^{\prime })(p\cdot p)}{(k\cdot
p)(k^{\prime }\cdot p)}\right) 
\]
See more details in section (6.7).

\subsection{Bremsstrahlung}

Bremsstrahlung is the emission of a free photon by an electron interacting
with the Coulomb field of a nucleus. It is thus analogous to Compton
scattering, where the initial photon (say $k$) is replaced by a classical
electromagnetic field\footnote{%
That's why in the formula (\ref{eq.2.1}) $P_{k}$ is put to 1.}

\begin{equation}
A_{k}=\frac{2\hbar c}{e^{2}}M_{k}e_{k}\cos (k\cdot x)  \label{eq.3.7}
\end{equation}

\begin{equation}
M_{k}=\frac{4\pi Ze^{2}}{k^{2}\hbar c}  \label{aux3}
\end{equation}

In a particular reference system, where $k_{0}=0$, $M_{k}$ is the Fourier
transform of the Coulomb potential $Ze^{2}/r$ of a nucleus. $e_{k}$ is again
a polarization vector.

To treat this problem, the strategy of Stueckelberg is to define what he
calls the generalized Klein-Nishina formula. To this end he replaces $V_{k}%
\sqrt{N_{k}}$ by $M_{k}$ ($k^{2}\neq 0,(e_{k}\cdot k)\neq 0$ ) everywhere in
the calculation for Compton scattering leading to the K-N formula. His aim
is then to deduce the formula for Bremsstrahlung from this generalized K-N
formula. In particular, he uses Eq. (\ref{eq.4.6}) whith one photon off
mass-shell.

Previously to Stueckelberg, the formula for Bremsstrahlung in the
electrostatic field was obtained by Bethe-Heitler 1934 and by Sauter 1934 in
second order of perturbation theory. On the other hand, Williams and v.
Weizs\"{a}cker 1934, using qualitative arguments, found an approximate
formula which for large initial energy $E_{e}$ of the electron and large
energy compared with $mc^{2}$of the emitted photon, agrees with the
Bethe-Heitler-Sauter formula. The idea of v. Weizs\"{a}cker-Williams,
inspired by Fermi 1924, goes as follows:

In the rest-system of the nucleus the Fourier transform of the static field
is given by

\begin{equation}
V^{L}=\frac{1}{(2\pi )^{3}}\int d^{4}k\delta (k_{0})M_{k}(e_{k}\cdot
k)e^{i(k\cdot x)}  \label{eq.6.1}
\end{equation}

Since $E_{e}\gg mc^{2}$ by assumption, the electron has a large velocity $%
\vec{v}$. In the rest- system of the electron, the partial waves of (\ref
{eq.6.1}) move with velocity $\vec{V}=$ -$\vec{v}$. $\left| \vec{V}\right| $
is almost equal to the light velocity $c$, and the partial waves for small $%
k^{2}$ are almost transverse (($e_{k}\cdot k)\cong $ 0). v.
Weizs\"{a}cker-Williams apply to these quasi-lightwaves the Klein-Nishina
formula for electrons at rest. The Bremsstrahlung, calculated in this frame,
appears as an incoherent sum of the scattering amplitudes of the individual
partial waves $k$. This can be justified by the fact that one averages over
wave-packets large in comparison with the nuclear field.

In his 1934 paper, Stueckelberg shows that the Bethe-Heitler formula for
Bremsstrahlung, exact to second order, can be deduced from the generalized
K-N formula in the same way as the approximation for large velocities was
deduced by v. Weizs\"{a}cker-Williams from the ordinary K-N formula\footnote{%
In Stueckelberg 1935b the same idea was used to obtain the formula for pair
creation by a fast electron in the field of a nucleus}.

Here too, Stueckelberg takes advantage of the manifest relativistic and
gauge invariance of his formalism, which allows him to go back and forth
between the nucleus rest-system (B-H-S formula) and the electron rest-system
(v. W-W formula). For example, to go from the former to the latter, a gauge
transformation is used from a polarization with only a time component, to
one with only space components. Finally, Stueckelberg shows that the
incoherent addition of the contributions of partial waves is rigorously
justified.

\subsection{Invariant averaging over photon polarization\protect\footnote{%
Remark on the intensity of the radiative scattering of moving free
electrons, Stueckelberg 1935a. }}

In section 6.4 we mentioned Stueckelberg's derivation of the Pauli formula
for Compton scattering by moving electrons. After the publication of
Stueckelberg's paper, Pauli pointed out to Stueckelberg that he first
averaged over the photon polarization in the rest system of the electron,
and only then Lorentz-transformed it back to the moving frame.

Stueckelberg replies to Pauli's objection by remarking that, since
unpolarized light is a Lorentz-invariant notion, the averaging is a
Lorentz-invariant operation and can therefore be performed in any reference
system.

Nevertheless, in the 1935 paper, Stueckelberg shows that one can average in
a way which exhibits manifest invariance from the beginning. Instead of
relating the polarization $e_{k}$ to the potential $A$ through 
\begin{equation}
A^{\mu }=e^{\mu }\exp i(k\cdot x)
\end{equation}
the idea is to use the anti-symmetric field-strength tensor $F$ and its
conjugate (dual) $\tilde{F}$ (corresponding to the usual electric and
magnetic fields). 
\begin{eqnarray}
F^{\mu \nu } &=&k^{\mu }e^{\nu }-k^{\nu }e^{\mu } \\
\tilde{F}_{\mu \nu } &=&\frac{1}{2}\epsilon _{\mu \nu \rho \sigma }F^{\rho
\sigma }  \nonumber
\end{eqnarray}

where $\epsilon _{\mu \nu \rho \sigma }$ is the usual antisymmetric symbol.
Consider the Lorentz force per unit charge $\epsilon ^{\mu }$ on the
electron with momentum $p\footnote{%
In particular, for $\mathbf{p=0,}$ this is just the electric field.}$. 
\begin{equation}
\epsilon ^{\mu }=F^{\mu \nu }p_{\nu }
\end{equation}
which can be normalized 
\begin{equation}
\frac{\epsilon ^{2}}{(k\cdot p)^{2}}=1
\end{equation}
and define also the dual expression 
\begin{equation}
\hat{\epsilon}^{\mu }=\tilde{F}^{\mu \nu }p_{\nu }
\end{equation}

Because of the following orthogonality properties 
\begin{equation}
\epsilon \cdot \hat{\epsilon}=\epsilon \cdot k=k\cdot \hat{\epsilon}=0
\end{equation}
the new four-vectors $\epsilon $ and $\hat{\epsilon}$ have the right
properties to express the polarization, since they are mutually orthogonal
and orthogonal to $k.$ On the other hand, with the old polarizations $e_{k},$
these orthogonality properties cannot be ensured in a Lorentz invariant way.

The squared matrix element for Compton scattering (\ref{eq.4.6} specialized
to $k^{2}=k\cdot e=0$ ) can now simply be written 
\begin{equation}
W=\frac{1}{2}\left\{ \frac{|\epsilon _{k^{\prime }}|}{|\epsilon _{k}|}+\frac{%
|\epsilon _{k}|}{|\epsilon _{k^{\prime }}|}\right\} -1+2\left( \frac{%
\epsilon _{k}\cdot \epsilon _{k^{\prime }}}{(k\cdot p)(k^{\prime }\cdot p)}%
\right) ^{2}.  \label{eq.35.21}
\end{equation}
Averaging the last term of this expression over the polarizations (using now
a special frame) gives 
\begin{equation}
\frac{1}{2}\sum_{\epsilon _{k}}\sum_{\epsilon _{k^{\prime }}}2\left( \frac{
\epsilon _{k}\cdot \epsilon _{k^{\prime }}}{(k\cdot p)(k^{\prime }\cdot p)}%
\right) ^{2}=1+\left( 1-\frac{(k\cdot k^{\prime })p^{2}}{(k\cdot
p)(k^{\prime }\cdot p)}\right) ^{2}
\end{equation}

and the same averaging doubles the other terms.

Now one finds again the result of Pauli 1933, namely 
\begin{equation}
\frac{1}{2}\sum_{\epsilon _{k}}\sum_{\epsilon _{k^{\prime }}}W=\frac{Dk_{0}}{%
D^{\prime }k_{0}^{\prime }}+\frac{D^{\prime }k_{0}^{\prime }}{Dk_{0}}-\sin
^{2}\theta
\end{equation}

\section{The response to Stueckelberg's paper}

Stueckelberg's 1934 paper, as well as others where he followed the same
method, did not get much attention. The first paper which referred to it was
Wannier (1935). He mentioned that his result for the trace over spinor
indices has also been obtained by Stueckelberg in his eq. (\ref{eq.4.6}).
But Stueckelberg's calculus is for instance not mentioned by Heitler 1936.
This could be an illustration of the difference of ''styles'' of Heitler and
Stueckelberg. While the first tries to get as simple expressions as
possible, ready for applications, the second is more interested in the
general structure of the theory with an emphasis on symmetries and other
fundamental principles. Strangely enough, even Wentzel, who suggested the
idea of Stueckelberg's work, never quoted it.

Pauli had certainly noticed the 1934 paper, since, as discussed above,
Stueckelberg 1935 wrote a sequel on a Lorentz-covariant polarization vector
answering the criticism of Pauli (see preceding section). Later on, in a
letter to Heisenberg (5 February 1937), Pauli drew Heisenberg's attention to
the 1934 paper\footnote{%
Zum Formalismus der St\"{o}rungstheorie m\"{o}chte ich Dich noch auf eine
Arbeit von Stueckelberg aufmerksam machen. Die Arbeit ist nicht sehr gut
geschrieben, aber die zu Grunde liegende Idee (die auf Wentzel
zur\"{u}ckgeht) scheint mir vern\"{u}nftig; Sie besteht darin, die
relativistische Invarianz dadurch in Evidenz zu setzen, dass man Raum und
Zeit ganz aus der Theorie entfernt und die Koeffizienten der \textit{vier-}
dimensionalen Fourier-Entwicklung der Wellenfunktion direkt untersucht.
(Pauli 1937).}:

\begin{quotation}
I would also like to draw your attention to a work of Stueckelberg
concerning the formalism of perturbation theory. The paper is not well
written, but the basic idea (due to Wentzel) seems reasonable to me; it
consists in making evident the relativistic invariance by eliminating space
and time completely from the theory and examining directly the coefficients
of the \textit{four}-dimensional Fourier expansion of the wave function.%
\footnote{%
On Wentzel's influence on Stueckelberg see section 6.1.}
\end{quotation}

After the 1939-45 war, a whole new era started in quantum field theory with
the renormalization program. In his report to the Solvay congress 1948, J.
R. Oppenheimer insists on the necessity to preserve covariance in all steps
of the calculation if one wants to eliminate the infinities. As an example
of such a covariant theory he quotes Stueckelberg's paper.

\begin{quotation}
Now it is true that the fundamental equations of quantum-electrodynamics are
gauge and Lorentz covariant. But they have in a strict sense no solutions
expansible in powers of $e$. If one wishes to explore these solutions,
bearing in mind that certain infinite terms will, in a later theory, no
longer be infinite, one needs a covariant way of identifying these terms,
and for that, not merely the field equations themselves, but the whole
method of approximation and solution must at all stages preserve covariance.
This means that the familiar Hamiltonian methods, which imply a fixed
Lorentz frame t=constant, must be renounced; neither Lorentz frame nor gauge
can be specified until after, in a given order in $e$, all terms have been
identified, and those bearing on the definition of charge and mass
recognized and relegated; then of course, in the actual calculation of
transition probabilities and the reactive corrections to them, or in the
determination of stationary states in fields which can be treated as static,
and in the reactive corrections thereto, the introduction of a definite
coordinate system and gauge for these no longer singular and completely
well-defined terms can lead to no difficulty.

It is probable that, at least to order $e^{2}$, more than one covariant
formalism can be developed. Thus Stueckelberg's four-dimensional
perturbation theory would seem to offer a suitable starting point, as also
do the related algorithms of Feynman\footnote{%
see the reprint in Schwinger 1958, p. 150.}.
\end{quotation}

In the same spirit F. J. Dyson (1949) comments in the notes added in proof
to his paper ''The Radiation Theories of Tomonaga, Schwinger, and Feynman'':
''A covariant perturbation theory similar to that of section III has
previously been developped by E.C.G. Stueckelberg'' referring to the 1934
paper and to Nature vol. 153 (1944), p. 143.

We have already mentioned Weisskopf's reminiscences (section 6.1). The
importance of \textit{manifest} Lorentz and gauge invariance for the
development of the theory in these times has been also emphasized by Pais
1986, p. 457. When commenting on Schwinger's Lorentz invariant 1948
calculation of various terms contributing to the Lamb shift, Pais
writes:''Schwinger's direct calculation of the electric term produced a most
unpleasant surprise, however: it was too small by a factor $1/3$ !''%
\footnote{%
Pais' source is here J. Schwinger in Brown and Hoddeson (1983), p. 329.}
Pais continues by quoting Schwinger:

\begin{quotation}
This difficulty is attributable to the incorrect transformation properties
of the electron self-energy in the conventional Hamiltonian treatment and is
completely removed in the covariant formalism now employed.\footnote{%
Schwinger (1949).}
\end{quotation}

Pais concludes with the following discussion:

\begin{quotation}
How can a fully covariant theory yield non-covariant results? Because during
the calculation one has to subtract infinity from infinity, which in general
is not a well-defined step. How can one hope to avoid non-covariant answers?
By computing in such a way that covariance is manifest at every stage; and
likewise for gauge invariance. Take, for example, Heisenberg and Pauli's
treatment of quantum electrodynamics in the Coulomb gauge which [...] may
not look covariant but is covariant nevertheless. It is not, however, 
\textit{manifestly} covariant at every stage. Thus the Coulomb gauge does
not lend itself (readily) to the evaluation of radiative corrections.
Likewise the second order perturbation formula [...] and its higher-order
partners, though actually covariant, are not manifestly so. One can,
however, cast them in an equivalent manifestly covariant form, as in fact
Stueckelberg had already shown in 1934.\footnote{%
In his book (p. 244-45), S. S. Schweber reports on the other hand this
revealing reaction of Schwinger when asked about the disparities between his
calculation of the Lamb shift and that of French and Weisskopf: ''Well, if
you do not keep the calculation explicitely covariant, anything can happen''.
\par
We shall address the general problem of renormalization in Quantum Field
Theory and Stueckelberg's contribution to it in a subsequent publication.}
\end{quotation}

Finally this opinion of Gell-Mann 1989 (p. 702):

\begin{quotation}
By about 1950 it was known that QED is renormalizable for charged spinor
particles [...] The second order renormalizability of the charge in QED had
been established in 1934 by Dirac and Heisenberg, and that of the mass by a
number of authors in 1948. Of those, the first ones to complete correct
relativistic calculations of the Lamb shift (Willis Lamb and Norman Kroll
and J. Bruce French and Victor Weisskopf) actually used the clumsy old
non-covariant method. The place where the new covariant methods played a
crucial role, particulary those of E.C.G. Stueckelberg and Richard P.
Feynman, which are still used today, was in permitting calculations to be
done quickly, especially to fourth and higher orders (which would have been
impractical with the old methods), and in making possible the proof of
renormalizability to all orders.
\end{quotation}

So, after all, why didn't people take advantage of Stueckelberg's method. ?
Several reasons could be proposed. The paper itself was certainly not
written in a transparent style (remember Weisskopf's and Pauli's comments).
His notations were clumsy (see our appendix). On the other hand, some of the
mathematical features of Stueckelberg's theory (contour integrals instead of
Dirac's delta function) could have made it unfamiliar to use. Also, more
generally, the paper didn't provide a new physical result but ''only'' a new
method for deriving results already obtained;\ no one realised (see
Weisskopf's reminiscences) that it could have been used to treat in a more
consistent way the unsolved divergence difficulties.

\section{Acknowledgments}

It is a pleasure to thank R. Casalbuoni, C, Gomez, K. von Meyenn, S. S.
Schweber, R. Stora, and M. Veltman for useful discussions; we also thank J.
Heilbron and A. Pais for their reading of the initial draft of this paper
and their helpful comments.

\section{Appendix}

In order to make easier the comparison between the various works considered
in this paper, we have adopted a set of conventions that we systematically
use when discussing the technical details. The necessary changes in
notations or conventions with respect to the original papers are of minor
importance, but for the sake of historical accuracy, we report below the
main alterations.

Four-vectors are written in normal type, with their spatial part noted in
boldface: and the time-component $a_{0}$; thus $a=(a_{1},a_{2},a_{3},a_{0})=(%
\mathbf{a,}a_{0})$. In particular, the space-time location of an event is
given by the four-vector $(\mathbf{x,}ct)$ where the time-component $%
x_{0}=ct $ and $c$ is the speed of light in vacuum. Our relativistic metric
is given by the matrix $g_{\mu \nu }$ with $g_{ii}=-g_{00}=1$, $i=1,2,3$.
All the authors we consider adopt instead the euclidean metric, introducing
for each time-component $a_{0}$ its euclidean counterpart $a_{4}=ia_{0}.$
Thus, the space-time scalar product between two four-vectors $a$ and $b$ is
in our notation $a\mathbf{\cdot }b=\mathbf{ab}-a_{0}b_{0}\equiv a_{\mu
}b^{\mu }$ where the sum convention has been used in the last expression.
Clearly, $a\mathbf{\cdot }b=\mathbf{ab}-a_{0}b_{0}=\mathbf{ab}+a_{4}b_{4}.$

An electromagnetic plane wave $\psi $ of angular frequency $\omega =2\pi \nu 
$, wavelength $\lambda ,$and direction of propagation given by the unit
vector $\mathbf{\hat{k}}$ is given by 
\[
\psi =\exp (i\frac{\omega }{c}(\mathbf{\hat{k}x-}ct))=\exp (ik\cdot x) 
\]
where one introduced the four-vector $k=(\mathbf{k},k_{0})=\frac{\omega }{c}(%
\mathbf{\hat{k},}1\mathbf{).}$ Similarly, using the Einstein-Planck-de
Broglie relations: 
\begin{eqnarray*}
\mathbf{p} &=&\hbar \mathbf{k} \\
E &=&\hbar \omega
\end{eqnarray*}
we associate to a particle with four-momentum $p=(\mathbf{p},\frac{E}{c})$
the matter-wave 
\[
\exp (\frac{i}{\hbar }(\mathbf{px-}Et)). 
\]

The wave vectors of the initial and final radiation are noted respectively $%
\mathbf{\hat{k}}$ and $\mathbf{\hat{k}}^{\prime },$ and in general
quantities related to the final state are primed.

The original notations of Stueckelberg appear quite unusual (which certainly
doesn't help the reader...). We present now a dictionary for going from his
to our notations.

Four momenta:

initial electron:$l^{0}\rightarrow p$

final electron: $l\rightarrow p^{\prime }$

initial photon: $k\rightarrow k$

final photon:$-p,$then $m\rightarrow k^{\prime }$

Photon polarization: $\sigma ^{k}\rightarrow e_{k}$

Photon state:$u(N)\rightarrow |N\succ $

Photon annihilation and creation operators: $\Gamma _{k},\Gamma
_{k}^{\dagger }\rightarrow a_{k},a_{k}^{\dagger }$

Electron spinor:$A(l^{0})\rightarrow u(p)$

Electron spinor on mass-shell:$B(l^{0})\rightarrow \omega (p)$

Adjoint spinor:$B^{\dagger }=B^{*}\gamma _{4}\rightarrow \bar{\omega}=\omega
^{\dagger }\gamma _{0}$


\section{Bibliography}

\begin{thebibliography}{Born$,$ Heisenberg and Jordan 1926 }
\bibitem[Bethe and Fermi 1932]{bf32}  H. Bethe and E. Fermi, \"{U}ber die
Wechselwirkung von zwei Elektronen, \textit{Zeit. f. Phys.}, vol. 77 (1932),
pp. 296-306.

\bibitem[\thinspace Bethe and Heitler 1934]{bh33}  H. Bethe and W. Heitler,
On the Stopping of Fast Particles and the Creation of Positive Electrons, 
\textit{Proc. Roy. Soc}. \textit{A}, vol. 146 (1933), p. 83-112.

\bibitem[Bjorken and Drell 1964]{bd64}  J. D. Bjorken and S. D. Drell, 
\textit{Relativistic Quantum Mechanics}, McGraw-Hill Book Company, 1964.

\bibitem[Bohr, Kramers and Slater 1924]{bks24}  N. Bohr, H. Kramers and J.
C. Slater, The Quantum Theory of Radiation, \textit{Phil. Mag}., vol.47
(1924), pp. 790-793.

\bibitem[Born and Jordan 1925]{bj25}  M. Born and P. Jordan, Zur
Quantenmechanik, \textit{Zeit. f. Phys.}, vol. 34 (1925), pp. 858-888.

\bibitem[Born$,$ Heisenberg and Jordan 1926 ]{bhj26}  M. Born, W. Heisenberg
and P. Jordan, Zur Quantenmechanik II, \textit{Zeit. f. Phys.}, vol. 35
(1926), p. 557-615.

\bibitem[Bothe 1923 ]{bot23}  W. Bothe, Ueber eine neue
Sekund\"{a}rstrahlung der R\"{o}ntgenstrahlen: I Mitteilung, \textit{Zeit.
f. Phys}., vol. 16 (1923), pp. 319-20.

\bibitem[Bothe and Geiger 1925 ]{botgei25}  W. Bothe and H. Geiger, Ueber
das Wesen des Comptoneffekts, \textit{Zeit. f. Phys}., vol. 32 (1925), pp.
639-663.

\bibitem[Breit 1926 ]{brei26}  G. Breit, A correspondence principle in the
Compton effect., \textit{Phys. Rev}., vol. 27, 1926, p. 362-372.

\bibitem[Bromberg 1977]{brom77}  J. Bromberg, Dirac's Quantum
Electrodynamics and the Wave-Particle Equivalence, in \textit{History of
20th Century Physics}, \textit{Varenna 1977}, C. Weiner (ed.), Academic
Press 1977, pp. 147-157.

\bibitem[Brown 1993]{brown93}  L. M. Brown, Introduction: Renormalization
1930-1950, in \textit{Renormalization. From Lorentz to Landau (and beyond)},
L. M. Brown (ed.), Springer 1993, pp. 1-27.

\bibitem[Brown and Hoddeson 1983 ]{brohodd83}  \textit{The birth of particle
physics}, L. M. Brown and L. Hoddeson (eds), Cambridge Univ. Press, 1983.

\bibitem[Cao 1997]{cao97}  T. Y. Cao, \textit{Conceptual developments of
20th century field theories}, Cambridge University Press 1997.

\bibitem[Cassidy 1981]{cass81}  D. C. Cassidy, Cosmic ray showers, high
energy physics, and quantum field theories: Programmatic interactions in the
1930s, \textit{Historical Studies in the Physical Sciences, }vol. 12 (1981),
pp. 1-39.

\bibitem[Casimir 1933]{casi33}  H. Casimir, Ueber die Intensit\"{a}t der
Streustrahlung gebundener Elektronen, \textit{Helv. Phys. Acta}, vol. 6
(1933), pp. 287-304

\bibitem[Compton 1921 ]{comp21}  A. H. Compton, The Softening of Secondary
X-rays, \textit{\ Nature}, vol. 109 (1921) pp. 366-367.

\bibitem[Compton 1923 ]{comp23}  A. H. Compton, A Quantum Theory of the
Scattering of X-rays by light Elements, \textit{Phys. Rev.}, vol. 21 (1923),
pp. 483-502.

\bibitem[Compton and Simon 1925 ]{compsim25}  A. H. Compton and A. W. Simon,
Directed Quanta of Scattered X-rays, \textit{Phys. Rev.}, vol. 26 (1925), p.
289-299.

\bibitem[Compton and Allison 1935]{compall35}  A. H. Compton and S. K.
Allison, \textit{X-rays in Theory and Experiment}, Macmillan and Co, London
, 1935.

\bibitem[Cushing 1990]{cus90}  J. T. Cushing, \textit{Theory construction
and selection in modern physics. The S Matrix}, Cambridge University Press,
1990.

\bibitem[Debye 1923 ]{deby25}  P. J. Debye, Zerstreuung von
R\"{o}ntgenstrahlen und Quantentheorie, \textit{Phys. Zeitschrift}, vol. 24
(1923), p. 161-166.

\bibitem[Dirac 1925 ]{dir25fun}  P. A. M. Dirac, The Fundamental Equations
of Quantum Mechanics, \textit{Proc. Roy. Soc. A}, vol. 109 (1925), pp.
642-653.

\bibitem[Dirac 1926a]{dir26a}  P. A. M. Dirac, Quantum Mechanics and a
Preliminary Investigation of Hydrogen Atom, \textit{Proc. Roy. Soc. A}, vol.
110 (1926), pp.561-579.

\bibitem[Dirac 1926b ]{dir26b}  P. A. M. Dirac, Relativity Quantum Mechanics
with an Application to Compton Scattering, \textit{Proc. Roy. Soc. A}, vol.
111 (1926), pp. 405-23.

\bibitem[Dirac 1926c ]{dir26c}  P. A. M. Dirac, On the Theory of Quantum
Mechanics, \textit{\ Proc. Roy. Soc. A}, vol. 112 (1926), pp. 661-77.

\bibitem[Dirac 1927a]{dir27a}  P. A. M. Dirac, The Compton Effect in Wave
Mechanics, \textit{\ \ Proceedings of the Cambridge Philosophical Society},
vol. 23, part V (1927), pp. 500-507.

\bibitem[Dirac 1927b ]{dir27b}  P. A. M. Dirac, The Quantum Theory of
Emission and Absorption of Radiation , \textit{Proc. Roy. Soc. A}, vol. 114
(1927), pp. 243-265.

\bibitem[Dirac 1927c ]{dir27c}  P. A. M. Dirac, The Quantum Theory of
Dispersion, \textit{\ Proc. Roy. Soc. A}, vol. 114 (1927), pp. 710-728.

\bibitem[Dirac 1928a ]{dir28a}  P. A. M. Dirac, The Quantum Theory of the
Electron, \textit{\ Proc. Roy. Soc. A}, vol. 117 (1928), pp. 610-624.

\bibitem[Dirac 1928b ]{dir28b}  P. A. M. Dirac, The Quantum Theory of the
Electron, Part II, \textit{Proc. Roy. Soc. A}, vol. 118 (1928), pp. 351-361.

\bibitem[Dirac 1930 ]{dir30}  P. A. M. Dirac, A theory of Electrons and
Protons, \textit{\ Proc. Roy. Soc. A, }vol. 126 (1932), pp. 360-365.

\bibitem[Dirac 1932]{dir32}  P. A. M. Dirac, Relativistic Quantum Mechanics, 
\textit{Proc. Roy. Soc. A}, vol. 136 (1932), pp. 453-464.

\bibitem[Dirac, Fock and Podolsky 1932 ]{dirfocpod32}  P. A. M. Dirac, V. A.
Fock\textit{\ }and B. Podolsky\textit{, }On quantum electrodynamics\textit{,
Physikalische Zeitschrift der Sowjetunion}, vol. 2, no. 6, (1932), pp.
468-79.

\bibitem[Dresden 1987]{dres87}  M. Dresden, \textit{H. A. Kramers. Between
Tradition and Revolution}, Springer Verlag, New York, 1987.

\bibitem[Dyson 1949 ]{dys49}  F. J. Dyson, The Radiation Theories of
Tomonaga, Schwinger, and Feynman, \textit{Phys. Rev.}, vol. 75 (1949), pp.
486-502.

\bibitem[Ekspong 1994]{ekspo94}  G. Ekspong, The Klein-Nishina Formula%
\textit{, }in: \textit{\ The Oskar Klein Memorial Lectures,} G. Ekspong
(ed), World Scientific, Singapore, 1994.

\bibitem[Enz 1997]{enz97}  \textit{Wolfgang Pauli und sein Wirken an der ETH
Z\"{u}rich}, C. P. Enz, B. Glaus and G. Oberkofler (eds), vdf
Hochschulverlag an der ETH Z\"{u}rich, 1997.

\bibitem[Fermi 1924]{ferm24}  E, Fermi, \"{U}ber die Theorie des Stosses
zwischen Atomen und elektrisch geladenen Teilchen, Zeit. f. Phys., vol. 29
(1924), p. 315-327.

\bibitem[Fermi 1932 ]{ferm32}  E, Fermi, Quantum Theory of Radiation, 
\textit{Rev. Mod. Phys}., vol. 4 (1932), pp. 87-132.

\bibitem[Feynman 1948 ]{feynpoco48}  R. P. Feynman in ''Conference on
Physics-Pocono Manor, Pennsylvania, 30 March-1 April 1948, sponsored by the
National Academy of Sciences''. According to Schweber (1994) p. 631 n143,
the Pocono conference notes were prepared informally by J. A. Wheeler. See
also Schweber (1994) pp. 436-445.

\bibitem[Feynman 1949]{feyn49}  R. P. Feynman, The Theory of Positrons, 
\textit{Phys Rev, }vol. 76 (1949), pp. 749-759, reprinted in Schwinger
(1958), pp.225-235.

\bibitem[Florance 1910 ]{floran10}  D. C. H. Florance, Primary and Secondary 
$\gamma $ Rays, \textit{Phil. Mag}., vol. 20 (1910), pp. 921-938.

\bibitem[Fock 1926a]{fock26a}  V. Fock, Zur Schr\"{o}dingerschen
Wellenmechanik, \textit{Zeit. f. Phys}., vol. 38 (1926), pp. 242-250.

\bibitem[Fock 1926b ]{fock26b}  V. Fock, \"{U}ber die invariante Form der
Wellen- und der Bewegungs- gleichungen f\"{u}r einen geladenen Massenpunkt, 
\textit{Zeit. f. Phys}., vol. 39 (1926), pp. 226-232.

\bibitem[Franz 1938]{franz38}  W. Franz, Die Streuung von Strahlung am
magnetischen Elektron, \textit{Ann. d. Phys.}, vol. 33 (1938), p. 689-707.

\bibitem[Gell-Mann 1989]{gellma89}  M. Gell-Mann, Progress in elementary
particle theory, 1950-1964, in: \textit{Pions to Quarks, }Fermilab Symposium
1985, L. M. Brown, M. Dresden and L. Hoddeson (eds), Cambridge Univ. Press,
1989.

\bibitem[Gordon 1926 ]{gord26}  W. Gordon, Der Comptoneffekt nach der
Schr\"{o}dingerschen Theorie, \textit{\textit{Zeit. f. Phys.}}, vol. 40
(1926), pp. 117-133.

\bibitem[Gray 1913 ]{gray13}  J. A. Gray, The Scattering and Absorption of
the $\gamma $ Rays of Radium, \textit{Phil. Mag}., vol. 26 (1913), pp.
611-623.

\bibitem[Heitler 1936]{heitl36}  W. Heitler, \textit{The quantum theory of
radiation}, Clarendon Press, Oxford,1936.

\bibitem[Heisenberg 1925 ]{heis25}  W. Heisenberg, Ueber quantentheoretische
Umdeutung kinematischer und mechanischer Beziehungen, \textit{Zeit. f. Phys}
., vol. 33 (1925), pp. 879-893.

\bibitem[Heisenberg 1934]{heis34}  W. Heisenberg, Bemerkung zur Diracschen
Theorie des Positrons, \textit{\textit{Zeit. f. Phys.}}, vol. 90 (1934), p.
209-231.

\bibitem[Heisenberg and Pauli 1929 ]{heispaul29}  W. Heisenberg and W.
Pauli, Zur Quantentheorie der Wellenfelder I, \textit{\textit{Zeit. f. Phys.}%
}, vol. 56 (1929), p. 1-61.

\bibitem[Heisenberg and Pauli 1930 ]{heispaul30}  W. Heisenberg and W.
Pauli, Zur Quantentheorie der Wellenfelder II, \textit{\textit{Zeit. f. Phys.%
}}, vol. 59 (1930), p. 168-190.

\bibitem[Hund 1967]{hund67}  F. Hund, \textit{Geschichte der Quantentheorie}%
, Bibliographisches Institut, Hochschultaschenb\"{u}cher Verlag, 1967.

\bibitem[Jammer 1966]{jam66}  M. Jammer, \textit{The Conceptual Development
of Quantum Mechanics}, McGraw-Hill Book Company 1966.

\bibitem[Jauch and Rohrlich 1955]{jaurohr55}  J. M. Jauch and F. Rohrlich, 
\textit{The Theory of Photons and Electrons}, Addison-Wesley, Cambridge (MA)
1955.

\bibitem[Jordan 1927a]{jor27a}  P. Jordan, Zur Quantenmechanik der
Gasentartung, \textit{Zeit. f. Phys}., vol. 44 (1927), pp. 473-480.

\bibitem[Jordan 1927b]{jor27b}  P. Jordan, Philosophical Foundations of
Quantum Theory, \textit{Nature}, vol. 119 (1927), pp. 566-569, 779.

\bibitem[Jost 1972]{jost72}  R. Jost, Foundation of Quantum Field Theory, in 
\textit{Aspects of Quantum Theory}, A. Salam and E. Wigner (eds), Cambridge
University Press 1972, pp. 61-77

\bibitem[Kikuchi 1931]{kiku31}  S. Kikuchi, Zur Theorie des Comptoneffektes, 
\textit{Zeit. f. Phys}., vol 68 (1931), pp. 803-812.

\bibitem[Klein 1926]{klein26}  O. Klein,\textit{\ }Quantentheorie und
f\"{u}nfdimensionale Relativit\"{a}tstheorie\textit{, Zeit. f. Phys}., vol.
37 (1926), pp. 895-906.

\bibitem[Klein 1927]{klein27}  O. Klein, Elektrodynamik und Wellenmechanik
vom Standpunkt des Korrespondenzprinzips, \textit{\textit{Zeit. f. Phys.}},
vol. 41 (1927), pp. 407-442.

\bibitem[Klein and Nishina 1929 ]{kleinishi29}  O. Klein and Y. Nishina,
Ueber die Streuung von Strahlung durch freie Elektronen nach der neuen
relativistischen Quantendynamik von Dirac, \textit{\textit{Zeit. f. Phys.}, }
vol. 52 (1929), pp. 853-868.

\bibitem[Klein 1968]{klein68}  O. Klein, From my life of physics, in \textit{%
From a life of physics}, A. Salam (ed.), World Scientific 1989, pp. 69-84.

\bibitem[Kragh 1984]{kragh84}  H. Kragh, Equation with many Fathers. The
Klein-Gordon equation in 1926, \textit{Am. J. Phys}., vol. 52 (1984), pp.
1024-1033.

\bibitem[Kragh 1990]{kragh90}  H. Kragh, \textit{Dirac. A scientific
Biography}, Cambridge University press 1990.

\bibitem[Kragh 1992]{kragh92}  H. Kragh, Relativistic Collisions: The Work
of Christian M\o ller in the Early 1930's, \textit{AHES, }vol. 43 (1992),
pp. 299-328

\bibitem[Kramers 1924]{kram24}  H. A. Kramers, The Law of Dispersion and
Bohr's Theory of Spectra, \textit{Nature}, vol. 113 (1924), pp. 673-676 and
The Quantum Theory of Dispersion, vol. 114 (1924), pp. 310-311

\bibitem[Kramers and Heisenberg 1925 ]{kramheis25}  H. A. Kramers and W.
Heisenberg, \"{U}ber die Streuung von Strahlung durch Atome, \textit{Zeit.
f. Phys}., vol. 31 (1925), p. 681.

\bibitem[Landsberg and Mandelstam 1928]{landsman28}  G. Landsberg and L.
Mandelstam, Eine neue Erscheinung bei der Lichtzerstreuung in Krystallen, 
\textit{Naturw}. vol. 16 (1928), pp. 557-558 and 772 (Berichtigung).

\bibitem[Mehra and Rechenberg 1982-87]{mehrech82}  J. Mehra and H.
Rechenberg, The Historical Development of Quantum Theory, volume I-V,
Springer 1982-87.

\bibitem[M\o ller 1931 ]{moell31}  Ch. M\o ller, \"{U}ber den Stoss zweier
Teilchen unter der Ber\"{u}cksichtigung der Retardation der Kr\"{a}fte, 
\textit{Zeit. f. Phys}., vol 70 (1931), pp. 786-795.

\bibitem[Morse and Stueckelberg 1929]{morstuck29}  P. M. Morse and E. C. G.
Stueckelberg, Diatomic Molecules According to the Wave Mechanics I:
Electronic Levels of the Hydrogen Molecular Ion, Phys. Rev., vol. 33 (1929),
pp. 932-947.

\bibitem[Murdoch 1987]{murd87}  D. Murdoch, \textit{Niels Bohr's Philosophy
of Physics}, Cambridge University Press 1987.

\bibitem[Nishina 1929a ]{nishi29a}  Y. Nishina,Die Polarisation der
Comptonstreuung nach der Diracschen Theorie des Elektrons, \textit{\textit{\
Zeit. f. Phys.}, }vol. 52 (1929), pp. 869-877.

\bibitem[Nishina 1929b]{nishi29b}  Y. Nishina, Polarisation of Compton
Scattering according to Dirac's New Relativistic Dynamics, \textit{Nature},
vol. 123 (1929), p. 349.

\bibitem[Oppenheimer 1930]{oppenh30}  J. R. Oppenheimer, Note on the Theory
of the Interaction of Field and Matter, \textit{Phys. Rev}., vol. 35 (1930),
pp. 461-477.

\bibitem[Oppenheimer 1948 ]{oppenh48}  J. R. Oppenheimer, Electron Theory,
Report to the Solvay Conference for Physics at Brussels, Belgium, September
27 to October 2, , 1948, pp. 269-279, reprinted in \textit{Quantum
electrodynamics}, J. Schwinger (ed.), Dover, New York,1958.

\bibitem[Pais 1982 ]{pais82}  A. Pais, \textit{Subtle is the Lord : the
Science and the Life of Albert Einstein, }Clarendon Press, New York, 1982.

\bibitem[Pais 1986 ]{pais86}  A. Pais, \textit{Inward Bound. Of Matter and
Forces in the Physical World, }Clarendon Press, Oxford 1986.

\bibitem[Pais 1993 ]{pais93}  A. Pais, \textit{Niels Bohr's times, in
physics, philosophy and polity, }Clarendon Press, Oxford, 1993.

\bibitem[Pauli 1933]{pauli33}  W. Pauli, Ueber die Intensit\"{a}t der
Streustrahlung bewegter freier Elektronen, \textit{Helv. Phys. Acta}, vol. 6
(1933) pp. 279-286.

\bibitem[Pauli 1937 ]{pauli37}  Letter to Heisenberg, date: 5 Febr. 1937, in:%
\textit{\ Wolfgang Pauli. Wissenschaftlicher Briefwechsel}, \textit{\ \ \
vol. II, 1930-1939}, K. von Meyenn (ed.), Springer Verlag, Berlin, 1985, p.
512-514.

\bibitem[Raman 1928 ]{raman28}  C. V. Raman, A change of Wave-length in
Light Scattering, \textit{Nature}, vol. 121 (1928), p. 619, and \textit{%
Indian J. Phys}., vol. 2 (1928), p. 387.

\bibitem[Raman and Krishnan 1928 ]{ramakrish28}  C. V. Raman and K. S.
Krishnan, \textit{Nature}, vol. 121 (1928), p. 501, and p. 711.

\bibitem[Rivier and Stueckelberg 1948 ]{rivstuck48}  D$.$ Rivier and E. C.
G. Stueckelberg, A Convergent Expression for the Magnetic Moment of the
Neutron, \textit{Phys. Rev.} vol 74 (1948), pp. 2 and 218

\bibitem[Rivier 1949 ]{rivi49}  D$.$ Rivier, Une m\'{e}thode
d'\'{e}limination des infinit\'{e}s en th\'{e}orie des champs
quantifi\'{e}s. Application au moment magn\'{e}tique du neutron, \textit{\
Helv. Phys. Acta, }vol. 22 (1949), pp. 265-318.

\bibitem[Roqu\'{e} 1992]{roquŽ92}  X. Roqu\'{e}, M\o ller Scattering: a
Neglected Application of Early Quantum Electrodynamics, AHES, vol. 44
(1992), pp. 187-264

\bibitem[Sauter 1934]{saut34}  F. Sauter, Ueber die Bremsstrahlung schneller
Elektronen, \textit{Ann. d. Phys.}[5] vol. 20 (1934), pp. 404-412.

\bibitem[Schr\"{o}dinger 1926c ]{schroed26c}  E. Schr\"{o}dinger,
Quantisierung als Eigenwertproblem (Dritte Mitteilung), \textit{Ann. d.
Phys. }(4), vol. 80 (1926), pp. 437-490.

\bibitem[Schr\"{o}dinger 1926d ]{schroed26d}  E. Schr\"{o}dinger,
Quantisierung als Eigenwertproblem (vierte Mitteilung), \textit{Ann. d.
Phys. }(4), vol. 81 (1926), pp. 109-139.

\bibitem[Schr\"{o}dinger 1927]{schroed27}  E. Schr\"{o}dinger,Ueber den
Comptoneffekt, \textit{Ann. d. Phys. }(4), vol. 82 (1927), pp. 257-264.

\bibitem[Schweber 1961 ]{schweb61}  S. S. Schweber, \textit{Relativistic
Quantum Field Theory}, Harper and Row, New York, 1961.

\bibitem[Schweber 1994 ]{schweb94}  S. S. Schweber, \textit{QED and the Men
who made it}, Princeton Univ. Press, Princeton 1994.

\bibitem[Schwinger 1949 ]{schwing49}  J. Schwinger, On Radiative Corrections
to Electron Scattering, \textit{Phys. Rev}., vol. 75 (1949), pp. 898-899,
reprinted in \textit{Quantum electrodynamics}, J. Schwinger (ed.), Dover,
New York,1958, pp. 143-144.

\bibitem[Schwinger 1958]{schwing58}  J. Schwinger (ed), \textit{Quantum
electrodynamics}, Dover, New York,1958.

\bibitem[Slater 1924 ]{slat24}  J. C. Slater, Radiation and Atoms, \textit{\
Nature}, vol. 113 (1924), pp. 307-308.

\bibitem[Smekal 1923 ]{smek23}  A. Smekal, Zur Quantentheorie der
Dispersion, \textit{Naturw}., vol. 11 (1923), pp. 873-875.

\bibitem[Smorodinskii 1987 ]{smorod87}  Ya. S. Smorodinskii, Some episodes, 
\textit{Usp. Fiz. Nauk}, vol. 153, (1987), pp. 187-190. English translation
in \textit{Soviet Physics Uspekhy} vol. 30, (1987), pp. 823-825.

\bibitem[Sommerfeld 1916]{somm16}  A. Sommerfeld, Zur Theorie des
Zeeman-Effekts der Wasserstofflinien mit einem Anhang \"{u}ber den
Stark-Effekt, \textit{Phys. Zs}., vol. 17 (1916), pp. 309-325.

\bibitem[Sopka 1991 ]{sopka91}  K. Sopka, \textit{Ernst C. G. Stueckelberg
in the United States}, unpublished report.

\bibitem[Stueckelberg 1932]{stuck32}  E. C. G. Stueckelberg, Theorie der
unelastischen St\"{o}sse zwischen Atomen, \textit{Helv. Phys. Acta}, vol. 5
(1932), pp. 369-422.

\bibitem[Stueckelberg 1934 ]{stueck34}  E. C. G. Stueckelberg,
Relativistisch invariante St\"{o}rungstheorie des Diracschen Elektrons, 
\textit{Ann. d. Phys.}, vol. 21 (1934), pp. 367-389 and 744.

\bibitem[Stueckelberg 1935a ]{stueck35a}  E. C. G. Stueckelberg, Bemerkungen
zur Intensit\"{a}t der Streustrahlung bewegter freier Elektronen, \textit{\
Helv. Phys. Acta}, vol. 8 (1935), pp. 197-204.

\bibitem[Stueckelberg 1935b ]{stueck35b}  E. C. G. Stueckelberg, Remarques
sur la production des paires d'\'{e}lectrons, \textit{Tagung der
Schweizerischen Physikalischen Gesellschaft}, Lausanne, 4-5 mai 1935, 
\textit{Helv. Phys. Acta}, vol. 8 (1935), pp. 324-326.

\bibitem[Stueckelberg 1935c ]{stueck35c}  E. C. G. Stueckelberg, Remarque
\`{a} propos des temps multiples dans la th\'{e}orie d'interaction des
charges entre elles,\textit{\ C. R. SPHN (Gen\`{e}ve)}, vol. 52 (1935) pp.
98-101.

\bibitem[Stueckelberg 1936]{stueck36}  E. C. G. Stueckelberg, Invariante
St\"{o}rungstheorie des Elektron-Neutrino Teilchens unter dem Einfluss von
elektromagnetischem Feld und Kernkraftfeld (Feldtheorie der Materie II), 
\textit{Helv. Phys. Acta}, vol. 9 (1936), pp. 533-554.

\bibitem[Stueckelberg 1938]{stueck38}  E. C. G. Stueckelberg, Die
Wechselwirkungskr\"{a}fte in der Elektrodynamik und in der Feldtheorie der
Kernkr\"{a}fte (Teil I, II, III), \textit{Helv. Phys. Acta}, vol. 11 (1938),
pp. 225-244, 299-328.

\bibitem[Stueckelberg 1939 ]{stueck39}  E. C. G. Stueckelberg, Sur
l'int\'{e}gration de l'\'{e}quation ($\sum_{1}^{4}\partial %
_{x_{i}}^{2}-l^{2})Q=-\rho $\ en utilisant la m\'{e}thode de Sommerfeld, 
\textit{C. R. SPHN (Gen\`{e}ve)}, vol. 56 (1939), pp. 43-45.

\bibitem[Stuewer 1975 ]{stuewer75}  R. H. Stuewer, \textit{The Compton
Effect: Turning Point in Physics, }Science History Publications, New York,
1975.

\bibitem[Tamm 1930 ]{tamm30}  Ig. Tamm, Ueber die Wechselwirkung der freien
Elektronen mit der Strahlung nach der Diracschen Theorie des Elektrons und
nach der Quantenelektrodynamik, \textit{\textit{Zeit. f. Phys.}}, vol. 62
(1930), pp. 545-568.

\bibitem[van der Waerden 1968]{vanderwaer68}  B. L. van der Waerden (ed.), 
\textit{Sources of Quantum Mechanics}, Dover Publications inc., New York,
1968.

\bibitem[Waller 1929 ]{wall29}  I. Waller, Die Streuung kurzwelliger
Strahlung durch Atome nach der Diracschen Strahlungstheorie\textit{, \textit{%
\ \ Zeit. f. Phys.}}, vol. 58 (1929), pp. 75-94.

\bibitem[Waller 1930 ]{wall30}  I. Waller, Die Streuung von Strahlung durch
gebundene und freie Elektronen nach der Diracschen relativistischen Mechanik%
\textit{, \textit{Zeit. f. Phys.}}, vol. 61 (1930), pp. 837-851.

\bibitem[Wannier 1935 ]{wanni35}  G. Wannier, Eine vereinfachte Ableitung
der Klein-Nishina-Formel, \textit{Helv. Phys. Acta, }vol. 8 (1935), pp.
665-673.

\bibitem[Weisskopf 1934]{weissko34}  V. Weisskopf, \"{U}ber die
Selbstenergie des Elektrons, \textit{\textit{Zeit. f. Phys.}}, vol. 89
(1934), pp. 27-39, and Berichtigung, \textit{\textit{Zeit. f. Phys.}}, vol.
90 (1934), pp. 817-818.

\bibitem[Weisskopf 1981 ]{weissko81}  V. Weisskopf, The Development of Field
Theory in the last Fifty Years, \textit{Physics Today}, Nov. 1981, pp. 69-85.

\bibitem[Weisskopf 1983 ]{weissko83}  V. Weisskopf, Growing up with field
theory, the development of quantum electrodynamics, in: \textit{The birth of
particle physics } (International Symposium on the History of Physics,
Fermilab 1980), L. M. Brown and L. Hoddeson (eds), Cambridge Univ. Press,
Cambridge 1983.

\bibitem[v. Weizs\"{a}cker 1934 ]{vweiz34}  C. F. v. Weizs\"{a}cker,
Austrahlung bei St\"{o}ssen sehr schneller Elektronen, \textit{\textit{Zeit.
f. Phys.}}, vol. 88 (1934), pp. 612-625.

\bibitem[Wenger 1986]{wenger86}  R. Wenger, \textit{Ernst C. G. Stueckelberg
von Breidenbach. Etude biographique}, Biblioth\`{e}que Section de Physique,
Universit\'{e} de Gen\`{e}ve, 1986.

\bibitem[Wentzel 1925]{wentz25}  G. Wentzel, Die Theorie des
Compton-Effektes. I., \textit{Phys. Zeitschrift}, vol. 26 (1925), pp.
436-454.

\bibitem[Wentzel 1926]{wentz26}  G. Wentzel, Die mehrfach periodischen
Systeme in der Quantenmechanik, \textit{Zeit. f. Phys.}, vol. 37 (1926), pp.
80-94.

\bibitem[Wentzel 1927]{wentz27}  G. Wentzel, Zur Theorie des Comptoneffekts, 
\textit{Zeit. f. Phys}., vol. 43 (1927), pp. 1-8 and 779-787.

\bibitem[Wentzel 1929]{wentz29}  G. Wentzel, Ueber den R\"{u}ckstoss beim
Comptoneffekt am Wasserstoffatom, \textit{Zeit. f. Phys}., vol. 58 (1929),
pp. 348-367.

\bibitem[Wentzel 1933]{wentz33}  G. Wentzel, Ueber die Eigenkr\"{a}fte der
Elementarteilchen I und II, \textit{Zeit. f. Phys}., vol. 86 (1933), pp.
479-494 and 635-645.

\bibitem[Wentzel 1934a]{wentz34a}  G. Wentzel, Ueber die Eigenkr\"{a}fte der
Elementarteilchen III, \textit{Zeit. f. Phys}., vol. 87 (1934), pp. 726-733.

\bibitem[Wentzel 1934b]{wentzl34b}  G. Wentzel, Zur Frage der Aequivalenz
von Lichtquanten und Korpuskelpaaren, Zeit. f. Phys., vol 92 (1934), pp.
337-358.

\bibitem[Wentzel 1960]{wentz60}  G. Wentzel, Quantum Theory of Fields (until
1947), in \textit{Theoretical Physics in the Twentieth Century}, M. Fierz
and V. F. Weisskopf (eds), Interscience Publishers Inc. 1960, pp. 48-77.

\bibitem[Williams 1934 ]{williams34}  E. J. Williams, Nature of the High
Energy Particles of Penetrating Radiation and Status of Ionization and
Radiation Formula, \textit{Phys. Rev.}, vol 45 (1934), p. 729-730.

\bibitem[Wilson 1923 ]{wilson23}  C. T. R. Wilson, Investigation on X-rays
and Beta Rays by the Cloud Method. Part I: X-rays, \textit{Proc. Roy. Soc. A}
, vol. 104 (1923) pp. 1-24.

\bibitem[Winans and Stueckelberg 1928]{winstuck28}  J. G. Winans and E. C.
G. Stueckelberg, The Origin of the Continous Spectrum of the Hydrogen
Molecule, Nat. Acad. Sci., vol. 14 (1928), pp. 867-873.
\end{thebibliography}
\end{document}